\documentclass[10pt]{iopart}
\usepackage{cite}
\usepackage{graphicx}
\usepackage{bm}
\usepackage{subcaption}
\usepackage[width=170mm,top=12mm,bottom=10mm,bindingoffset=6mm]{geometry}
\usepackage{float}

\usepackage[symbol]{footmisc}

\usepackage{amsmath}

\newcommand{\gauss}{\text{gauss}}

\usepackage[dvipsnames]{xcolor}
\usepackage[linktocpage]{hyperref}
\hypersetup{
    colorlinks=true,
    linkcolor=BrickRed,     
    urlcolor=BrickRed,
    citecolor=BrickRed
    }
    
\usepackage{listings}
\usepackage{xcolor}
\definecolor{codegreen}{rgb}{0,0.6,0}
\definecolor{codegray}{rgb}{0.5,0.5,0.5}
\definecolor{codepurple}{rgb}{0.58,0,0.82}
\definecolor{backcolour}{rgb}{0.95,0.95,0.92}
\lstdefinestyle{mystyle}{
    backgroundcolor=\color{backcolour},   
    commentstyle=\color{codegreen},
    keywordstyle=\color{magenta},
    numberstyle=\tiny\color{codegray},
    stringstyle=\color{codepurple},
    basicstyle=\ttfamily\footnotesize,
    breakatwhitespace=false,         
    breaklines=true,                 
    captionpos=b,                    
    keepspaces=true,                 
    numbers=left,                    
    numbersep=5pt,                  
    showspaces=false,                
    showstringspaces=false,
    showtabs=false,                  
    tabsize=2
}
\makeatletter
\newrobustcmd{\fixappendix}{%
  \patchcmd{\l@section}{1.5em}{7em}{}{}%
  \patchcmd{\l@subsection}{2.3em}{7em}{}{}%
}
\makeatother

\begin{document}

\title[SymbolFit: Automatic Parametric Modeling with Symbolic Regression]{SymbolFit: Automatic Parametric Modeling with Symbolic Regression}

\author{Ho Fung Tsoi$^{1}$, Dylan Rankin$^{1}$, Cecile Caillol$^{2}$, Miles Cranmer$^{3}$, Sridhara Dasu$^{4}$, Javier Duarte$^{5}$, Philip Harris$^{6,7}$, Elliot Lipeles$^{1}$, Vladimir Loncar$^{6}$\footnote{Also at Institute of Physics Belgrade, Serbia}}

\address{$^{1}$University of Pennsylvania, USA \\
$^{2}$European Organization for Nuclear Research (CERN), Switzerland \\
$^{3}$University of Cambridge, UK \\
$^{4}$University of Wisconsin-Madison, USA \\
$^{5}$University of California San Diego, USA \\
$^{6}$Massachusetts Institute of Technology, USA \\
$^{7}$Institute for Artificial Intelligence and Fundamental Interactions, USA}

\ead{ho.fung.tsoi@cern.ch}
\vspace{3pt}


\begin{abstract}
\normalsize
We introduce $\tt{SymbolFit}$\footnote{An API is available at \url{https://github.com/hftsoi/symbolfit}}, a framework that automates parametric modeling by using symbolic regression to perform a machine-search for functions that fit the data while simultaneously providing uncertainty estimates in a single run.
Traditionally, constructing a parametric model to accurately describe binned data has been a manual and iterative process, requiring an adequate functional form to be determined before the fit can be performed.
The main challenge arises when the appropriate functional forms cannot be derived from first principles, especially when there is no underlying true closed-form function for the distribution.
In this work, we develop a framework that automates and streamlines the process by utilizing symbolic regression, a machine learning technique that explores a vast space of candidate functions without requiring a predefined functional form because the functional form itself is treated as a trainable parameter, making the process far more efficient and effortless than traditional regression methods.
We demonstrate the framework in high-energy physics experiments at the CERN Large Hadron Collider (LHC) using five real proton-proton collision datasets from new physics searches, including background modeling in resonance searches for high-mass dijet, trijet, paired-dijet, diphoton, and dimuon events.
We show that our framework can flexibly and efficiently generate a wide range of candidate functions that fit a nontrivial distribution well using a simple fit configuration that varies only by random seed, and that the same fit configuration, which defines a vast function space, can also be applied to distributions of different shapes, whereas achieving a comparable result with traditional methods would have required extensive manual effort.
\end{abstract}

\begin{figure}[H]
\centering
\includegraphics[width=1\textwidth]{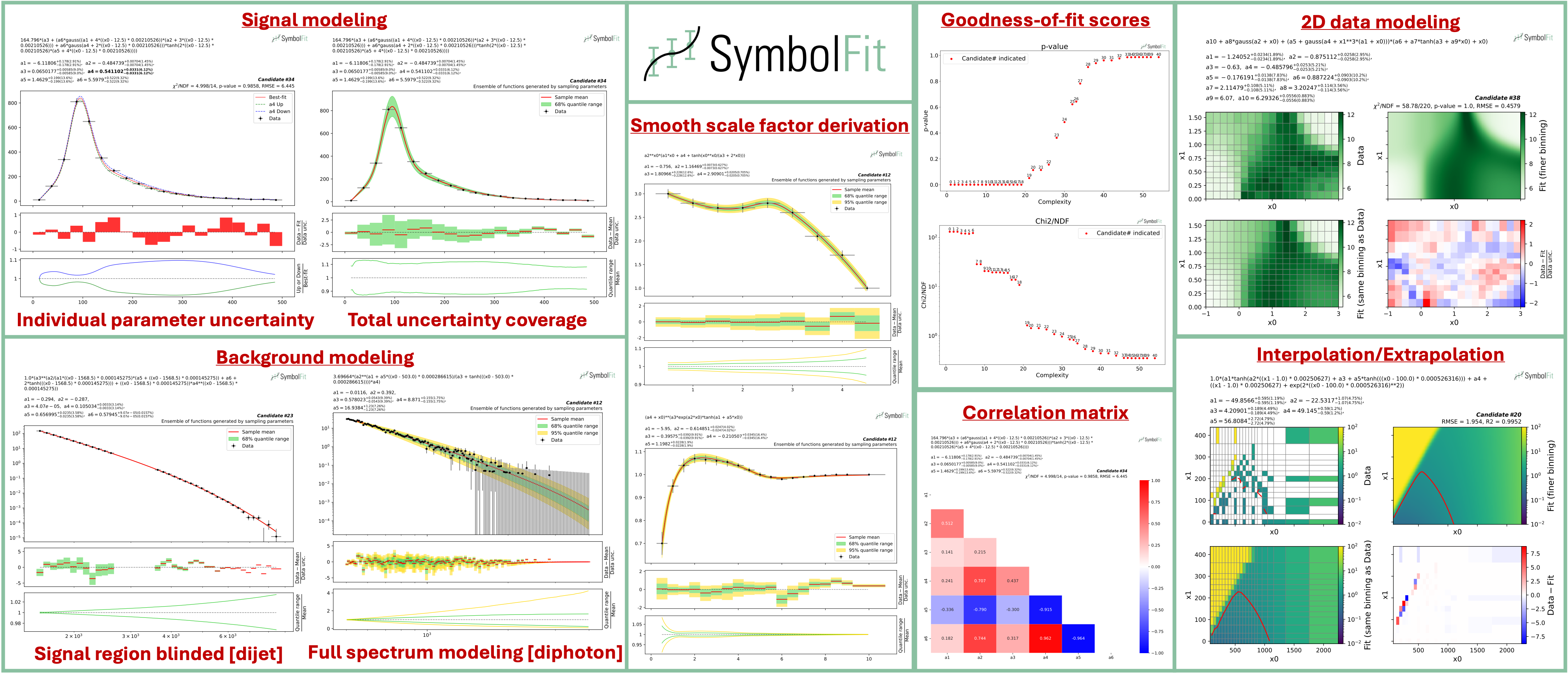}
\end{figure}

%
%
%
%
%

\clearpage

\tableofcontents

\section{Introduction}
\label{sec:introduction}

Traditional parametric modeling methods, such as polynomial regression, require specifying and fixing an adequate functional form before fitting the data.
Identifying suitable functional forms for distributions with arbitrary shapes is often challenging and time-consuming, as, in most cases, these functions cannot be derived from first principles and must be determined through trial and error.
Instead, symbolic regression (SR) is a more flexible and powerful technique that performs a $\textit{machine-search}$ for functions that best fit the data.
In SR, the functional form itself is treated as a trainable parameter that is dynamically adjusted throughout the fitting process, eliminating the need to predefine an exact function--an empirical task that is often challenging.
We refer the reader to Ref.~\cite{srbench,doi:10.1126/sciadv.aay2631,Keren_2023,lacava2021contemporarysymbolicregressionmethods,cranmer2023interpretable,Tsoi:2024ypg,Davis_2023,PhysRevC.108.L021901,Wadekar_2023,Lemos:2022cdj,Delgado:2021cuw,Tsoi:2023isc,Butter:2021rvz,Shao:2021qoa,Schmidt2009,gplearn,Operon,Virgolin_2021} for a review of the subject and some recent works.

\begin{figure}[!t]
     \centering
     \begin{subfigure}[b]{0.38\textwidth}
         \centering
         \includegraphics[width=\textwidth]{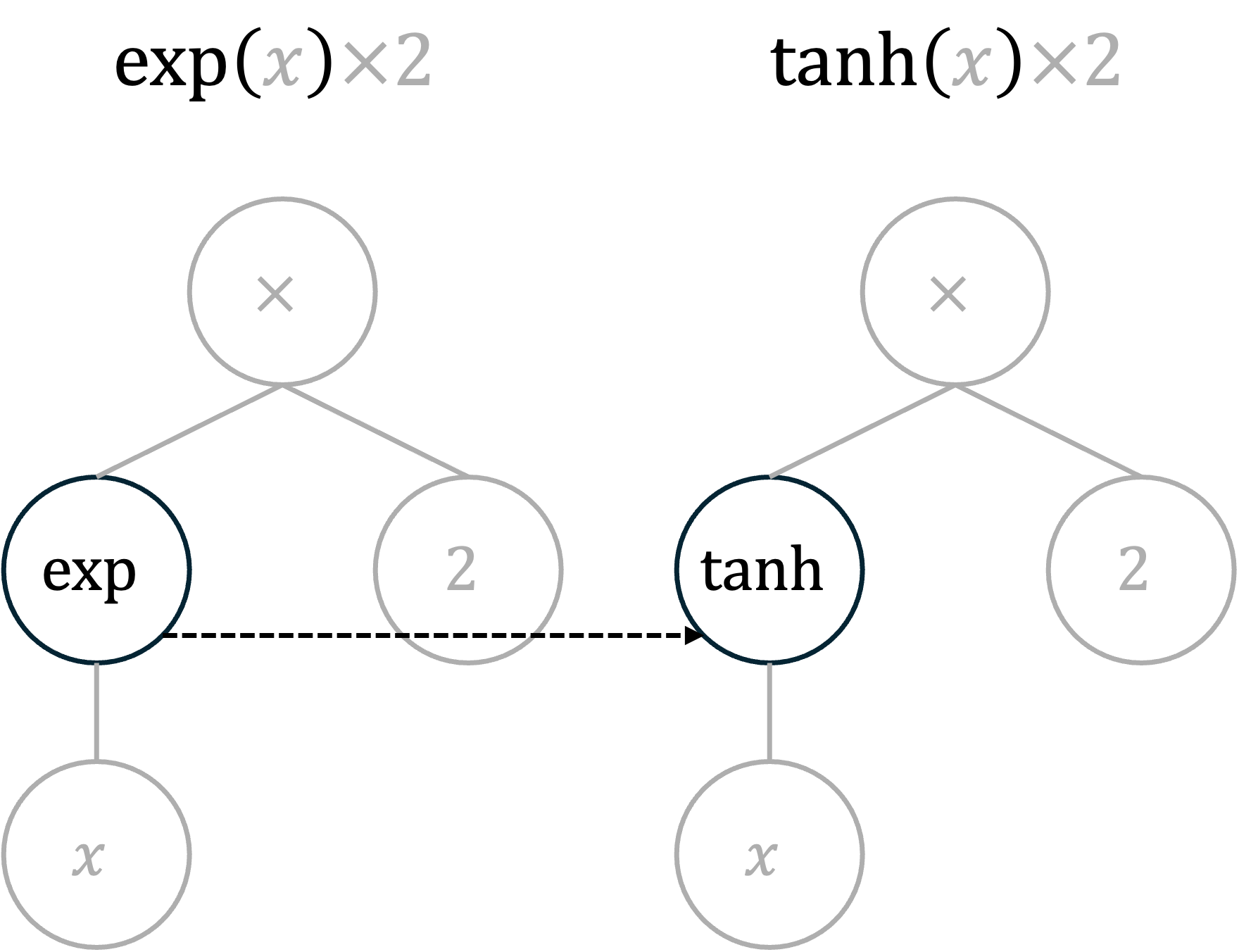}
         \caption{Node mutation.}
     \end{subfigure}
     \hspace{0.3cm}
     \begin{subfigure}[b]{0.58\textwidth}
         \centering
         \includegraphics[width=\textwidth]{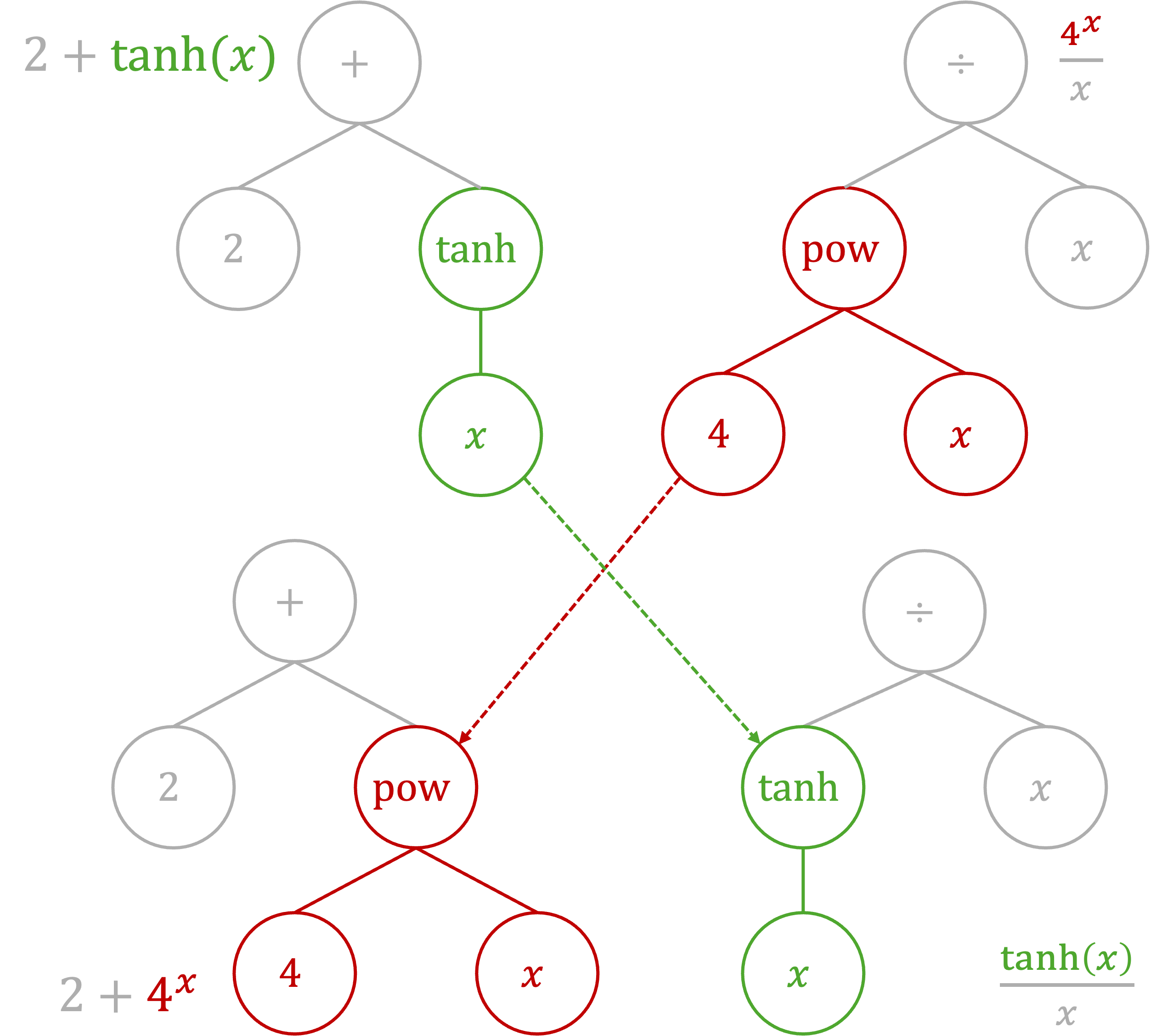}
         \caption{Subtree crossover.}
     \end{subfigure}\vspace{0.2cm}
        \caption{Genetic programming approach to symbolic regression. Functions are represented by expression trees. New functions are generated through mutation of tree nodes (left) and crossover between subtrees (right).
        }
        \label{fig:sr}
\end{figure}

Genetic programming~\cite{Koza1994} is a popular approach to SR~\cite{Schmidt2009,gplearn,Operon,Virgolin_2021,cranmer2023interpretable}.
In this approach, a function is represented as an expression tree, where the building blocks--mathematical operators, variables, and constants--are denoted as nodes, connected to represent their algebraic relations.
Different functional forms are generated through the evolution of these expression trees, where tree nodes are randomly selected and changed (mutation), and subtrees from different candidates are swapped to create new candidates (crossover), as illustrated in Fig.~\ref{fig:sr}.
As a result, the functional forms evolve during the fitting process, guiding the model toward convergence.
Instead of predefining the final functional form, SR algorithms based on genetic programming need far less prior knowledge about the functions themselves.
Only the constraints for constructing the expression trees need to be specified, such as the allowable mathematical operators ($+$, $\times$, $/$, $\text{pow}$, $\sin(\cdot)$, $\exp(\cdot)$, etc.).
This flexibility eliminates the need to know the exact fitting function beforehand or to fine-tune one empirically.

Our primary application focus is the intermediary stage of data analysis in high-energy physics (HEP) experiments at the CERN Large Hadron Collider (LHC), where parametric functions are constructed to model data distributions and subsequently used for downstream statistical inference.
In these analyses, uncertainty modeling is necessary, as the uncertainties associated with the parametric functions propagate to the final physics results.
Standard SR algorithms generate best-fit functions but do not inherently provide uncertainty estimates.
Our framework bridges this gap by automatically re-optimizing and estimating uncertainties for all candidate functions found by SR.
This functionality is critical, as parametric models without well-defined uncertainties cannot be used in the statistical inference workflows within HEP.
In the following section, we identify two analysis scenarios in which parametric modeling is traditionally used.
We discuss the limitations of current methods in these contexts and how SR, by eliminating the need to predefine a functional form, can offer a more flexible and efficient alternative.

We also emphasize that, unlike SR applications in other domains, where the primary goal is often interpretability, such as direct extraction of physical laws from data (e.g., \cite{cranmer2023interpretable,Lemos:2022cdj}), our objective is different.
We focus on constructing valid and flexible parametric models for statistical inference, rather than deriving interpretable expressions to uncover underlying physical laws from data.
Our goal is to model arbitrary distribution shapes, as commonly encountered in LHC analyses, where no single underlying physical law governs the data.
In such cases, interpretability is not a relevant requirement.
Instead, the key criteria for a suitable parametric function in this setting are 1) it should smoothly and accurately describe the shape of the distribution, and 2) its associated uncertainty should be well-behaved and capable of capturing the uncertainty in the data.

\section{Challenges in traditional methods}

\subsection{Scenario 1: signal and background modeling}
\label{sec:scenario1}
When analyzing proton-proton collision data at the LHC in the search for new physics signatures, the data are typically binned and presented as histograms representing physical observables, such as the invariant mass of the final-state particles.
Each bin records the observed or expected number of collision events with mass values within that range.
To search for new physics signatures, which are often hypothesized as narrow and small peaks over a smoothly falling background in the invariant mass distribution, parametric functions are required to model both the signal and background based on these binned distributions.
These models are then used to perform hypothesis testing.
    
In the traditional approach to parametric modeling, one typically relies on manually guessing the appropriate family of functions that might describe the shape of the distribution.
Although these distributions often represent physical observables, they are usually obtained after applying a series of selection cuts on various variables, which can introduce arbitrary shape effects into the final distributions being modeled.
As a result, these distributions generally do not have a known underlying true function, making it impossible to derive a suitable functional form from first principles and leaving empirical constructions as the only option.
In some cases, when a suitable function cannot be found to describe the distribution after many trials, one is forced to compromise by adjusting the analysis strategy, say splitting the main distribution into multiple sub-distributions and fitting them separately, which could lead to a more complicated combined likelihood function.
This empirical approach has been the standard strategy in the HEP community, requiring significant manual effort to craft a candidate function and iteratively fine-tune it.
    
For example, a search for new physics in high-mass trijet events performed by the CMS experiment~\cite{trijet} modeled the background by fitting the trijet invariant mass distribution, $m_{\text{jjj}}$, using three families of empirical functions.
One of these functions takes the form:
\begin{equation}
    \label{eq:trijet-n}
    f(x;N)=\frac{p_0(1-x)^{p_1}}{x^{\sum_{i=2}^{N}p_i\log^{i-2}(x)}},
\end{equation}
where $x=m_{\text{jjj}}/\sqrt{s}$ is a dimensionless variable ($\sqrt{s}$ is the center-of-mass energy of the collisions), $p_{i}$ are free parameters, and $N$ is a hyperparameter for the function form.
The function was fitted multiple times with different trial $N$ values, and the optimal value was determined through a separate statistical test, such as an F-test~\cite{fisher}.

Note that the functional form in Eq.~\ref{eq:trijet-n} was constructed empirically, rather than derived from first principles, to reproduce the observed spectrum.
This reflects the fact that the trijet distribution arises from events that have passed through multiple stages selection, including triggering, reconstruction, and optimization, rather than being determined by a single underlying physics law.
The same is true for other similar analyses at the LHC.

The challenge lies in the need to empirically craft a specific functional form, such as Eq.~\ref{eq:trijet-n}, for each individual distribution.
These empirical functions are tailored to the particular distribution being fitted, making them rigid and potentially ineffective if there are slight changes in the data.
For instance, variations in final-state objects, event selection strategies, or detector conditions during data collection can all introduce arbitrary modifications in the shape of the distribution.
In such cases, function families that worked for past datasets may no longer be effective for future datasets, even within the same analysis channel, and an empirical searching for suitable functions must be repeated.

Analyses at the LHC have traditionally relied on this empirical fitting method when modeling signal and background processes from binned data.
Examples include the milestone analyses that led to the discovery of the Higgs boson in 2012~\cite{ATLAS:2012yve,CMS:2012qbp,CMS:2013btf}, as well as some recent results from CMS searches for high-mass resonances in dijet~\cite{dijet}, paired-dijet~\cite{paired-dijet}, trijet~\cite{trijet}, diphoton~\cite{diphoton}, and dimuon~\cite{dimuon} events.

In this context, SR has the potential to transform the approach to parametric modeling.
By conducting a $\textit{machine-search}$ for suitable functional forms, SR significantly reduces the manual effort required in the modeling process, providing a more efficient and adaptive alternative to traditional methods.

\subsection{Scenario 2: derivation of smooth descriptions from binned data}

When predicting signal and background processes using simulation, there is always some degree of mismatch with the observed data, which may result from inaccuracies in theoretical predictions, mis-modeling of detector effects, or measurements errors.
These discrepancies are corrected by applying data-to-simulation scale factors (measured from isolated control regions) to the simulated events, ensuring that the simulation provides a more realistic representation of the observed data.
Examples include jet energy scale corrections parameterized by jet $p_{\text{T}}$ and $\eta$\footnote{Common coordinate system used to define particle kinematics in collider physics: $p_{\text{T}}$ is transverse momentum and $\eta$ is the pseudorapidity angle.}~\cite{CMS:2016lmd}, heavy-flavor jet tagging efficiency corrections parameterized by the jet $p_{\text{T}}$ and $\eta$~\cite{CMS:2017wtu}, hadronic tau identification efficiency corrections parameterized by the tau $p_{\text{T}}$, $\eta$, and decay modes~\cite{CMS:2018jrd}.

These scale factors are typically derived from binned data and applied as binned weights, resulting in coarse-grained corrections.
When smooth scale factors are desired, the process often follows the same empirical approach as described in Sec.~\ref{sec:scenario1}, facing the same limitations discussed earlier.
In cases where the scale factor is parameterized by more than one variable, it becomes even more challenging to empirically construct an adequate functional form, forcing one to rely on coarse-grained corrections.

Another common scenario involves data-driven background estimation methods, where transfer factors are derived to estimate events in the signal region based on those in the sideband region.
For example, in a search for a boosted Higgs boson decaying to b quarks performed by the CMS experiment~\cite{htobb}, the QCD multijet background was estimated from observed data, where the transfer factor was parameterized and empirically constructed as the sum of products of Bernstein polynomials:
\begin{equation}
    f(x_0,x_1)=\bigg(\sum_{\mu=0}^{n_{x_0}}\sum_{\nu=0}^{n_{x_1}}a_{\mu,\nu}b_{\mu,n_{x_0}}(x_0)b_{\nu,n_{x_1}}(x_1)\bigg)\times g(x_0,x_1),
\end{equation}
where $b_{n.N}(x)$ is the $n$-th Bernstein basis polynomial of degree $N$, $a_{\mu,\nu}$ are parameters to be extracted from a fit to observed data, and $g(x_0,x_1)$ is a function fitted separately to simulated events.
The degrees of the Bernstein polynomials, $n_{x_0}$ and $n_{x_1}$, are determined separately using an F-test.

By using SR, these empirical steps for deriving smooth scale factors can be significantly simplified into a single SR fit, without knowledge of the final functional form.

\subsection{An alternative method: Gaussian process regression}

An alternative fitting method is Gaussian process regression (GPR), which has been explored for these scenarios~\cite{Frate:2017mai,Gandrakota:2022wyl,Xu:2901208}.
GPR models the dependent variable as following a Gaussian distribution at each point along the independent variable.
The smoothness of the probability function is controlled by a chosen covariance kernel between bins.
As a result, GPR provides a probabilistic prediction, yielding both a smooth mean function and a variance function, which define a very generic distribution of functions that describe the data instead of a single exact function.

Fitting a GPR model to $n$ data points requires inverting an $n\times n$ covariance matrix, which scales with a time complexity of $\mathcal{O}(n^3)$~\cite{Swiler_2020}.
This can become computationally prohibitive, especially for datasets with more than one independent variable.
Additionally, integrating Bayesian GPR outputs into the standard HEP search framework requires subtle treatments~\cite{Frate:2017mai}, whereas SR directly provides explicit function templates that can be straightforwardly integrated into existing workflows.

Despite the potential of alternative methods like GPR, but due to the limitations described above, the empirical method remains the primary approach to parametric modeling within the HEP community.
There is currently a lack of an efficient framework or a package based on an alternative method that can be readily used out-of-the-box.

\section{Proposed solution with symbolic regression}

For the scenarios discussed above, we propose using SR to replace traditional methods, shifting the paradigm of parametric modeling in HEP.

In this paper, we introduce a Python API\footnote{\url{https://github.com/hftsoi/symbolfit}} for the $\tt{SymbolFit}$ framework, which interfaces with $\tt{PySR}$~\cite{cranmer2023interpretable} (a high-performance SR library) and $\tt{LMFIT}$~\cite{newville_2015_11813} (a nonlinear least-square minimization library), aimed at $\textit{automating}$ parametric modeling of binned data using SR.
We demonstrate the effectiveness of the framework in two common HEP applications: parametric modeling of signal and background, and the derivation of smooth scale factors.
These applications are validated using five real datasets from new physics searches at the CERN LHC, along with several toy datasets.
The key features of the framework are summarized below.
\begin{itemize}
    \item \textbf{Pre-determined functional forms are no longer needed.}
    With SR, only minimal $\textit{constraints}$ are required to define the function space, such as specifying the allowed mathematical operators ($+$, $\times$, $/$, $\text{pow}$, $\sin(\cdot)$, $\exp(\cdot)$, etc.).
    This does not demand extensive and prior knowledge of the final functions that describe the distribution.
    The search for suitable functions is automatically performed by machine, eliminating the empirical and manual process.
    We show that a simple SR configuration can flexibly fit a wide variety of distribution shapes.

    \item \textbf{Generating multiple candidate functions per fit.}
    SR based on genetic programming generates and evolves successive generations of functions, producing a batch of candidate functions in each search iteration.
    The same search configuration can be repeated with different random seeds to explore different suitable functions from the vast function space.
    This flexibility in generating a variety of candidate functions allows for adaptability across different downstream tasks.

    \item \textbf{Inclusion of uncertainty measure.}
    While SR algorithms alone are dedicated to function searching and do not inherently provide any uncertainty estimation, our framework bridges the gap by incorporating a re-optimization process for the candidate functions.
    This step improves the best-fit models and generates uncertainty estimation needed to access the modeling reliability.
    
    \item \textbf{Modeling of multi-dimensional data.}
    The framework easily accommodates modeling data with multiple variables, which is particularly useful in HEP scenarios where scale factors are sometimes parameterized by more than one variable.

\end{itemize}

Moreover, the framework is designed to automate the process as much as possible, minimizing manual effort.
Results are evaluated and plotted automatically, which are saved in readable formats such as CSV and PDF files, including diagnostic plots that allow users to visually assess the fit quality and data comparison.
The candidate functions generated can be seamlessly integrated into downstream statistical inference tools commonly used in HEP, such as $\tt{Combine}$~\cite{CMS:combine} and $\tt{pyhf}$~\cite{pyhf,pyhf_joss}, as the output models are in identical representation to those from traditional methods--closed-form functions.
The efficiency of SR in generating a wide range of well-fitted functions per fit also allows flexible modeling as the function choice can be treated as a source of systematic uncertainty through the discrete profiling method~\cite{Dauncey:2014xga}.

The rest of the paper is structured as follows.
Sec.~\ref{sec:method} describes the $\tt{SymbolFit}$ framework.
Sec.~\ref{sec:demonstration} presents demonstrations using a real LHC dataset as well as several toy datasets.
Sec.~\ref{sec:summary} provides a summary of the work.
More demonstrations are presented in \ref{apd:more-examples}.

\section{Method}
\label{sec:method}

\begin{figure}[!t]
\centering
\includegraphics[width=1\textwidth]{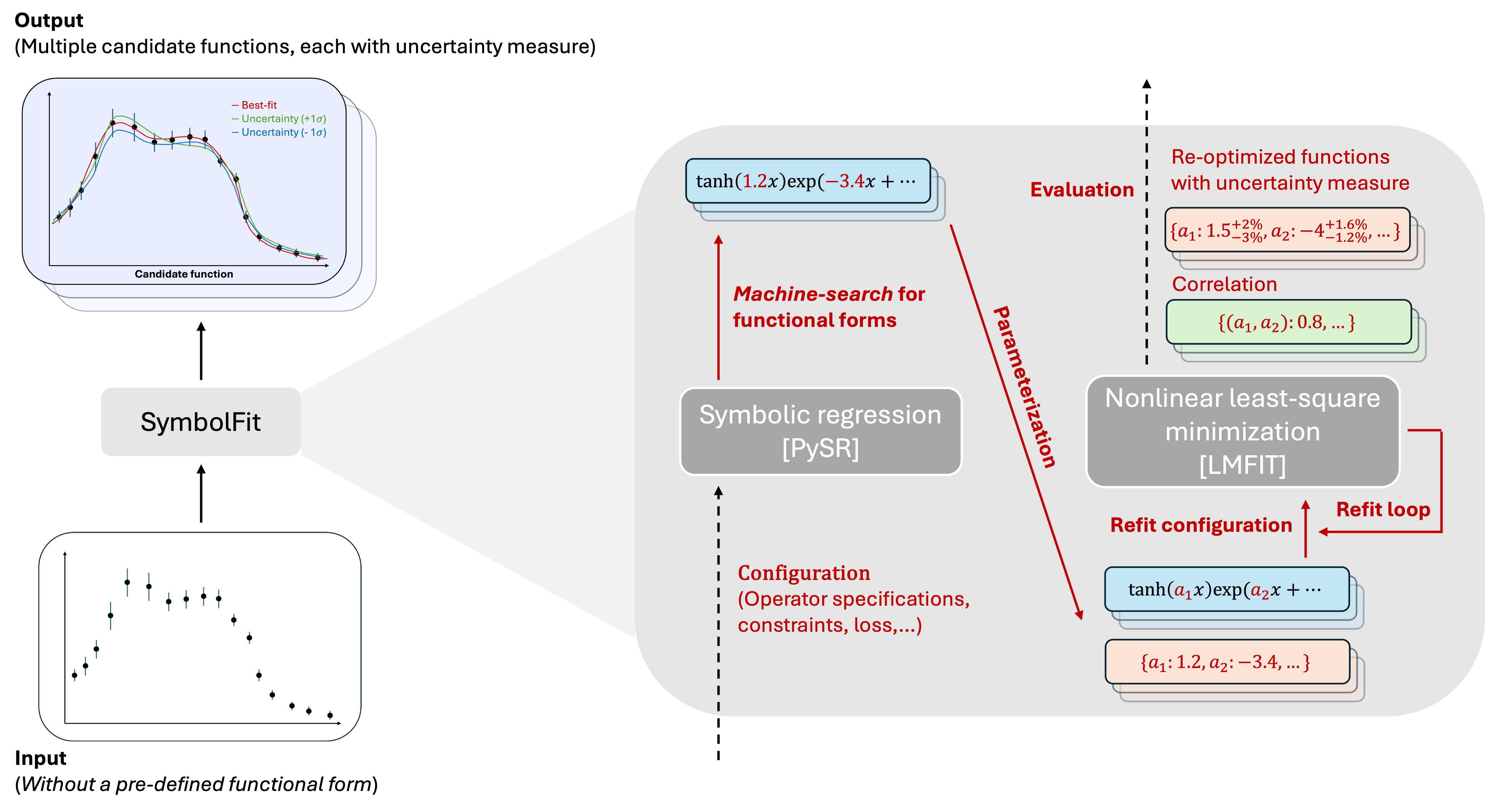}
\caption{A schematic sketch of the internal steps within the $\tt{SymbolFit}$ framework illustrates how it interfaces with $\tt{PySR}$~\cite{cranmer2023interpretable} and $\tt{LMFIT}$~\cite{newville_2015_11813} to automate parametric modeling using SR.
The process begins with an input dataset that does not require a predefined functional form.
Functional forms are generated using SR, parameterized, and then re-optimized through standard nonlinear least-square minimization.
The output is a batch of candidate functions, each with associated uncertainty estimates.}
\label{fig:schematic}
\end{figure}

The $\tt{SymbolFit}$ framework is illustrated in Fig.~\ref{fig:schematic} and explained in the following.
\begin{enumerate}
    \item \textbf{Input data.}
    We consider the input dataset $\{(\bm{x}^{i}, y^{i}, y^{i}_{\text{up}}, y^{i}_{\text{down}})\}_{i=1}^{n}$, where $\bm{x}^{i}$ represents one or more independent variables, $y^{i}$ is the dependent variable with associated uncertainties $y^{i}_{\text{up/down}}$ at one standard deviation, and $n$ is the number of data points.
    
    In the context of binned histograms, which are commonly used in HEP data analysis, there are $n$ bins.
    Here, $\bm{x}^{i}$ represents the center of the $i$-th bin, and $y^{i}$ is the bin content, representing the number of events within the bin.
    The associated uncertainties $y^{i}_{\text{up/down}}$ account for measurement errors or modeling inaccuracies.
    
    \item \textbf{Symbolic regression.}
    The core of the framework is to leverage SR to perform a $\textit{machine-search}$ for suitable functions to model the data, without predefining a functional form.
    We utilize $\tt{PySR}$~\cite{cranmer2023interpretable}, a Python library for genetic programming-based SR, which is highly configurable in defining the function space for the search.
    The configuration process is highly simplified, requiring only the specification of allowed mathematical operators ($+$, $\times$, $/$, $\text{pow}$, $\sin(\cdot)$, $\exp(\cdot)$, etc.) and the constraints for the functional form.
    The objective of the search is to minimize:
    \begin{equation}
        \label{eq:chi2}
        \chi^2 \equiv \sum_{i=1}^{n} \bigg(\frac{f(\bm{x}^{i}) - y^i}{y_{\text{up}}^{i}\mathbf{1}_{f(\bm{x}^{i}) - y^i\geq 0} + y_{\text{down}}^{i}\mathbf{1}_{f(\bm{x}^{i}) - y^i< 0}}\bigg)^2,
    \end{equation}
    where $f$ is the candidate function.
    Since $\tt{PySR}$ uses a multi-population strategy to evolve and select functions, each run generates a batch of candidate functions.
    These functions are then re-optimized in subsequent steps to improve the fit and provide uncertainty estimates.
    
    \item \textbf{Parameterization.}
    SR algorithms search for exact functions but do not inherently provide any uncertainty measures.
    However, uncertainty estimation is essential in HEP data analysis to gauge the reliability of the observation and prediction.
    To address this, we freeze the functional forms found by SR and then re-optimize all constants in each function using standard nonlinear minimization techniques.
    The uncertainties in these re-optimized constants are used as the uncertainty measure for the candidate functions.
    
    First, within each candidate function, the constants are automatically identified and parameterized as $\{a_1,a_2,...\}$, with the original values stored as initial values for the re-optimization process.
    
    \item \textbf{Re-optimization fit (ROF).}
    To perform ROF of the candidate functions, we utilize $\tt{LMFIT}$~\cite{newville_2015_11813}, a nonlinear least-square minimization library, to perform a second-fit for the parameters while keeping the functional forms fixed.
    The objective is to minimize $\chi^2$ defined in Eq.~\ref{eq:chi2}.
    The parameterized functions are parsed to identify the set of parameters to be varied, and initially, all parameters are allowed to vary in the fit.
    
    In some cases, the minimization may fail to converge due to a too complex objective function.
    To handle these cases, a loop for ROF is implemented in the framework.
    This loop progressively reduces the number of degrees of freedom (NDF) by freezing more parameters to their initial values until the fit succeeds and all relative errors are below a pre-defined threshold.

    Finally, the candidate functions are evaluated and ranked in the outputs.

\end{enumerate}

To summarize, $\tt{SymbolFit}$ automates all these steps in the modeling process, including the computation of various goodness-of-fit scores and the evaluation of correlations between the parameters.
This integration streamlines the workflow and minimizes manual intervention while providing full information for downstream statistical analysis.
The computation time of the workflow is primarily due to the search for functional forms, for which we utilize a highly optimized SR algorithm $\tt{PySR}$.
As a result, the process is not computationally intensive and can be flexibly configured.
As the function space is usually huge even when under constrained, one can repeat the fit with the same configuration but with a different random seed to obtain a different batch of candidate functions.

\section{Demonstrations}
\label{sec:demonstration}

We demonstrate the effectiveness of our framework using five real LHC datasets from new physics searches published recently, as well as several toy datasets.

The LHC datasets consist of real proton-proton collision data at a center-of-mass energy of $\sqrt{s}=13$ TeV, collected by the CMS experiment during Run 2.
These datasets cover various search channels: dijet~\cite{dijet} in Sec.~\ref{sec:dijet}, diphoton~\cite{diphoton} in \ref{sec:diphoton}, trijet~\cite{trijet} in \ref{sec:trijet}, paired-dijet~\cite{paired-dijet} in \ref{sec:paired-dijet}, and dimuon~\cite{dimuon} in \ref{sec:dimuon}.
Each dataset consists of 1D binned data of the invariant mass of the respective objects, where smooth background predictions are obtained through parametric modeling and then tested for excess events indicative of new physics. 
In all these analyses, CMS reported no evidence of new physics is observed in the data.
Therefore, for our demonstrations, we assume that each invariant mass spectrum contains no signal.
We also perform experiments to validate the SR outputs for signal extraction in these LHC datasets, and details of these steps are given in Sec.~\ref{sec:dijet}.

In addition to the real LHC datasets, we also generate toy datasets for demonstration purposes.
While SR has been shown to successfully identify the correct underlying function from noisy data~\cite{srbench,cranmer2023interpretable}, here we focus on toy data generated by hand without an underlying function to illustrate SR's capability in modeling arbitrary distribution shapes.
Toy Dataset 1, presented in Sec.~\ref{sec:toy-dataset-1}, is a 1D binned distribution featuring a sharp peak and a high-end tail.
Toy Dataset 2, presented in Sec.~\ref{sec:toy-dataset-2}, consists of three 1D binned distributions with various shapes.
Toy Dataset 3, presented in Sec.~\ref{sec:toy-dataset-3}, consists of three 2D binned distributions.

\subsection{Toy Dataset 1 (1D) [signal modeling]}
\label{sec:toy-dataset-1}
\begin{figure}[!t]
\centering
\includegraphics[width=0.5\textwidth]{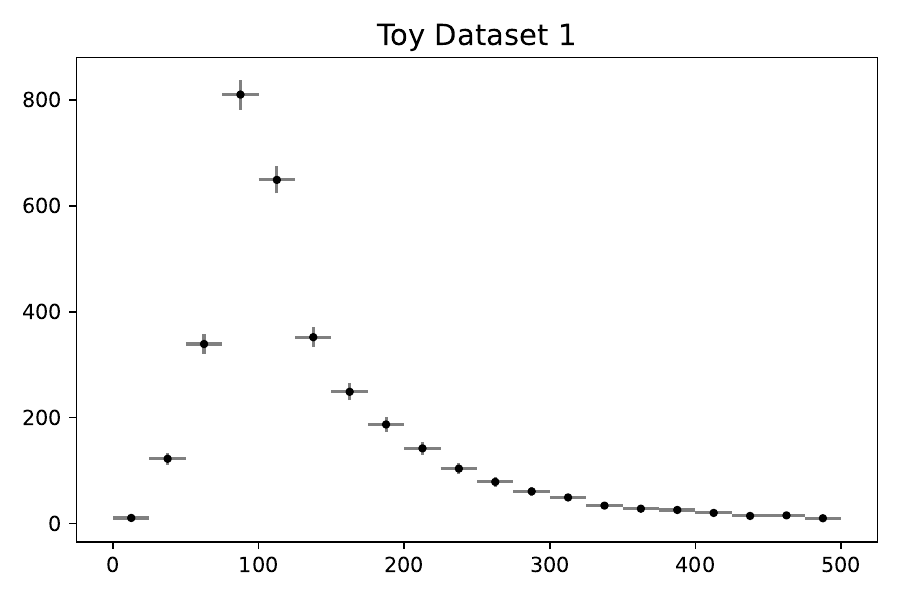}
\caption{Toy Dataset 1: a 1D binned dataset with uncertainties represented by vertical error bars. The data points are manually generated without reference to an underlying function.}
\label{fig:toy_dataset_1_data}
\end{figure}

Fig.~\ref{fig:toy_dataset_1_data} shows Toy Dataset 1, which consists of a 1D binned distribution.
The distribution has a shape characterized by a sharp peak and a high-end tail, and it is generated by hand without reference to an underlying function.

In spite of the lack of a true functional form, we demonstrate that our framework using SR can replace the empirical process with minimal efforts.
This can be seen from the simple $\tt{Python}$ snippet shown in List.~\ref{config-toy1}\footnote{The option definitions can be found at \url{https://github.com/MilesCranmer/PySR} and Ref.~\cite{cranmer2023interpretable}.}, which defines the function space and configures $\tt{PySR}$ to perform a $\textit{machine-search}$ for functions to fit this dataset.
In this example, the maximum function complexity is set to 60 to constrain the model size, ensuring that the total number of nodes in an expression tree does not exceed 60, with each operator equally weighted with a complexity of one.
The allowed operators include two binary operators ($+$ and $\times$) and three unary operators ($\exp(\cdot)$, $\tanh(\cdot)$, and a custom-defined $\text{gauss}(\cdot)\equiv\exp(-(\cdot)^2)$).
Constraints on operator nesting are imposed to prohibit scenarios like $\tanh(\tanh(\cdot))$.
The loss function used is $\chi^2$.

\begin{lstlisting}[language=Python, caption=The $\tt{Python}$ code snippet that configures $\tt{PySR}$ to search for candidate functions for Toy Datasets 1 and 3., label=config-toy1]
from pysr import PySRRegressor
import sympy

pysr_config = PySRRegressor(
    model_selection = "accuracy",
    timeout_in_seconds = 60*100,
    niterations = 200,
    maxsize = 60,
    binary_operators = ["+", "*"],
    unary_operators = [
        "exp",
        "gauss(x) = exp(-x*x)",
        "tanh"
    ],
    nested_constraints = {
        "tanh":   {"tanh": 0, "exp": 0, "gauss": 0, "*": 2},
        "exp":    {"tanh": 0, "exp": 0, "gauss": 0, "*": 2},
        "gauss":  {"tanh": 0, "exp": 0, "gauss": 0, "*": 2},
        "*":      {"tanh": 1, "exp": 1, "gauss": 1, "*": 2}
    },
    extra_sympy_mappings={"gauss": lambda x: sympy.exp(-x*x)},
    loss = "loss(y, y_pred, weights) = (y - y_pred)^2 * weights"
)
\end{lstlisting}

\begin{figure}[!t]
     \centering
     \begin{subfigure}[b]{0.45\textwidth}
         \centering
         \includegraphics[width=\textwidth]{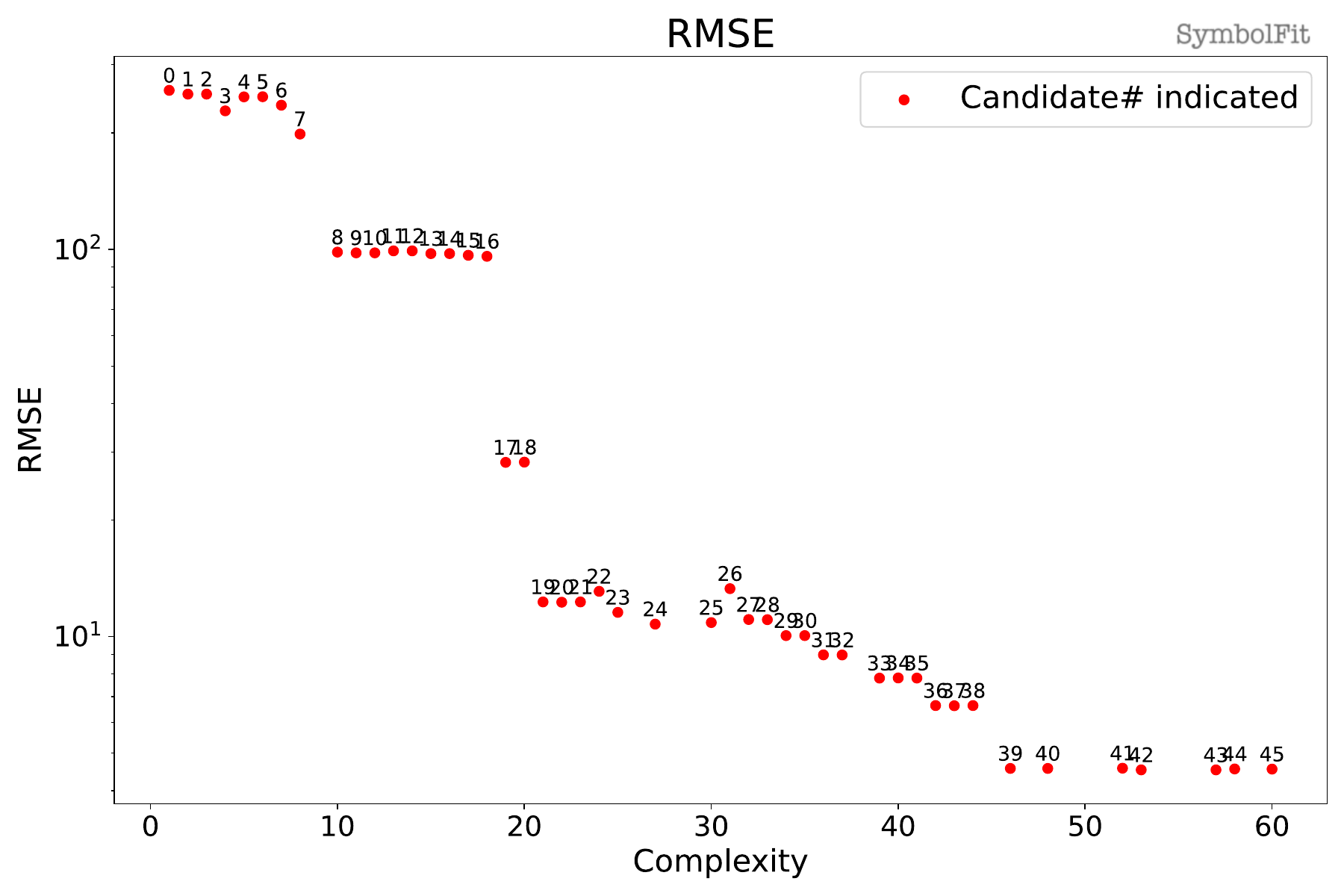}
         \caption{Root-mean-square error (RMSE).}
     \end{subfigure}
     \begin{subfigure}[b]{0.45\textwidth}
         \centering
         \includegraphics[width=\textwidth]{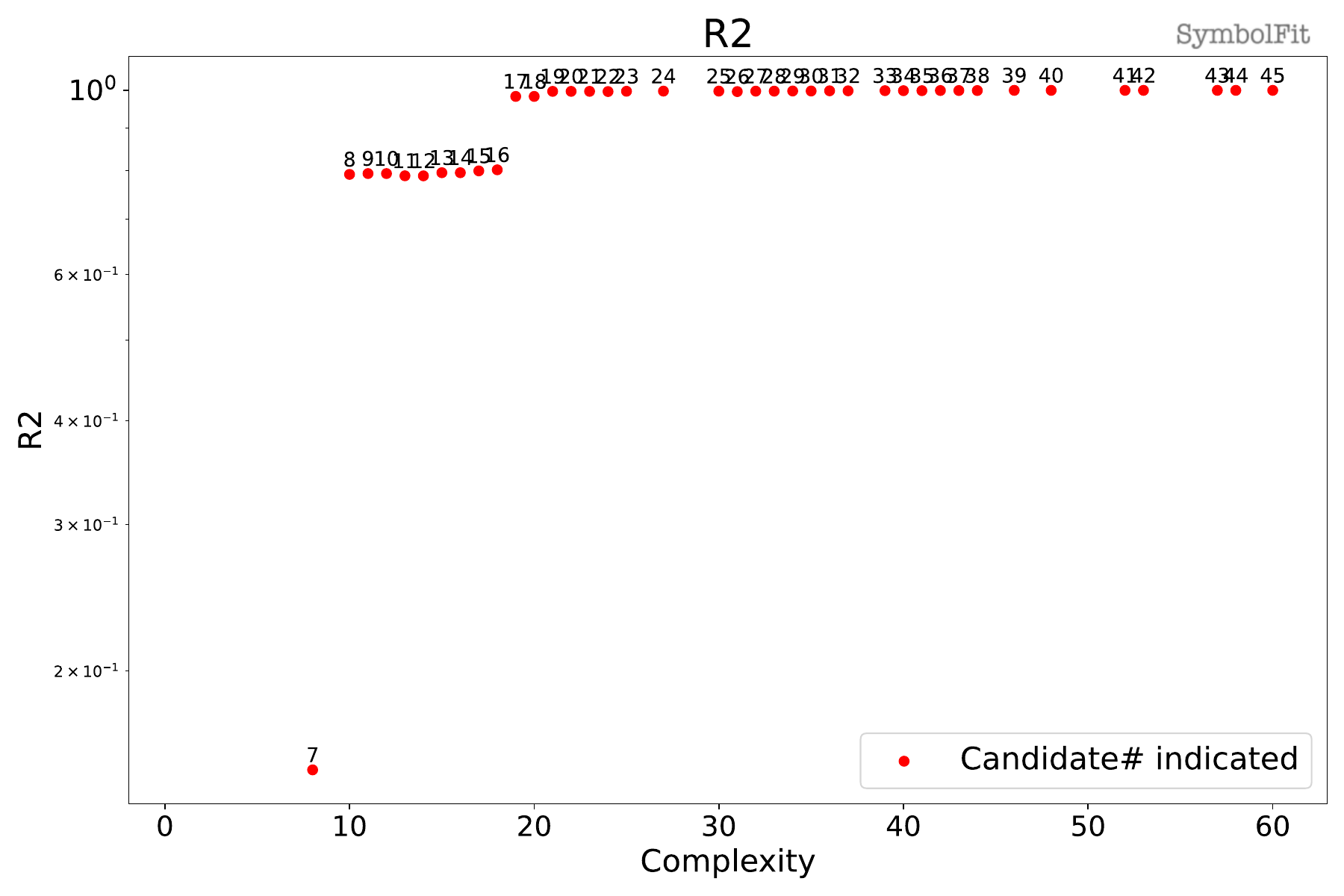}
         \caption{Coefficient of determination (R2).}
     \end{subfigure}\\\vspace{0.5cm}
     \begin{subfigure}[b]{0.45\textwidth}
         \centering
         \includegraphics[width=\textwidth]{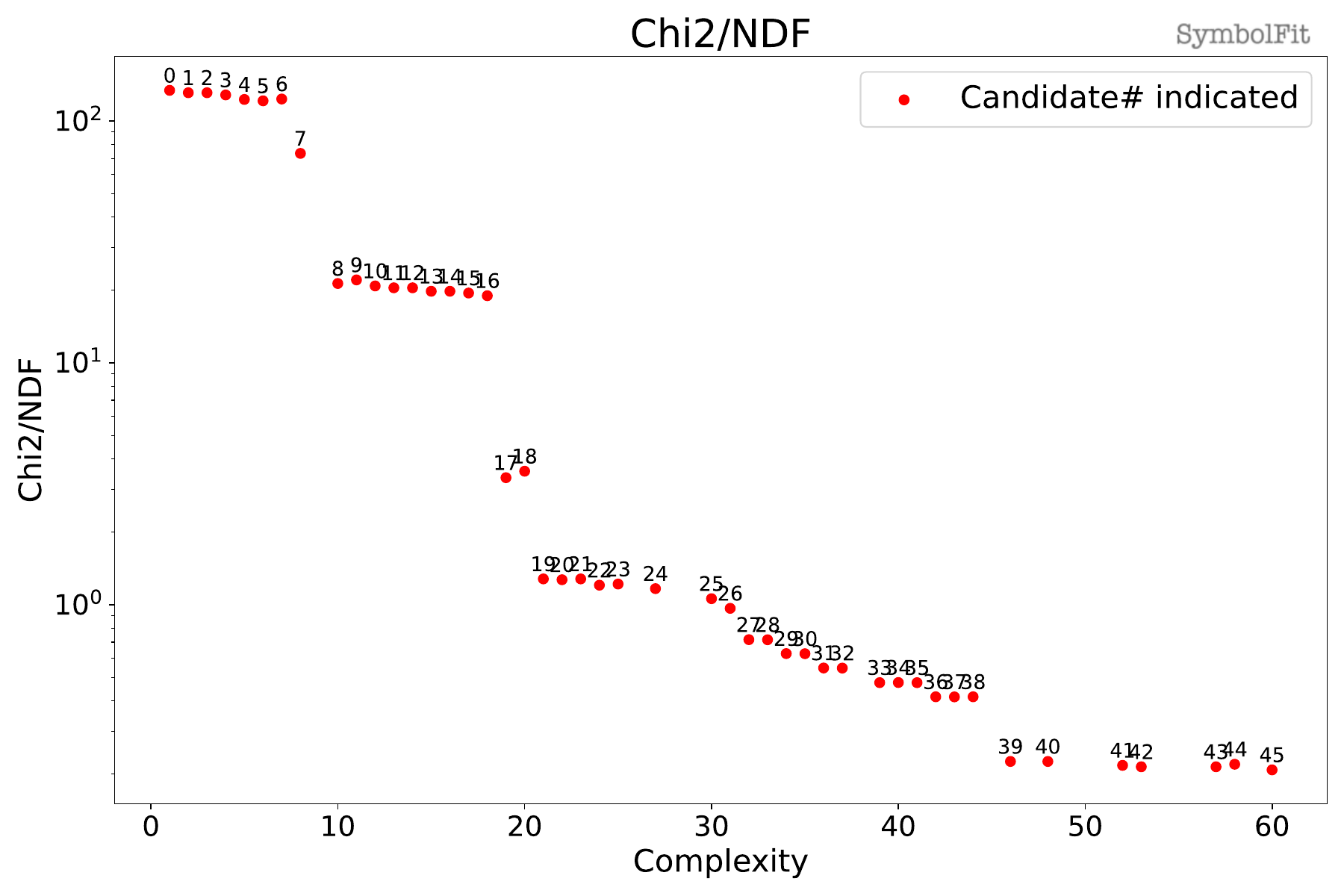}
         \caption{$\chi^2/\text{NDF}$.}
     \end{subfigure}
     \begin{subfigure}[b]{0.45\textwidth}
         \centering
         \includegraphics[width=\textwidth]{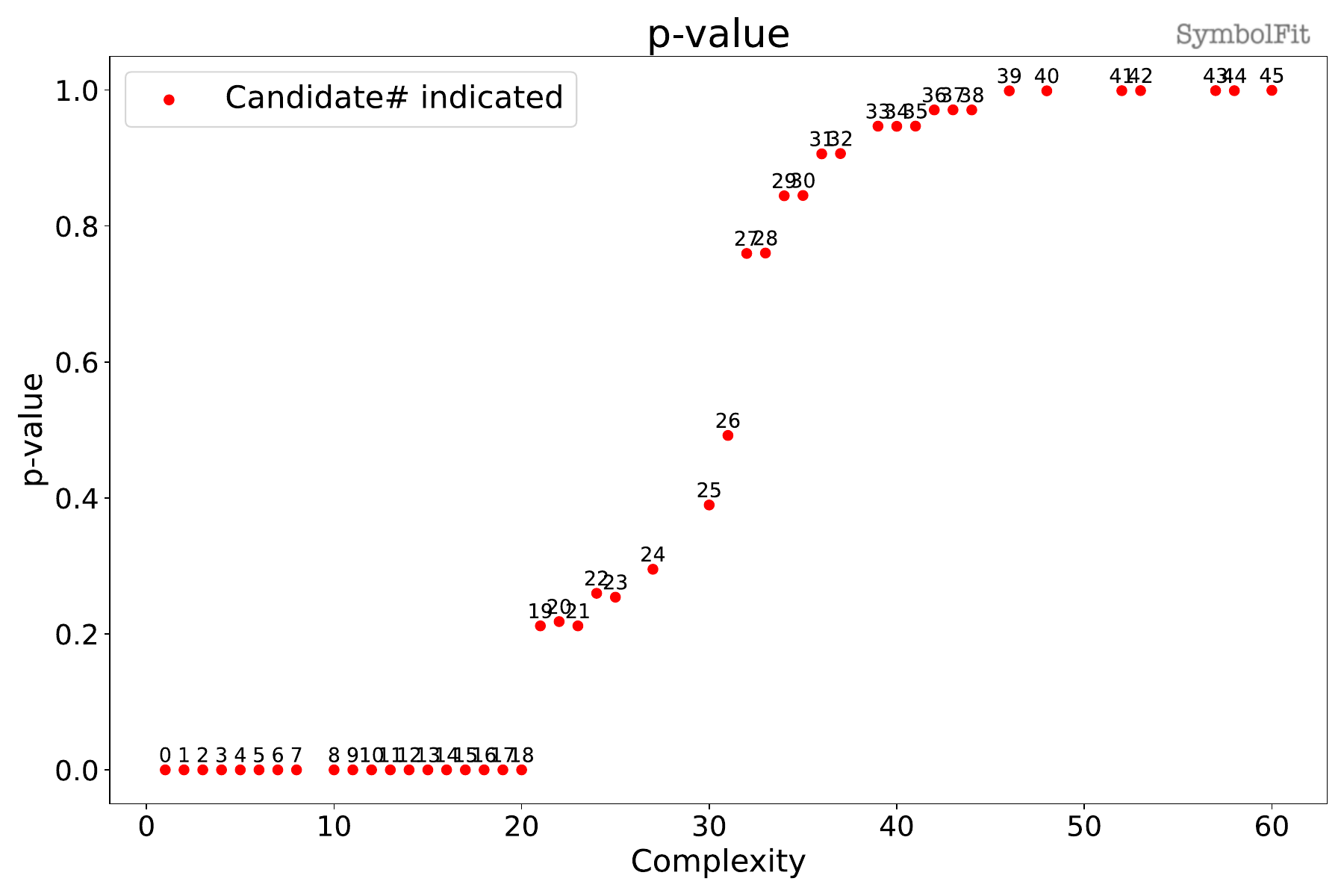}
         \caption{p-value.}
     \end{subfigure}\vspace{0.5cm}
     
    \caption{Goodness-of-fit scores vs. function complexity.
    A total of 46 candidate functions (labeled \#0--\#45) were obtained from a single fit on Toy Dataset 1 in Sec.~\ref{sec:toy-dataset-1}.
    Candidate functions \#10, \#17, \#27, and \#38 are listed in Tab.~\ref{tab:toy-dataset-1}, and their combined uncertainty coverage is presented in Fig.~\ref{fig:toy_dataset_1_sampling}.
    Specifically, individual parameter variations for candidate function \#38 are shown in Fig.~\ref{fig:toy_dataset_1_candidate}, with its parameter correlation matrix shown in Fig.~\ref{fig:toy_dataset_1_candidate_correlation}.}
    \label{fig:toy_dataset_1_gof}
\end{figure}

We run $\tt{SymbolFit}$ with the $\tt{PySR}$ configuration shown in List.~\ref{config-toy1}.
In particular, the parameter ``maxsize'' controls the maximum complexity of the functions, which is up to the user to choose if simpler or more complex functions are desired.
This single fit generates 46 candidate functions (labeled \#0 to \#45), with function complexity values ranging from 1 to 60.
These complexity values provide a rough estimate of the model size and are computed before any algebraic simplification.
Four goodness-of-fit scores, including the $\chi^2/\text{NDF}$ and p-value, are plotted against function complexity in Fig.~\ref{fig:toy_dataset_1_gof}.
As expected, more complex functions tend to offer better fits to the data, as their higher complexity makes them more expressive.
The variety of functions produced provides flexibility, allowing one to choose a candidate function based on the desired fit quality for downstream tasks.
For example, nine of the 46 candidate functions are listed in Tab.~\ref{tab:toy-dataset-1}, showing $\chi^2/\text{NDF}$ values above 1, around 1, and below 1, offering a range of fit qualities

In general, the $\chi^2/\text{NDF}$ score improves significantly after ROF compared to the original function found from SR.
This is because SR algorithms are typically focused on finding optimal functional forms rather than fine-tuning a specific function.
As a result, the ROF step improves the constants in each function to achieve a better fit and provides uncertainty estimates for the parameters.

\begin{figure}[!t]
     \centering
     \begin{subfigure}[b]{0.44\textwidth}
         \centering
         \includegraphics[width=\textwidth]{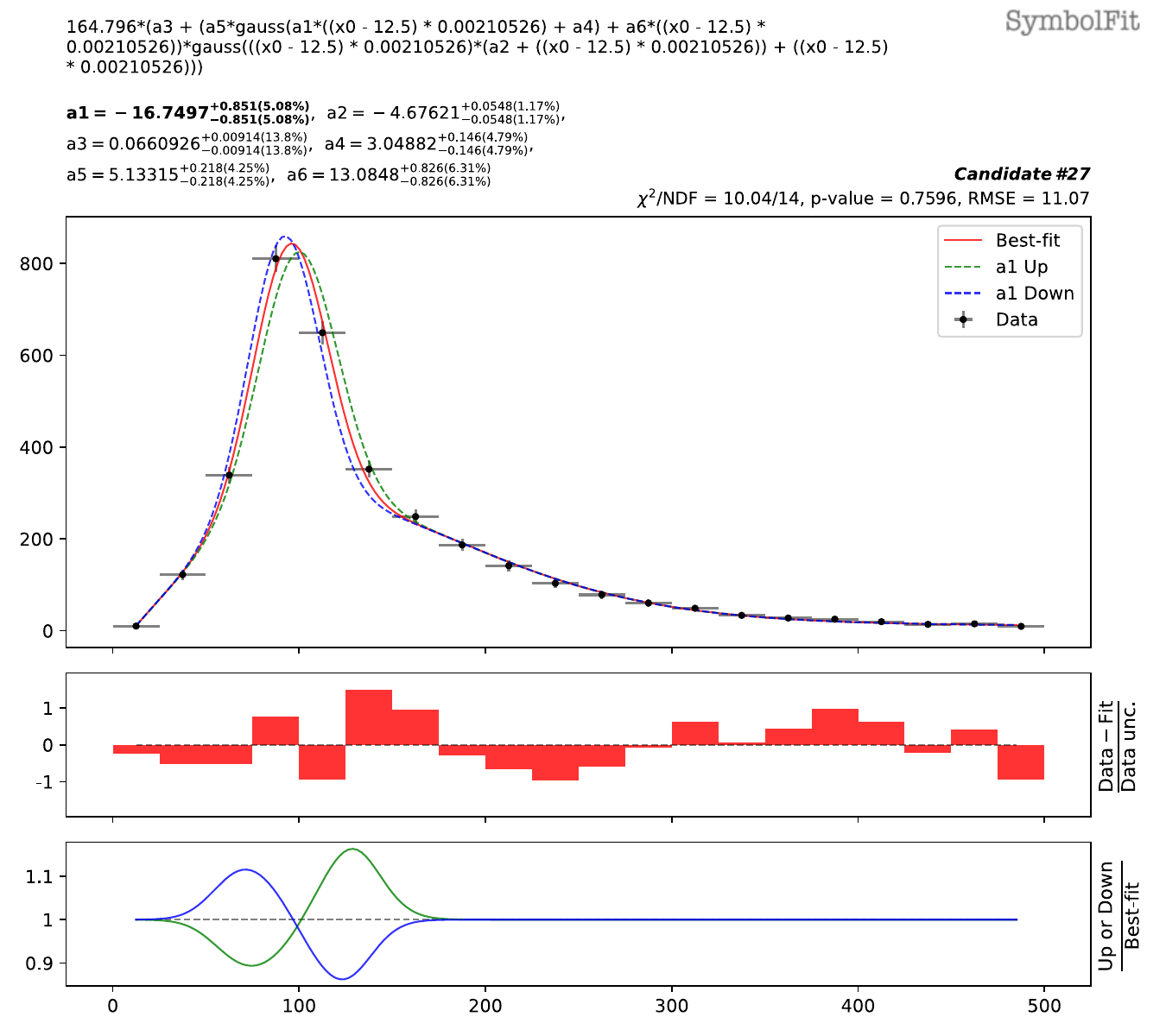}
         \caption{$\pm1$ std. variations of $a_1$.}
     \end{subfigure}
     \begin{subfigure}[b]{0.44\textwidth}
         \centering
         \includegraphics[width=\textwidth]{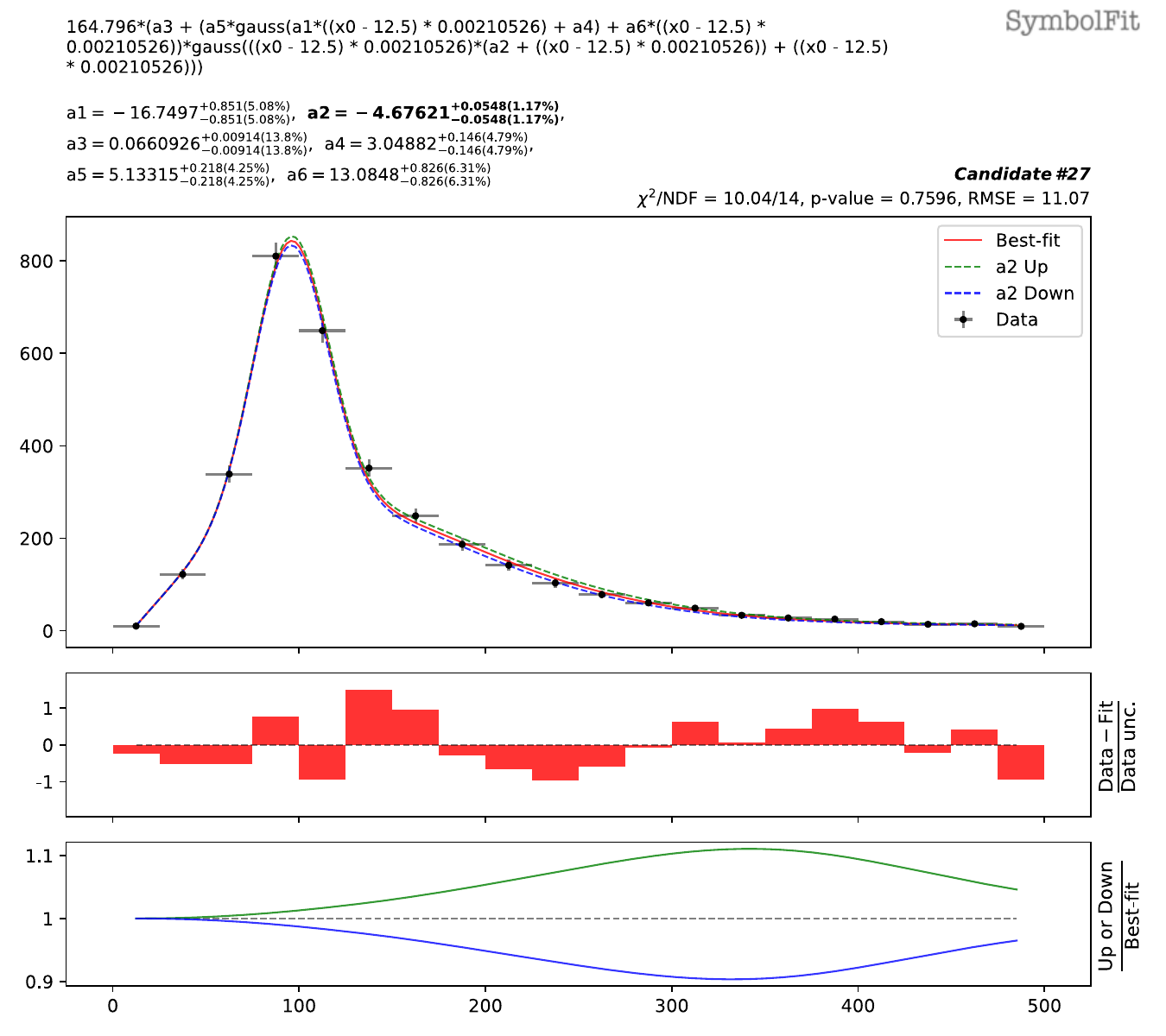}
         \caption{$\pm1$ std. variations of $a_2$.}
     \end{subfigure}\\\vspace{0.3cm}
     \begin{subfigure}[b]{0.44\textwidth}
         \centering
         \includegraphics[width=\textwidth]{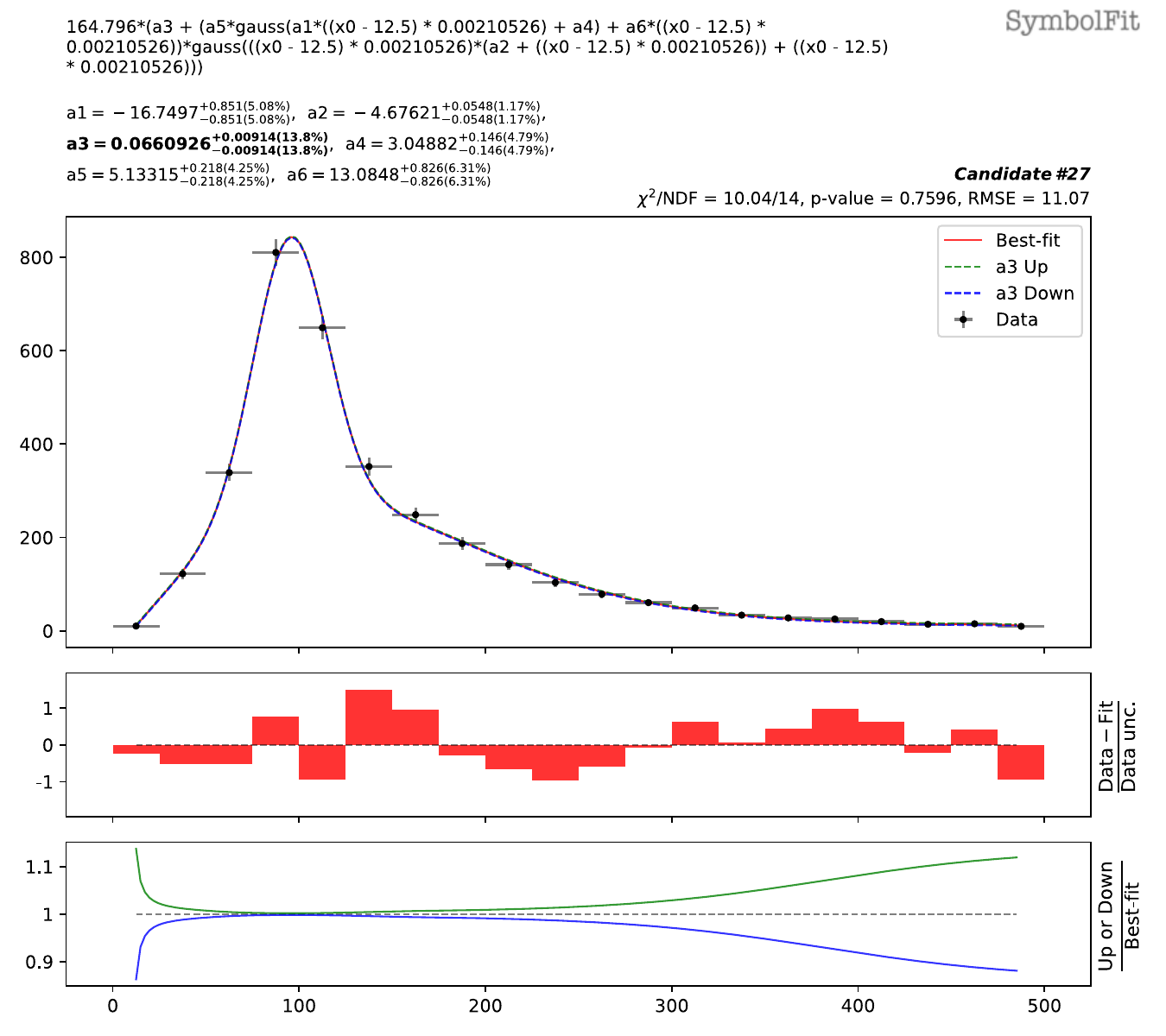}
         \caption{$\pm1$ std. variations of $a_3$.}
     \end{subfigure}
     \begin{subfigure}[b]{0.44\textwidth}
         \centering
         \includegraphics[width=\textwidth]{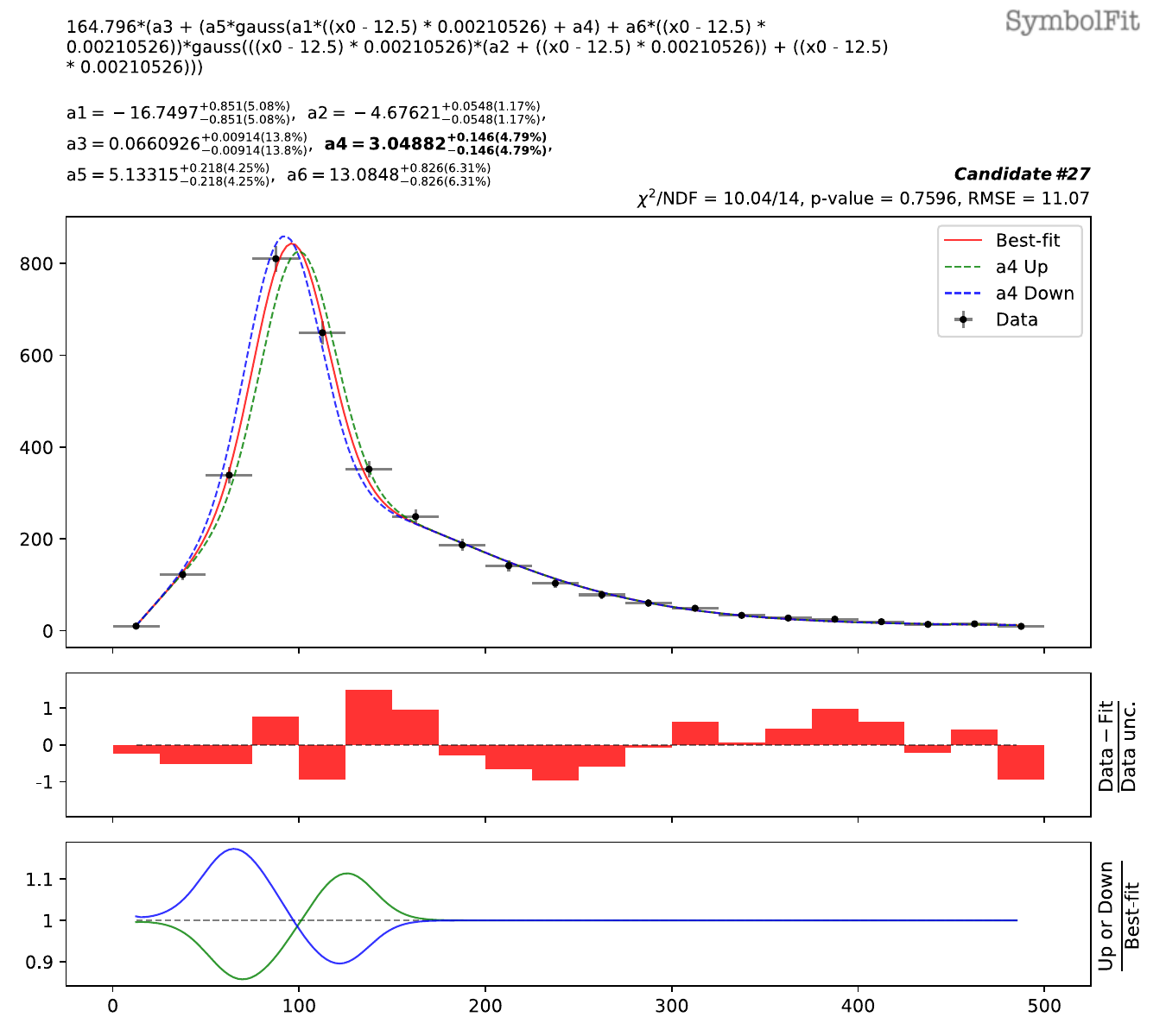}
         \caption{$\pm1$ std. variations of $a_4$.}
     \end{subfigure}\\\vspace{0.3cm}
     \begin{subfigure}[b]{0.44\textwidth}
         \centering
         \includegraphics[width=\textwidth]{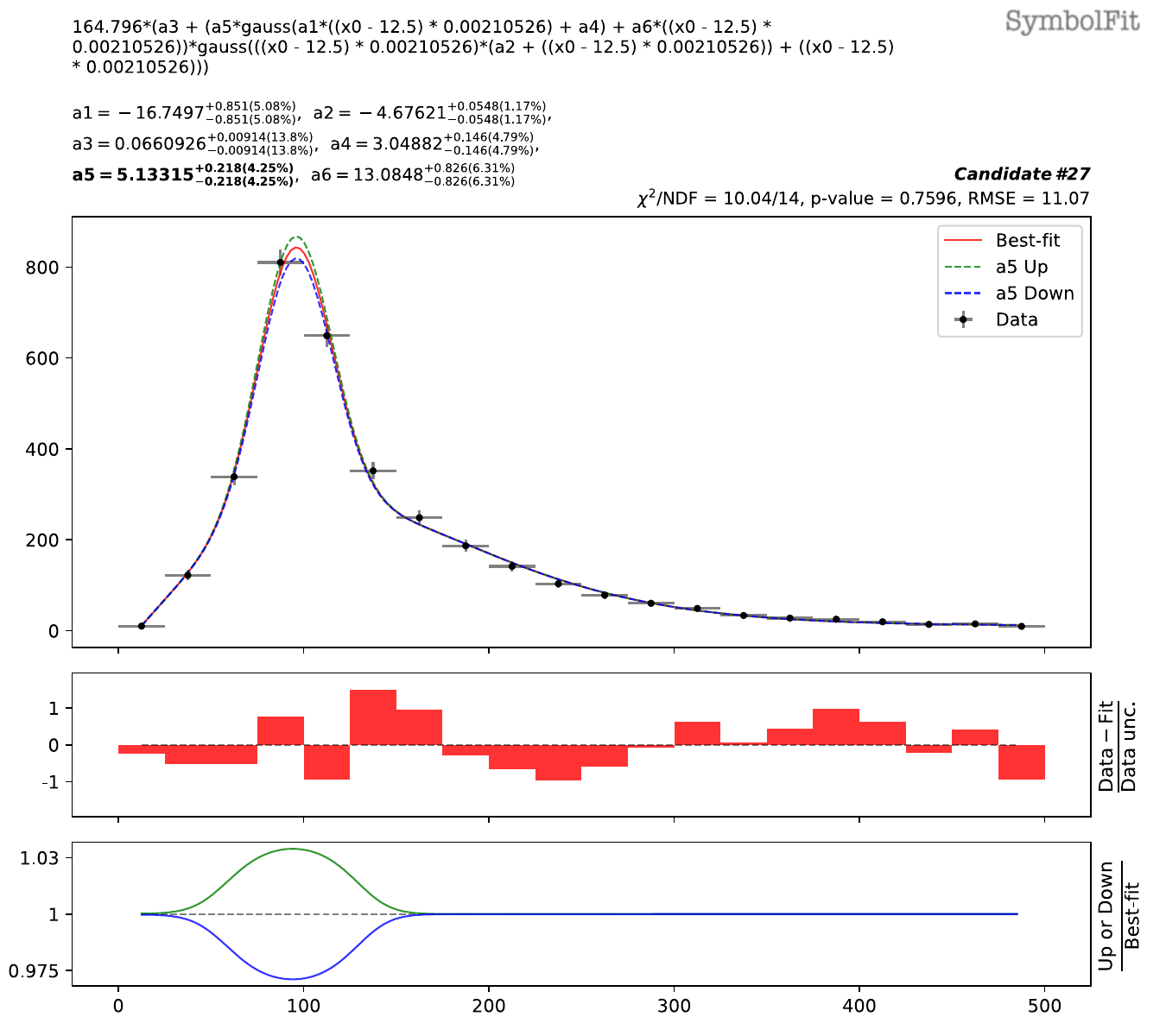}
         \caption{$\pm1$ std. variations of $a_5$.}
     \end{subfigure}
     \begin{subfigure}[b]{0.44\textwidth}
         \centering
         \includegraphics[width=\textwidth]{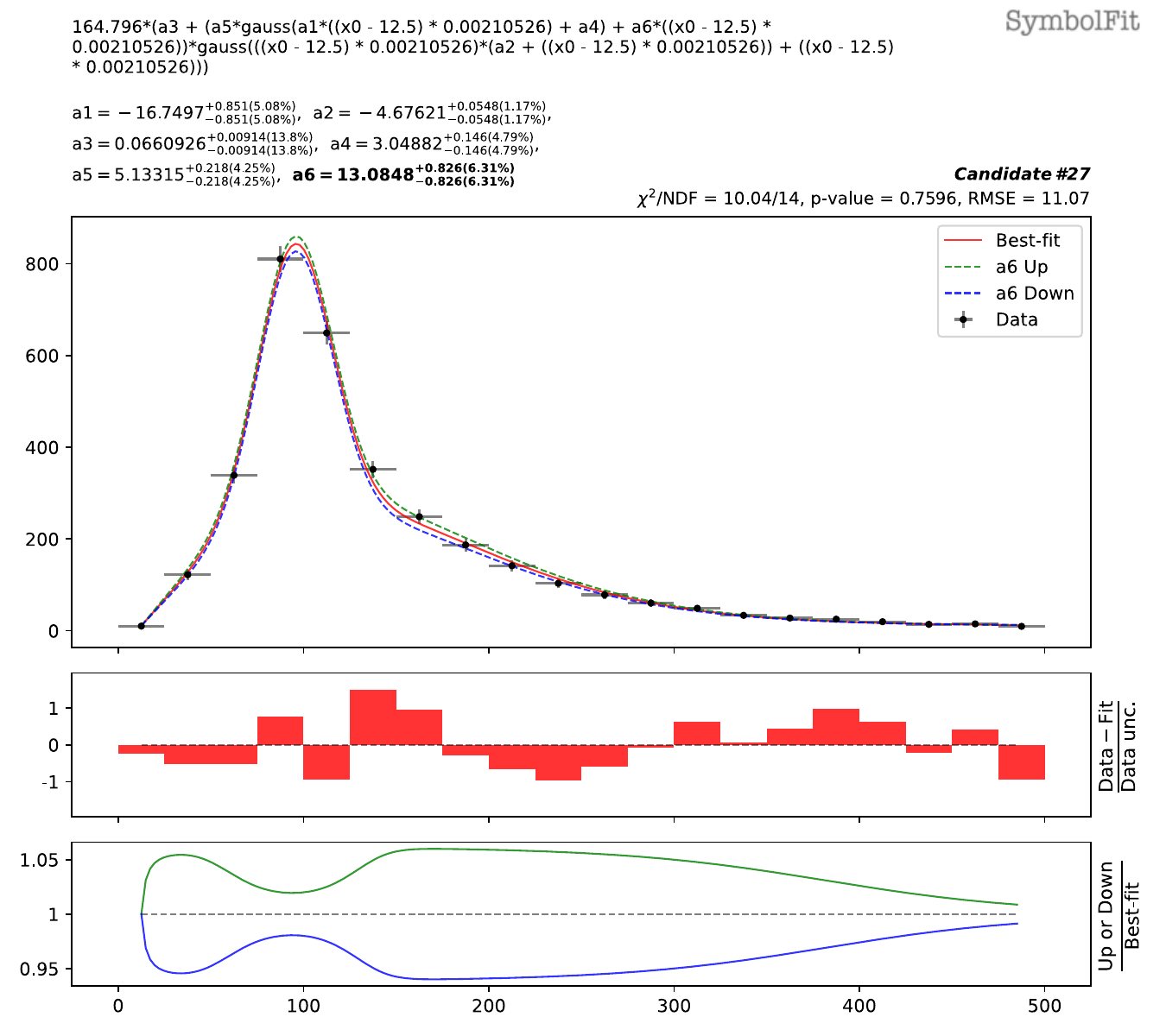}
         \caption{$\pm1$ std. variations of $a_6$.}
     \end{subfigure}
        \caption{Individual parameter variations in candidate function \#27 from a fit to Toy Dataset 1.
        The parameterized form of this function is shown at the top of each subfigure, along with the best-fit values of the parameters and their associated uncertainties.
        Each subfigure shows the same function, but with one parameter shifted by its $\pm 1$ standard deviation (green/blue), while the other parameters remain fixed at their best-fit values.
        The function with all parameters held at their best-fit values is plotted in red and compared to the data, represented by black data points.
        The middle panel shows the weighted residual error: $\frac{\text{Data}-\text{Fit}}{\text{Data unc.}}$.
        The bottom panel shows the ratio of the function with the uncertainty variations to the best-fit function.
        }
        \label{fig:toy_dataset_1_candidate}
\end{figure}

\begin{figure}[!t]
     \centering
     \begin{subfigure}[b]{1\textwidth}
         \centering
         \includegraphics[width=\textwidth]{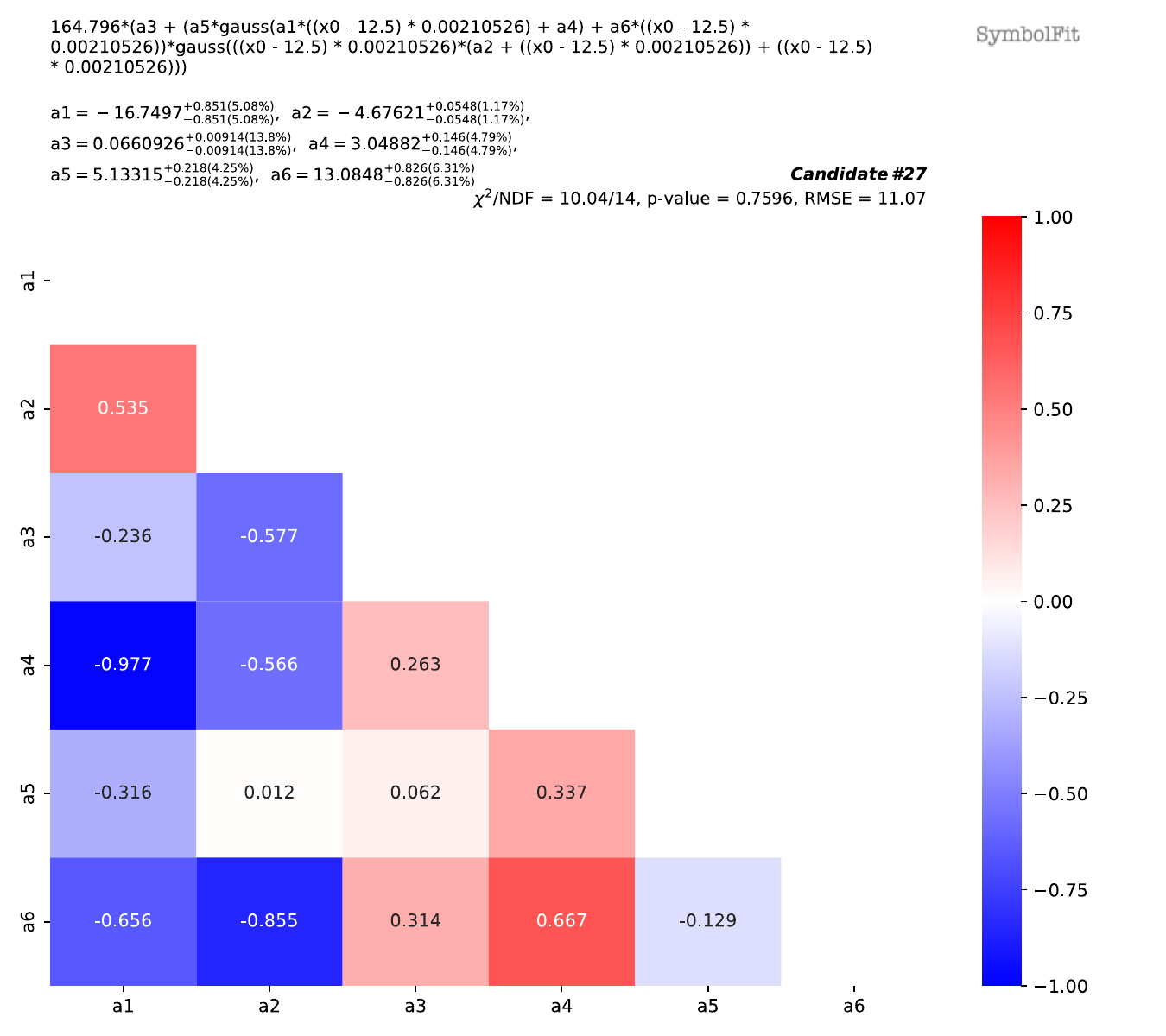}
     \end{subfigure}
     
        \caption{Correlation matrix for the parameters of candidate function \#27 from a fit to Toy Dataset 1 (see Tab.~\ref{tab:toy-dataset-1} and Fig.~\ref{fig:toy_dataset_1_candidate}). The parameter uncertainties and their correlation define the uncertainty model associated with the function.
        }
        \label{fig:toy_dataset_1_candidate_correlation}
\end{figure}

As an example, Fig.~\ref{fig:toy_dataset_1_candidate} shows candidate function \#27.
This candidate function has six parameters, with uncertainty variations for each plotted separately.
The correlation matrix for these parameters is shown in Fig.~\ref{fig:toy_dataset_1_candidate_correlation}.
The correlation matrix, along with the parameter uncertainties, defines the uncertainty model associated with the function and is required for propagating uncertainties to the final statistical inference results.

\begin{figure}[!t]
     \centering
     \begin{subfigure}[b]{0.495\textwidth}
         \centering
         \includegraphics[width=\textwidth]{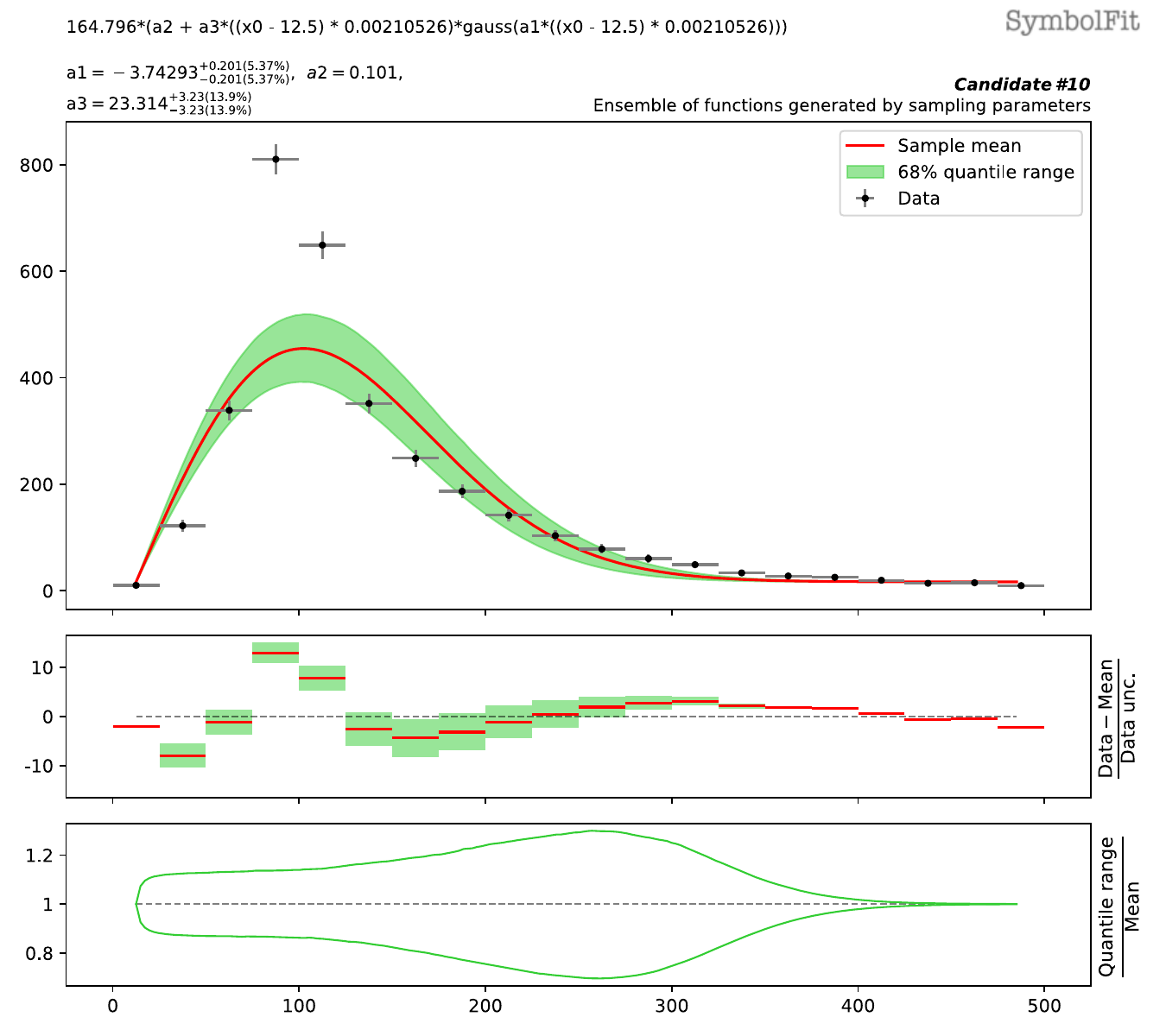}
         \caption{Candidate function \#10.}
     \end{subfigure}
     \begin{subfigure}[b]{0.495\textwidth}
         \centering
         \includegraphics[width=\textwidth]{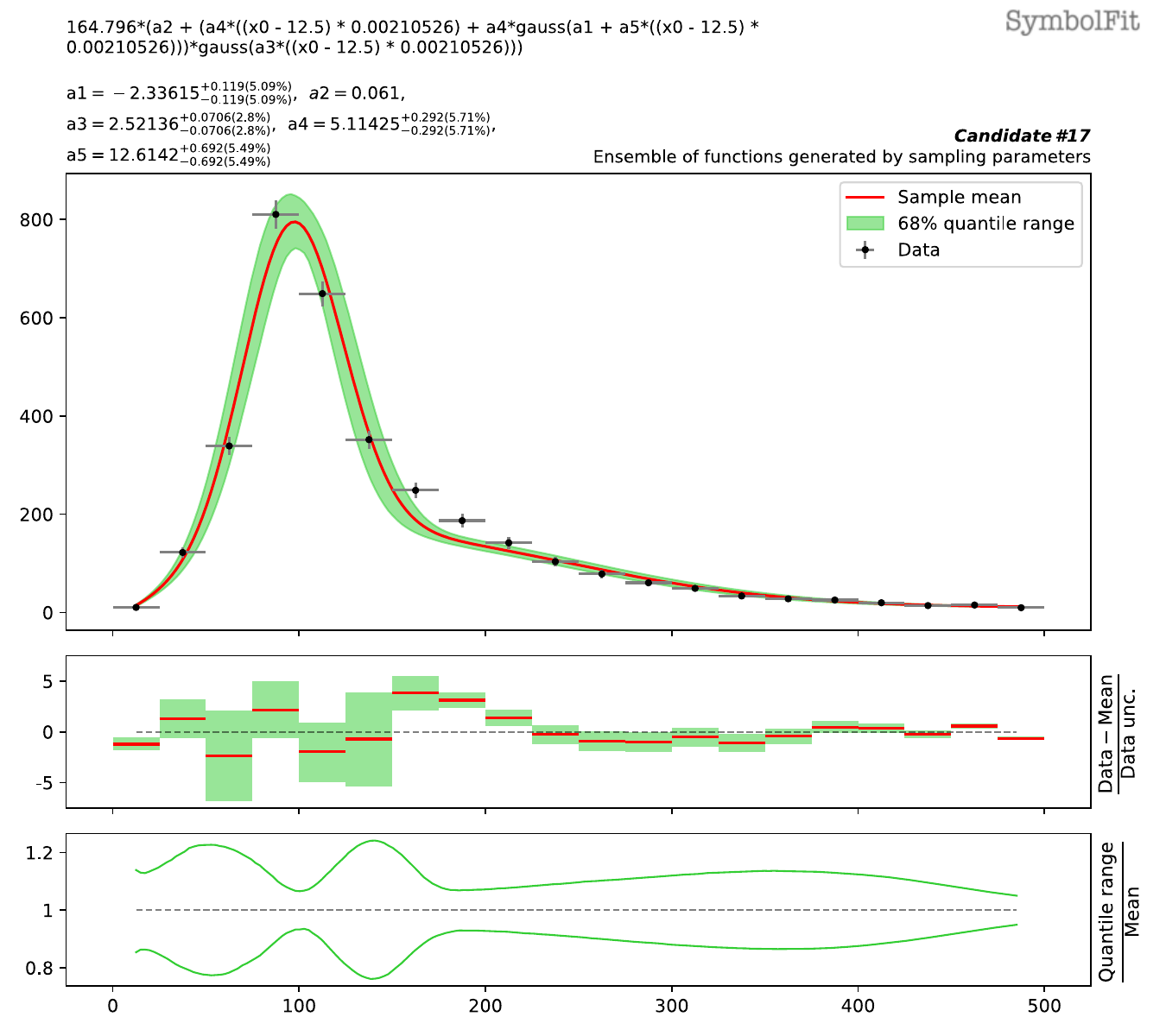}
         \caption{Candidate function \#17.}
     \end{subfigure}\\\vspace{0.5cm}
     \begin{subfigure}[b]{0.495\textwidth}
         \centering
         \includegraphics[width=\textwidth]{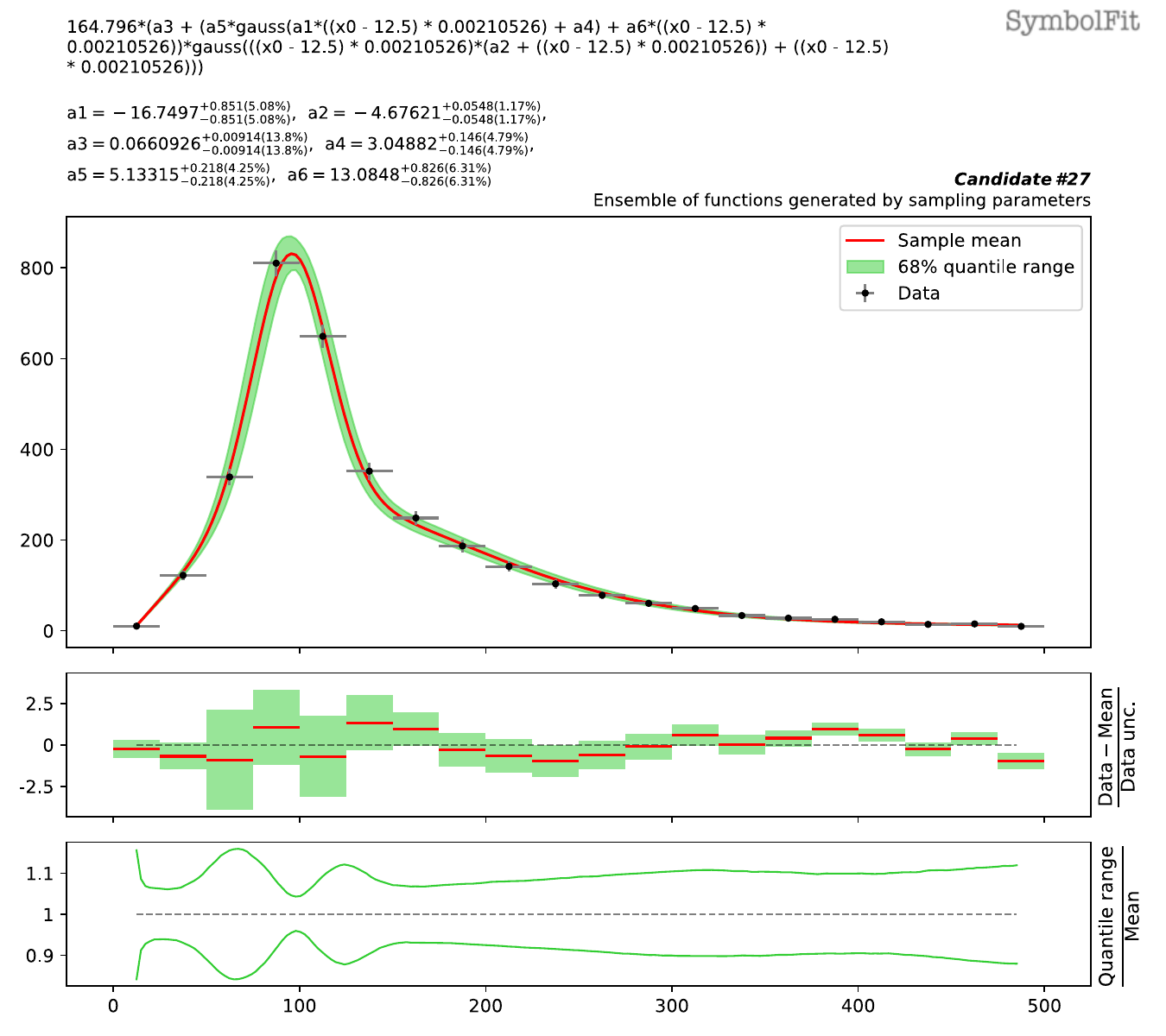}
         \caption{Candidate function \#27.}
     \end{subfigure}
     \begin{subfigure}[b]{0.495\textwidth}
         \centering
         \includegraphics[width=\textwidth]{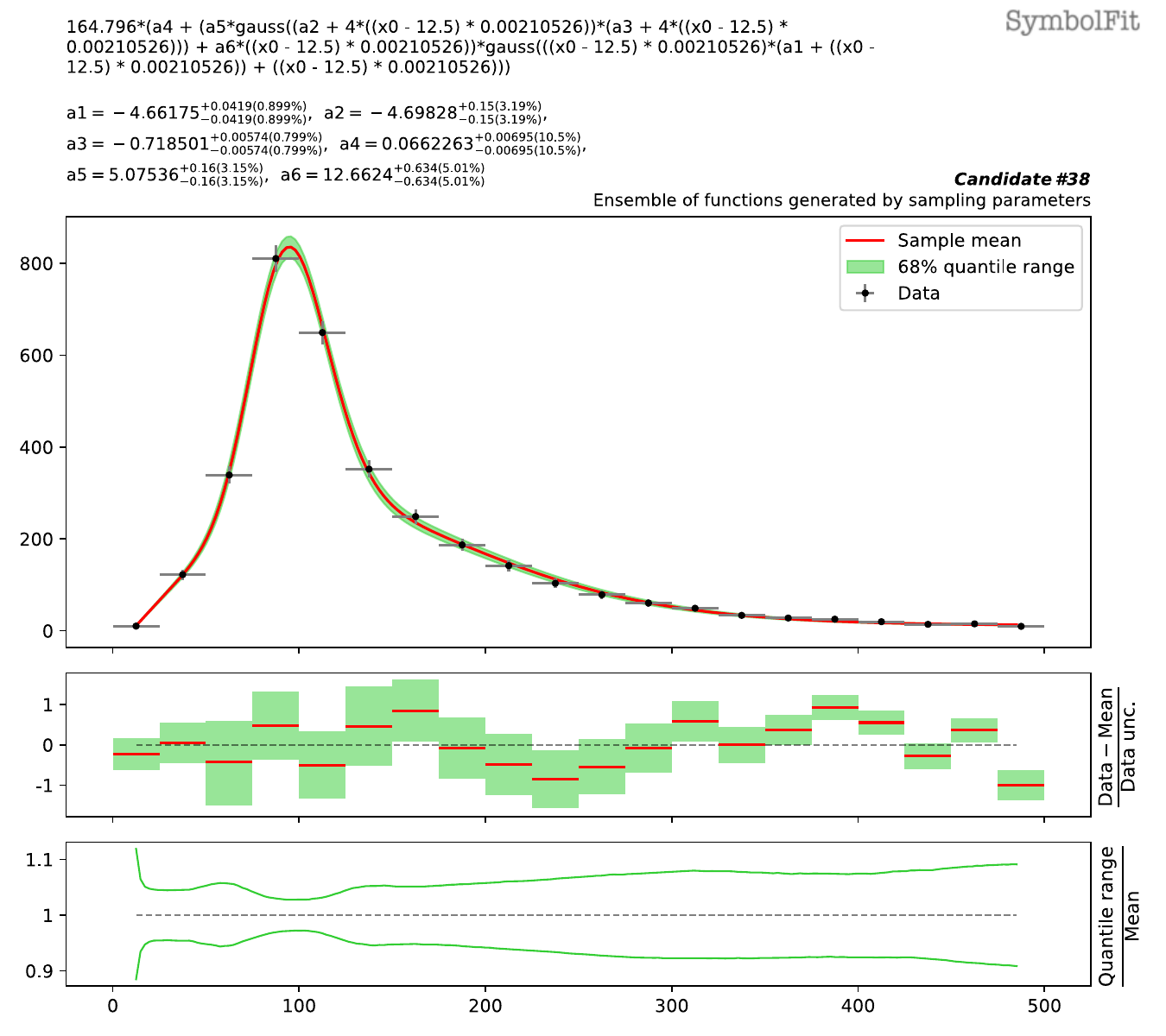}
         \caption{Candidate function \#38.}
     \end{subfigure}\vspace{0.5cm}
     
        \caption{Convergence of candidate functions to the data (Toy Dataset 1), from lower to higher function complexity values.
        To visualize the total uncertainty coverage of each candidate function, the green band in each subfigure represents the 68\% quantile range ($\pm 1 \sigma$) of functions obtained by sampling parameters, taking into account the best-fit values and the covariance matrix within a multidimensional normal distribution.
        The red line denotes the mean of the function ensemble.
        At the top of each subfigure, the candidate function and the fitted parameters are shown.
        The middle panel shows the weighted residual error: $\frac{\text{Data}-\text{Mean}}{\text{Data unc.}}$.
        The bottom panel shows the ratio of the 68\% quantile range to the mean.
        }
        \label{fig:toy_dataset_1_sampling}
\end{figure}

To compare various candidate functions obtained from a single fit in the framework, Fig.~\ref{fig:toy_dataset_1_sampling} shows four candidate functions with a range of fit scores from very low to very high.
This demonstrates that within the variety of functions generated per fit, there is a convergence from poorer to better-fitted functions, providing flexibility in choices.
To illustrate the uncertainty coverage for a candidate function, an ensemble of functions is generated by sampling parameters from a multidimensional normal distribution, considering their best-fit values and covariance matrix.
The 68\% quantile range of the function ensemble is shown for each candidate function, representing the total uncertainty coverage by simultaneously accounting for uncertainties in all parameters.
These computations are all automatically performed within the SymbolFit framework.

\begin{table*}[!t]
\caption{Nine examples from the 46 candidate functions obtained from a single fit for Toy Dataset 1.
These functions were fitted to a scaled dataset (to enhance fit stability and prevent numerical overflow), which can be rescaled to describe the original dataset using the transformation: $f(x)\rightarrow 165\times f(0.00211(x - 12.5))$.
The comparison between the $\chi^2/\text{NDF}$ scores before and after the ROF step is presented.
The total uncertainty coverage of candidate functions \#10, \#17, \#27, \#38 is shown in Fig.~\ref{fig:toy_dataset_1_sampling}.
Individual parameter variations for candidate function \#27 are plotted in Fig.~\ref{fig:toy_dataset_1_candidate}.
The function complexity values, providing a rough estimate of the model size, are computed before algebraic simplification.
Numerical values are rounded to three significant figures for display purposes.}
\label{tab:toy-dataset-1}
\centering
\resizebox{\textwidth}{!}{
\begin{tabular}{c|l|c|c|c|c}\hline
    \textbf{Complexity} & \textbf{Candidate function} & \textbf{\# param.} & \textbf{$\chi^2/\text{NDF}$} & \textbf{$\chi^2/\text{NDF}$} & \textbf{p-value} \\ 

    & (after ROF) & & (before ROF) & (after ROF) & (after ROF) \\ \hline
    & & & & & \\
    12 \textbf{(\#10)} & $0.101 + 23.3\gauss(-3.74x)x$ & 2 & 374.3 / 18 & 374.3 / 18 & $10^{-68}$ \\
    & & & & & \\ \hline
    & & & & & \\
    19 \textbf{(\#17)} & $0.061 + (5.11x + 5.11\gauss(-2.34 + 12.6x))\gauss(2.52x)$ & 4 & 54.06 / 16 & 53.59 / 16 & $10^{-06}$ \\
    & & & & & \\ \hline
    & & & & & \\
    22 (\#20) & $0.0837 + (4.76\gauss(-15.7x + 2.84) + 10.4\tanh(x))\gauss(2.99x)$ & 6 & 48.47 / 14 & 17.76 / 14 & 0.218 \\
    & & & & & \\ \hline
    & & & & & \\
    27 (\#24) & $(4.79\gauss((-5.5 + 2x)(-0.538 + 3x)) + 10.2x)\gauss(-3.03x)$ & 6 & 16.88 / 14 & 16.31 / 14 & 0.2951 \\
    & $+0.0841$ & & & & \\
    & & & & & \\ \hline
    & & & & & \\
    31 (\#26) & $(4.9x + 4.9\gauss(-2.79 + 15.4x) + 4.9\tanh(x))\gauss(3x) + $ & 4 & 15.79 / 16 & 15.45 / 16 & 0.4919 \\
    & $0.0789\gauss(x)\exp(x)$ & & & & \\
    & & & & & \\ \hline
    & & & & & \\    
    32 \textbf{(\#27)} & $(5.13\gauss(-16.7x + 3.05) + 13.1x)\gauss(x(-4.68 + x) + x)$ & 6 & 12 / 14 & 10.04 / 14 & 0.760 \\
    & $+ 0.0661$ & & & & \\
    & & & & & \\ \hline
    & & & & & \\
    37 (\#32) & $(5.07\gauss((-4.42 + 2x)(-0.724 + 4x)) + a5\tanh(x) + 7.79x)\times$ & 6 & 8.359 / 14 & 7.655 / 14 & 0.9065 \\
    & $\gauss(x(-4.65 + x) + x)+0.066$ & & & & \\
    & & & & & \\ \hline
    & & & & & \\
    44 (\#38) & $(5.08\gauss((-4.7 + 4x)(-0.719 + 4x)) + 12.7x)\times$ & 6 & 6.278 / 14 & 5.826 / 14 & 0.971 \\
    & $\gauss(x(-4.66 + x) + x) + 0.0662$ & & & & \\
    & & & & & \\ \hline
    & & & & & \\
    52 (\#41) & $0.0657 + (5\gauss((-4.96 + 6x)(-0.712 + 4x)) + 12.4\tanh(x))\times$ & 6 & 3.564 / 14 & 3.032 / 14 & 0.999 \\
    & $\gauss(x(-4.6 + x) + x)-0.00624x$ & & & & \\
    & & & & & \\ \hline
    
    \end{tabular}%
    }
\end{table*}

\clearpage

\subsection{CMS dijet dataset (1D) [background modeling]}
\label{sec:dijet}

CMS performed a search for high-mass dijet resonances using proton-proton collision data at a center-of-mass energy of $\sqrt{s}=13$ TeV and reported no significant deviations from the Standard Model prediction~\cite{dijet}.
The dataset for the dijet spectrum is publicly available on HEPDATA at Ref.~\cite{hepdata.dijet}.
In the analysis, CMS used an empirical 4-parameter function to model the background contribution in the distribution of the dijet invariant mass, $m_{\text{jj}}$:
\begin{equation}
    \label{eq:dijet}
    f(x)=\frac{p_0(1-x)^{p_1}}{x^{p_2+p_3\ln(x)}},
\end{equation}
where $x=m_{\text{jj}}/\sqrt{s}$ is dimensionless and $p_{\{0,1,2,3\}}$ are free parameters.
While this function fits the current dijet spectrum reasonably well, it may be too rigid to accommodate potential future changes in the dijet spectrum due to modifications in analysis strategies or detector performance.

For our experiments, we use the $\tt{PySR}$ configuration shown in List.~\ref{config-lhc} to fit the dijet dataset.
The main difference between List.~\ref{config-lhc} and List.~\ref{config-toy1} used in Sec.~\ref{sec:toy-dataset-1} is that it does not explicitly include a Gaussian operator, as the mass spectrum assumes no peaks in the background.
This same SR configuration is also applied to the other LHC datasets as well as Toy Dataset 2, generating a range of well-fitted functions for each case despite their very different distribution shapes, demonstrating the flexibility of the SR approach.

\begin{lstlisting}[language=Python, caption=The $\tt{Python}$ code snippet configures $\tt{PySR}$ to search for candidate functions for the dijet dataset. This same configuration is used for the other four LHC datasets and Toy Dataset 2 with variations in the maximum complexity values., label=config-lhc]
from pysr import PySRRegressor

pysr_config = PySRRegressor(
    model_selection = "accuracy",
    timeout_in_seconds = 60*100,
    niterations = 200,
    maxsize = 80,
    binary_operators = [
        "+", "*", "/", "^"
                     ],
    unary_operators = [
        "exp",
        "tanh",
    ],
    nested_constraints = {
        "exp":    {"exp": 0, "tanh": 0, "*": 2, "/": 1, "^": 1},
        "tanh":   {"exp": 0, "tanh": 0, "*": 2, "/": 1, "^": 1},
        "*":      {"exp": 1, "tanh": 1, "*": 2, "/": 1, "^": 1},
        "^":      {"exp": 1, "tanh": 1, "*": 2, "/": 1, "^": 0},
        "/":      {"exp": 1, "tanh": 1, "*": 2, "/": 0, "^": 1},
    },
    loss="loss(y, y_pred, weights) = (y - y_pred)^2 * weights",
)
\end{lstlisting}

To validate the SR approach in background modeling and signal extraction, we start with the original dijet spectrum and generate pseudodata by injecting a small and narrow Gaussian signal, $s_0 \frac{1}{\sqrt{2\pi}s_2}\exp\big(-\frac{(x-s_1)^2}{2s_2^2}\big)$, centered at $m_{\text{jj}}=3.1$ TeV ($s_1$), with a width of 0.2 TeV ($2s_2$) and a signal strength of $s_0=10$.
The injected signal intensity per bin is perturbed by 10\% random noise.

To model the background, we first blind the signal region by masking the $m_{\text{jj}}$ bins near the injected signal peak, specifically between 2.7 and 3.5 TeV, and then perform fits to this blinded pseudodata.
We conduct three separate $\tt{SymbolFit}$ runs on the blinded pseudodata, using the same fit configuration but with different random seeds.
This demonstrates that a variety of well-fitted functions can be obtained from the same fit configuration, given the vast function space in the SR search.
From each of the three fits, one candidate function is selected, referred to as ``SR model 1'', ``SR model 2'', and ``SR model 3'', respectively.
These three background models are then compared with the empirical function in Eq.~\ref{eq:dijet} used by CMS, referred to as the ``empirical model (CMS)''.

\begin{figure}[!t]
     \centering
     \begin{subfigure}[b]{0.495\textwidth}
         \centering
         \includegraphics[width=\textwidth]{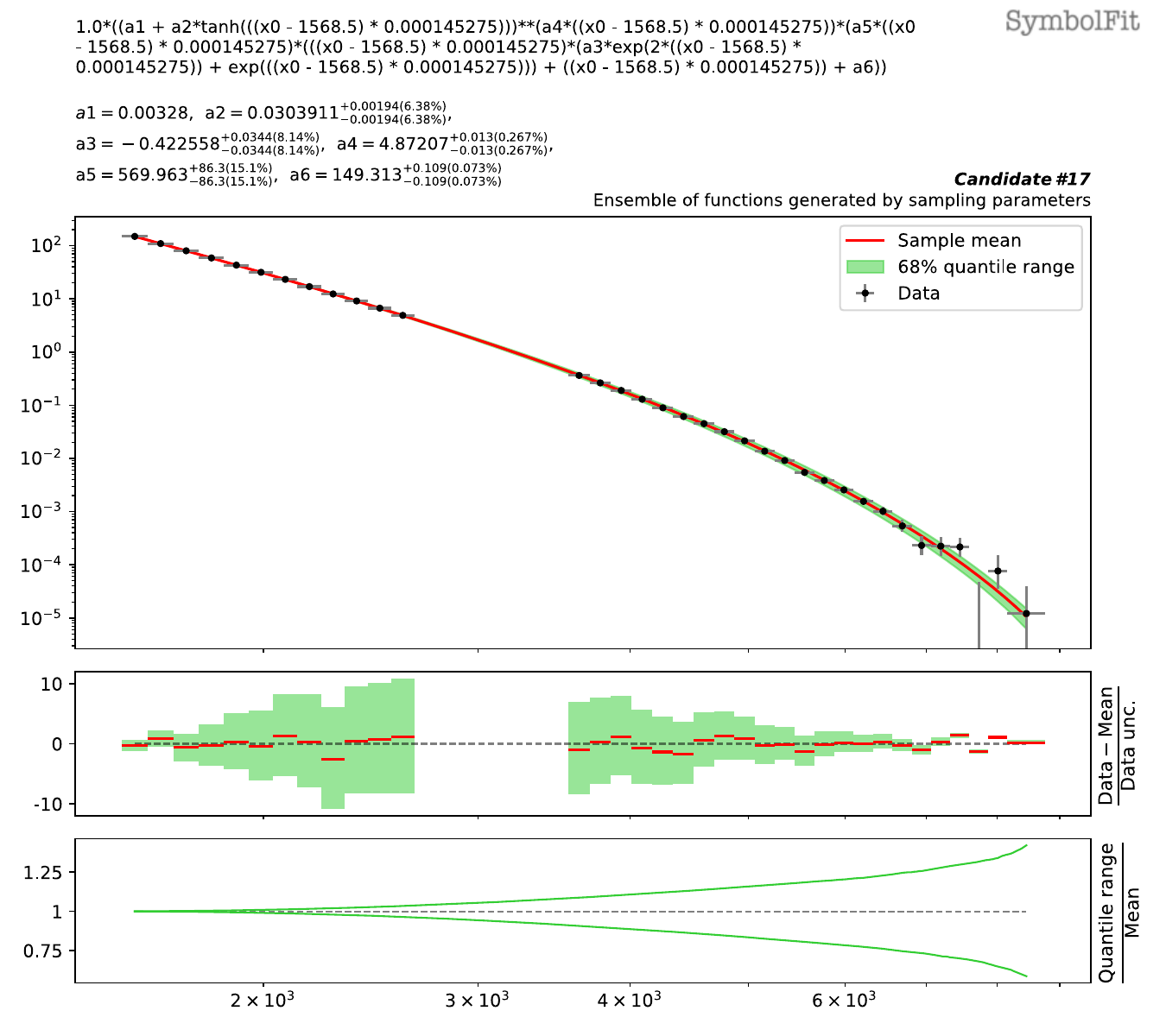}
         \caption{SR model 1.}
     \end{subfigure}
     \begin{subfigure}[b]{0.495\textwidth}
         \centering
         \includegraphics[width=\textwidth]{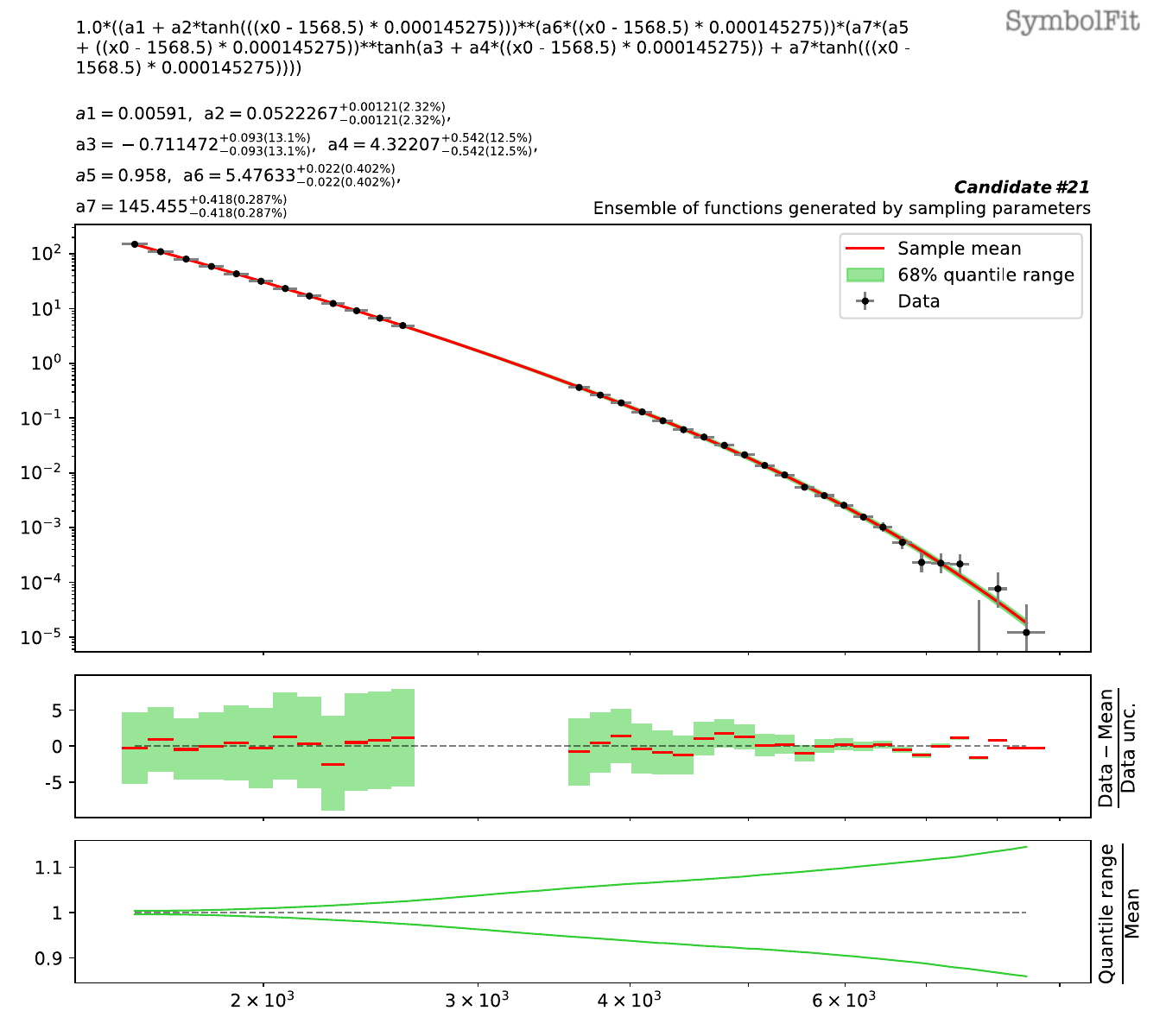}
         \caption{SR model 2.}
     \end{subfigure}\\\vspace{0.5cm}
     \begin{subfigure}[b]{0.495\textwidth}
         \centering
         \includegraphics[width=\textwidth]{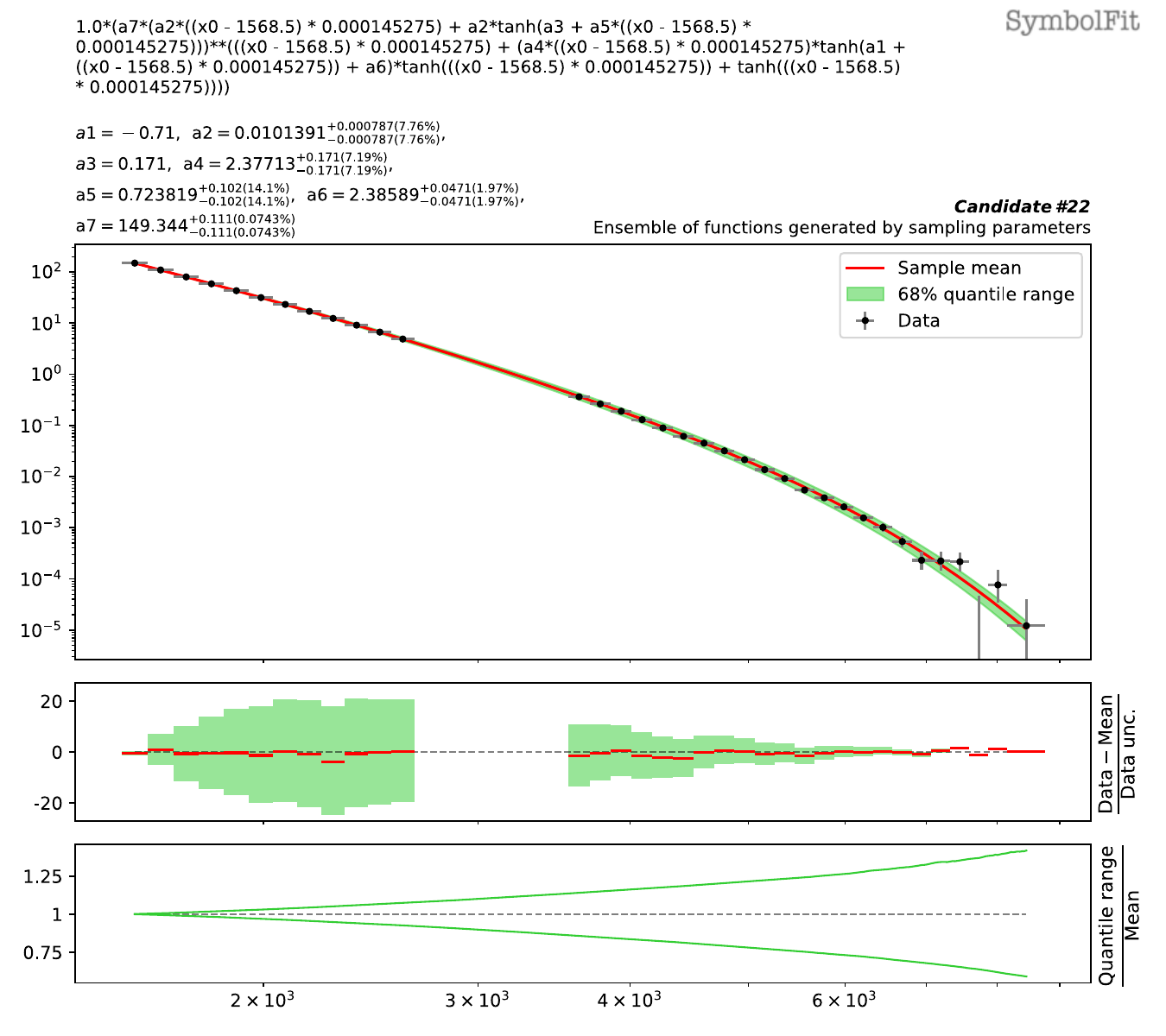}
         \caption{SR model 3.}
     \end{subfigure}\vspace{0.5cm}
     
    \caption{
        The three SR models fitted to the pseudodata of the dijet spectrum with the signal region blinded (see Tab.~\ref{tab:dijet_candidates}).
        To visualize the total uncertainty coverage of each candidate function, the green band in each subfigure represents the 68\% quantile range of functions obtained by sampling parameters, taking into account the best-fit values and the covariance matrix within a multidimensional normal distribution.
        The red line denotes the mean of the function ensemble.
        At the top of each subfigure, the candidate function and the fitted parameters are shown.
        The middle panel shows the weighted residual error: $\frac{\text{Data}-\text{Mean}}{\text{Data unc.}}$.
        The bottom panel shows the ratio of the 68\% quantile range to the mean.}
    \label{fig:dijet_sampling}
\end{figure}

\begin{figure}[!t]
     \centering
     \begin{subfigure}[b]{1\textwidth}
         \centering
         \includegraphics[width=\textwidth]{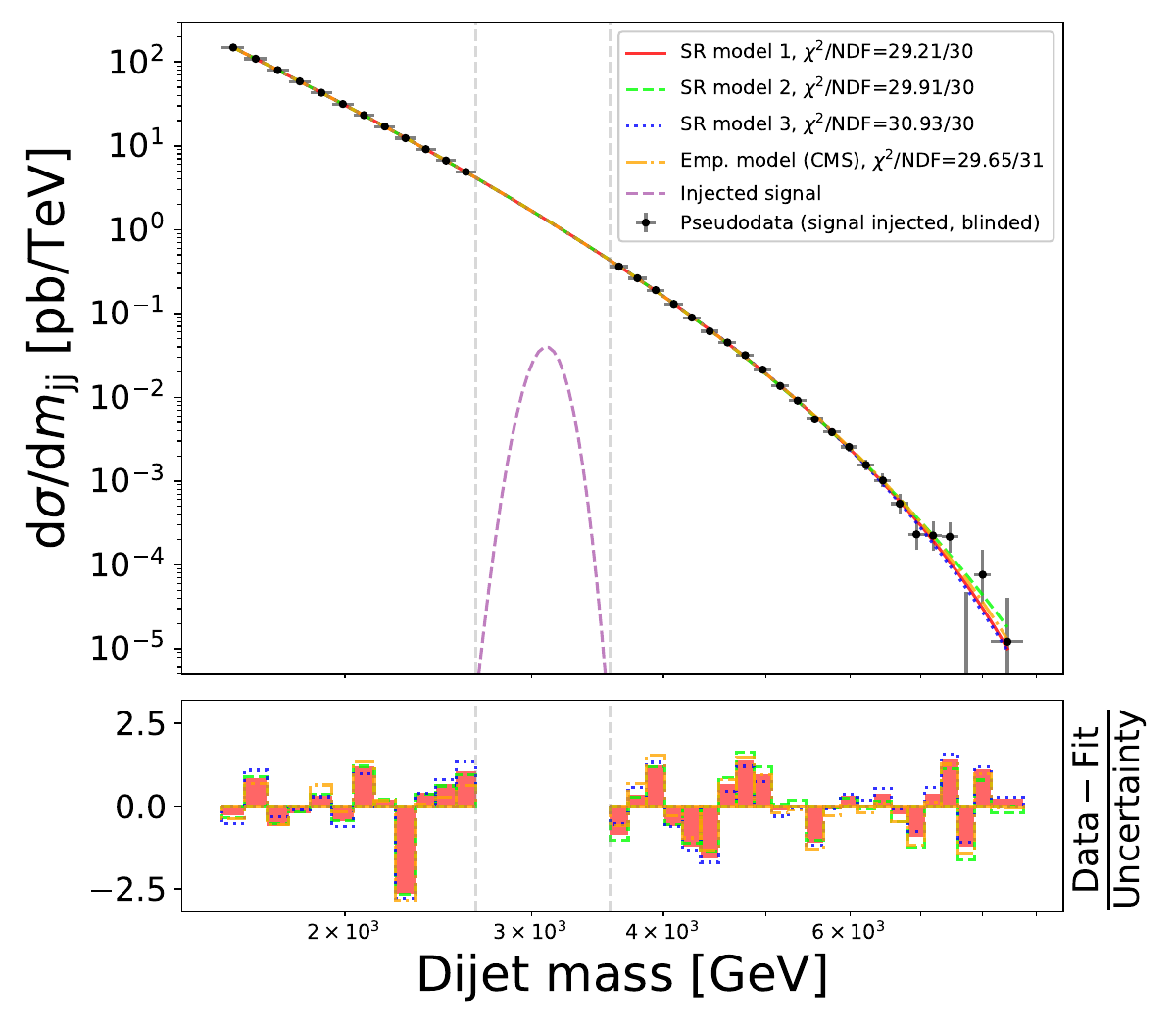}
     \end{subfigure}
     
    \caption{Pseudodata of the dijet spectrum with the injected signal shown in the blinded signal region. The three SR models (see Tab.~\ref{tab:dijet_candidates}) are compared against the empirical model used by CMS. The lower panel shows the residual error per bin, measured in units of the data uncertainty. It can be seen that the three SR models, generated easily from three separate fits using the same simple fit configuration with different random seeds, yields results that are readily comparable to the CMS empirical model that would have required extensive manual effort to obtain.}
    \label{fig:dijet_blinded}
\end{figure}

Tab.~\ref{tab:dijet_candidates} lists the three SR models, each obtained from a fit initialized with a different random seed.
The $\chi^2/\text{NDF}$ scores improve significantly after the ROF step compared to the original functions returned by $\tt{PySR}$, with the final scores close to 1.
The three background models fit the blinded pseudodata well, as shown in 
Fig.~\ref{fig:dijet_sampling} for the total uncertainty coverage and Fig.~\ref{fig:dijet_blinded} for a comparison with the empirical model used by CMS.

\begin{table*}[!t]
\caption{The candidate functions are obtained from three separate fits using the same fit configuration but with different random seeds, fitted to the pseudodata of the dijet spectrum with the (injected) signal region blinded.
The fits were performed on a scaled dataset (to enhance fit stability and prevent numerical overflow), and the functions can be transformed back to describe the original spectrum using the transformation: $x\rightarrow 0.000145(x - 1568.5)$.
These functions are plotted and compared with the blinded pseudodata in Fig.~\ref{fig:dijet_blinded}.
Numerical values are rounded to three significant figures for display purposes.}
\label{tab:dijet_candidates}
\centering
\resizebox{\textwidth}{!}{
\begin{tabular}{c|l|c|c|c|c}\hline
    & \textbf{Candidate function} & \textbf{\# param.} & \textbf{$\chi^2/\text{NDF}$} & \textbf{$\chi^2/\text{NDF}$} & \textbf{p-value} \\ 

    & (after ROF) & & (before ROF) & (after ROF) & (after ROF) \\ \hline

    & & & & & \\
    SR model 1 & $(570x(x(-0.423\exp(2x) + \exp(x)) + x) + 149)\times$ & 5 & 400.5 / 30 & 29.21 / 30 & 0.507 \\
    & $(0.00328 + 0.0304\tanh(x))^{4.87x}$ & & & & \\
    & & & & & \\ \hline
    & & & & & \\
    SR model 2 & $(145(0.958 + x)^{\tanh(-0.711 + 4.32x)} + 145\tanh(x))\times$ & 5 & 103.8 / 30 & 29.91 / 30 & 0.47 \\
    & $(0.00591 + 0.0522\tanh(x))^{5.48x}$ & & & & \\
    & & & & & \\ \hline
    & & & & & \\
    SR model 3 & $\text{pow}(149(0.0101x + 0.0101\tanh(0.171 + 0.724x)),$ & 5 & 214.8 / 30 & 30.93 / 30 & 0.419 \\
    & $x + (2.38x\tanh(-0.71 + x) + 2.39)\tanh(x) + \tanh(x))$ & & & & \\
    & & & & & \\ \hline
    
    \end{tabular}%
    }
\end{table*}

Once the background models are established, we incorporate a parameterized Gaussian signal template into each model $f(x)$:
\begin{equation}
    \label{eq:sb_model}
    f(x) + s_0 \frac{1}{\sqrt{2\pi}s_2}\exp\bigg(-\frac{(x-s_1)^2}{2s_2^2}\bigg).
\end{equation}
In the following analysis, when the model is fitted to the unblinded pseudodata with $s_0=0$ held fixed, it is referred to as a background-only (b-only) fit.
When $s_0$ is allowed to vary, it is referred to as a signal-plus-background (s+b) fit.

\begin{figure}[!t]
     \centering
     \begin{subfigure}[b]{0.495\textwidth}
         \centering
         \includegraphics[width=\textwidth]{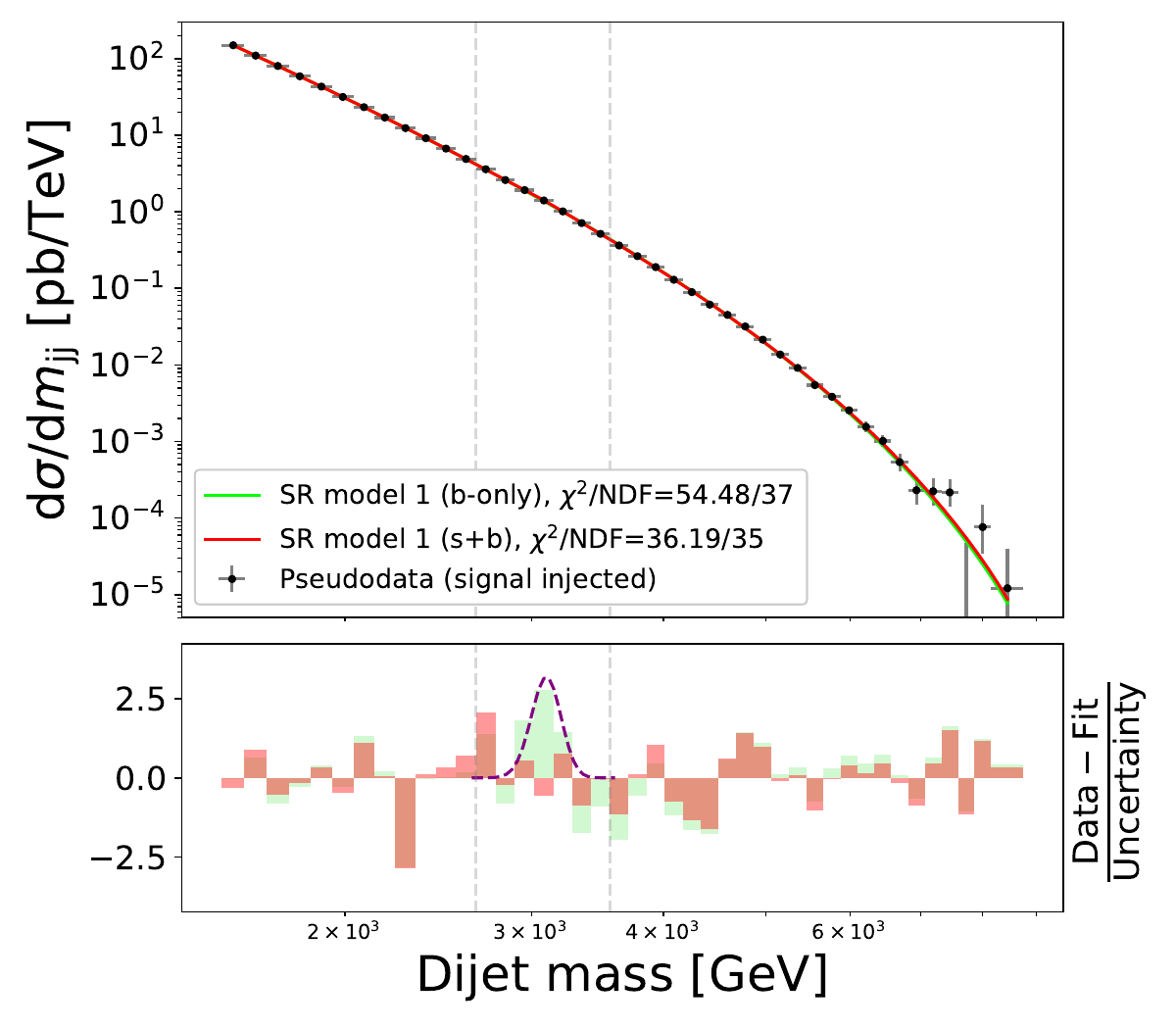}
         \caption{SR model 1.}
     \end{subfigure}
     \begin{subfigure}[b]{0.495\textwidth}
         \centering
         \includegraphics[width=\textwidth]{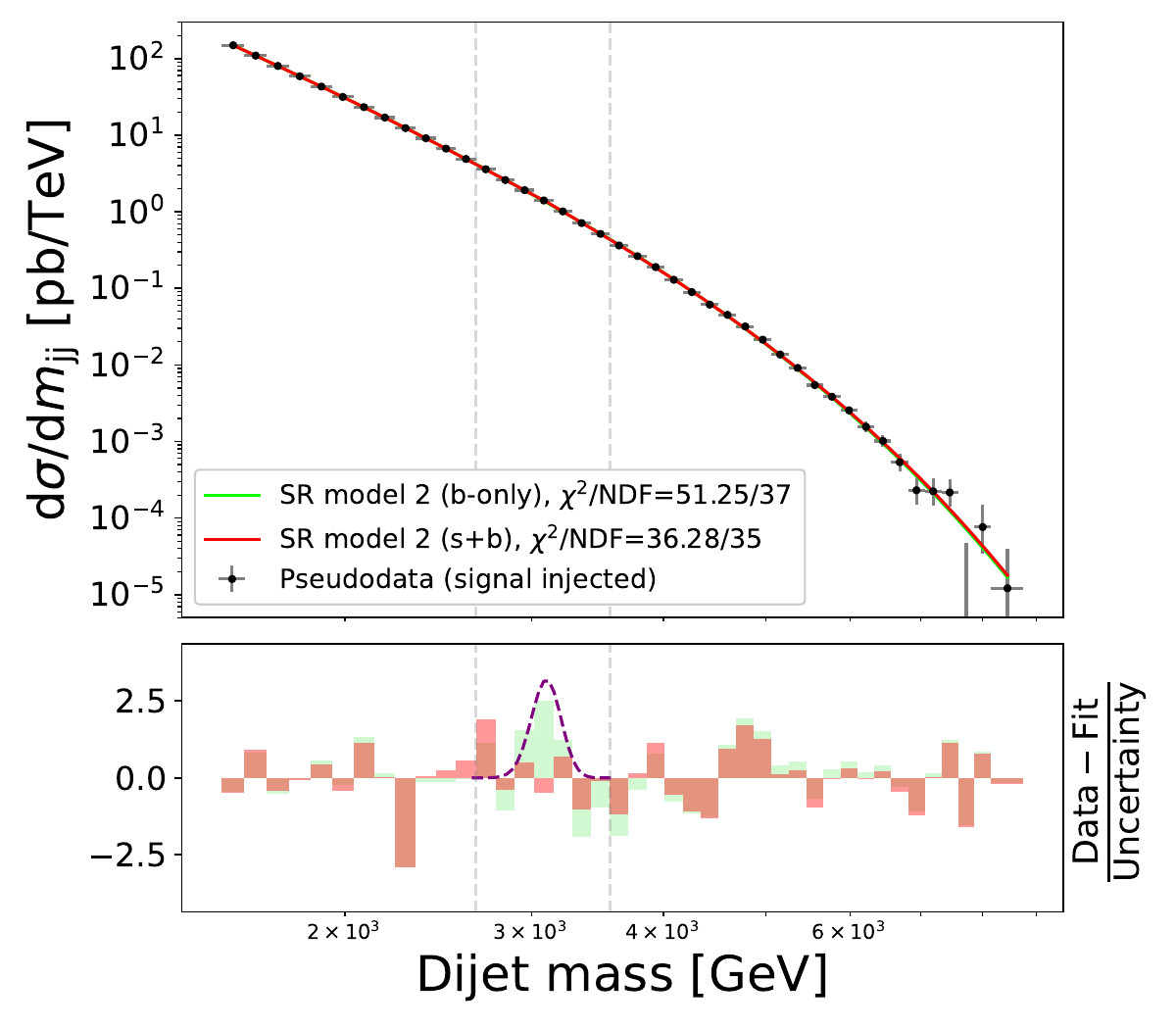}
         \caption{SR model 2.}
     \end{subfigure}\\\vspace{0.5cm}
     \begin{subfigure}[b]{0.495\textwidth}
         \centering
         \includegraphics[width=\textwidth]{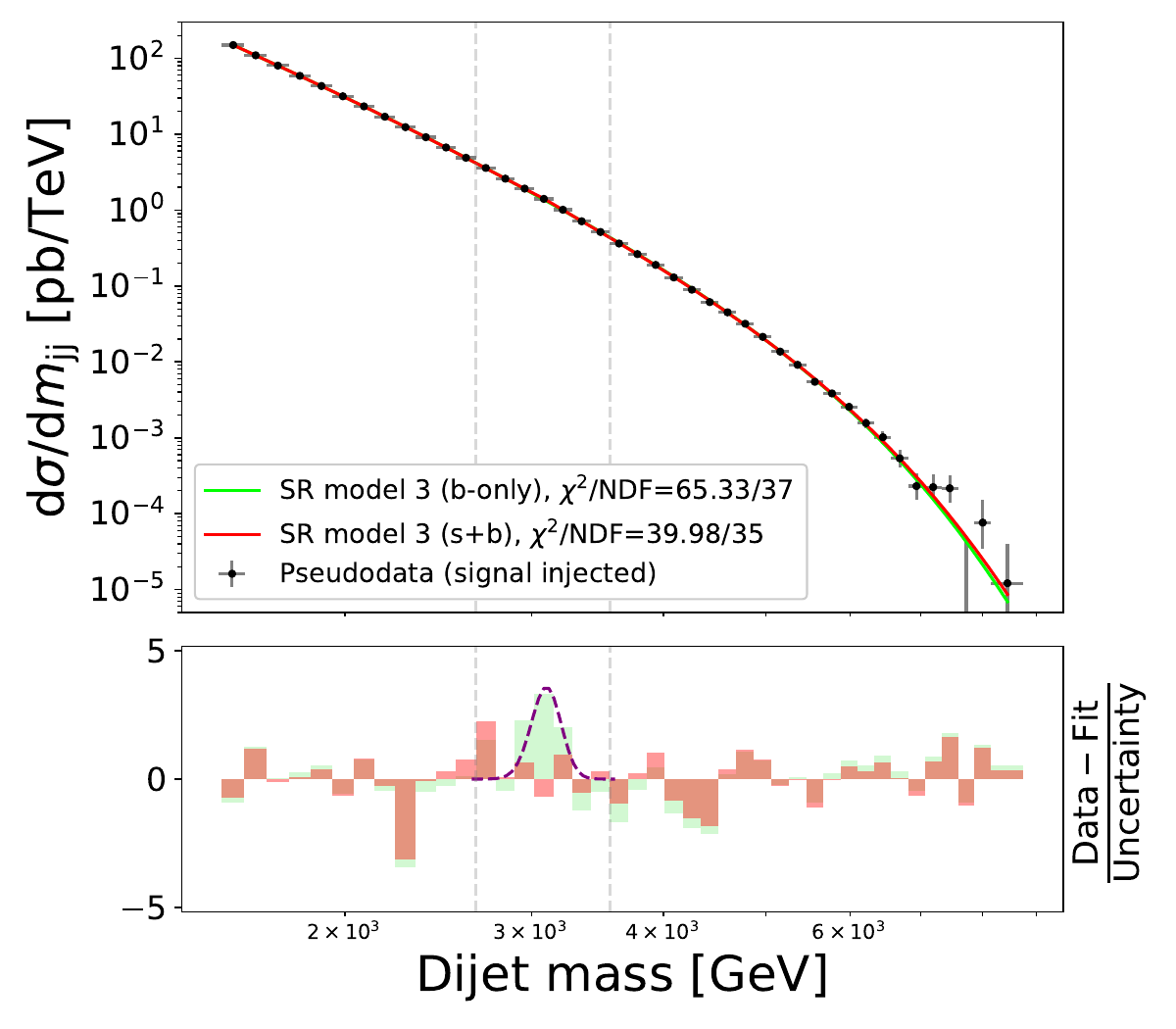}
         \caption{SR model 3.}
     \end{subfigure}
     \begin{subfigure}[b]{0.495\textwidth}
         \centering
         \includegraphics[width=\textwidth]{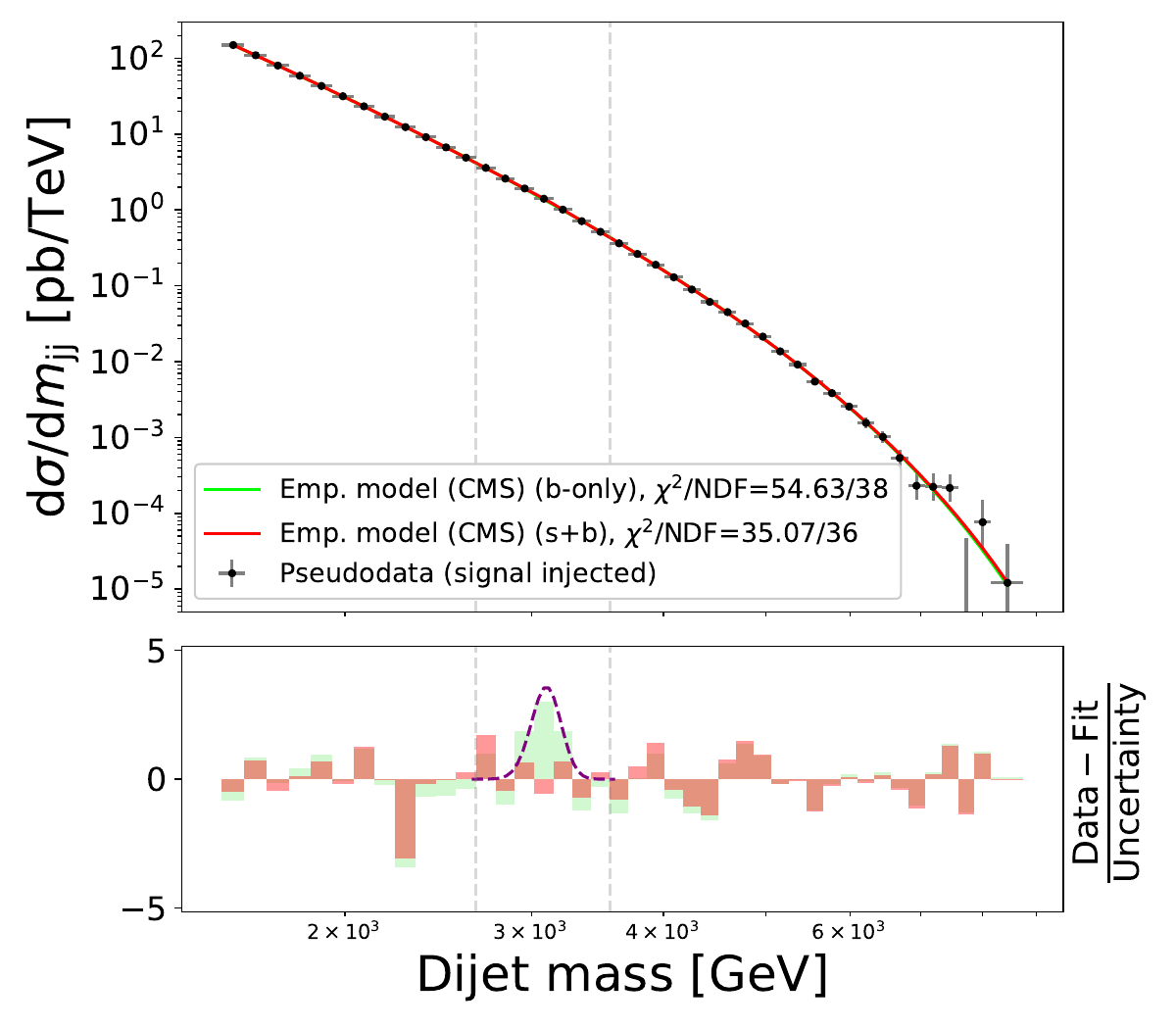}
         \caption{Empirical model (CMS).}
     \end{subfigure}\vspace{0.5cm}
     
    \caption{Comparison of the b-only fits and the s+b fits to the unblinded pseudodata of the dijet spectrum. The lower panel shows the residual error per bin, measured in units of the data uncertainty. The shape of the injected signal is also shown.}
    \label{fig:dijet_unblinded}
\end{figure}

Now, we unblind the pseudodata and perform b-only fits and s+b fits on the full dijet spectrum.
Since the pseudodata contain an injected signal, we expect to observe an excess of events over the background model around the signal location, provided the background is properly modeled and not overly fitted.
When performing the s+b fits, we expect that the excess of events observed in the b-only fits will diminish as the signal is accounted for by the model template.
The results of the b-only and s+b fits for each model are compared and shown in Fig.\ref{fig:dijet_unblinded}.
In all three SR models, as well as the CMS empirical model, the excess of events over the background around the injected signal location observed in the b-only fits is reduced in the s+b fits, demonstrating that the models are sensitive to the injected signal.
Tab.~\ref{tab:dijet_chi2} lists the $\chi^2/\text{NDF}$ scores for each model, indicating the fit performance in response to the presence of the injected signal.

\begin{table*}[!t]
\caption{Comparison of the $\chi^2/\text{NDF}$ scores from three types of fits to the dijet dataset: the b-only fits to the blinded pseudodata , the b-only fits to the unblinded pseudodata, and the s+b fits to the unblinded pseudodata.
The background models used for the fits are listed in Tab.~\ref{tab:dijet_candidates}, and the fits are shown in Fig.~\ref{fig:dijet_blinded} (blinded) and Fig.~\ref{fig:dijet_unblinded} (unblinded).}
\label{tab:dijet_chi2}
\centering
\resizebox{\textwidth}{!}{
\begin{tabular}{l|c|c|c}\hline
    & \textbf{$\chi^2/\text{NDF}$ (b-only, blinded)} & \textbf{$\chi^2/\text{NDF}$ (b-only, unblinded)} & \textbf{$\chi^2/\text{NDF}$ (s+b, unblinded)} \\ \hline
    
    SR model 1 & 29.21 / 30 = 0.974 & 54.48 / 37 = 1.47 & 36.19 / 35 = 1.03 \\
    & & & \\ \hline

    SR model 2 & 29.91 / 30 = 0.997 & 51.25 / 37 = 1.39 & 36.28 / 35 = 1.04 \\
    & & & \\ \hline

    SR model 3 & 30.93 / 30 = 1.03 & 65.33 / 37 = 1.77 & 39.98 / 35 = 1.14 \\
    & & & \\ \hline

    Emp. model (CMS) & 29.65 / 31 = 0.956 & 54.63 / 38 = 1.44 & 35.07 / 36 = 0.974 \\
    & & & \\ \hline
    
    \end{tabular}%
    }
\end{table*}

Additionally, to assess whether the SR models can extract the injected signals, we generate multiple sets of pseudodata by injecting Gaussian signals with varying mean values between 2980 to 3150 GeV and signal strengths ranging from 2 to 38.
We then perform the s+b fits to extract the corresponding signal parameters.
Fig.~\ref{fig:dijet_pair_scan} shows the extracted signal parameters (mass and strength) plotted against their injected values.
All three SR models are capable of extracting the correct signal parameter values within reasonable uncertainties.
They perform comparably to the empirical model used by CMS and, in some cases, yield more accurate fitted values, demonstrating that functions obtained from SR are effective for such tasks.

We demonstrated that our framework can easily generate multiple suitable candidate functions using a very simple fit configuration.
By simply changing the random seed, additional candidate functions can be obtained as needed, and these functions are readily comparable to the empirical function obtained by CMS, which required extensive manual and iterative effort.
This shows that our method is preferable to traditional methods, as the labor-intensive step of finding adequate functional forms is now automated using machine learning.

\begin{figure}[!t]
     \centering
     \begin{subfigure}[b]{0.6\textwidth}
         \centering
         \includegraphics[width=\textwidth]{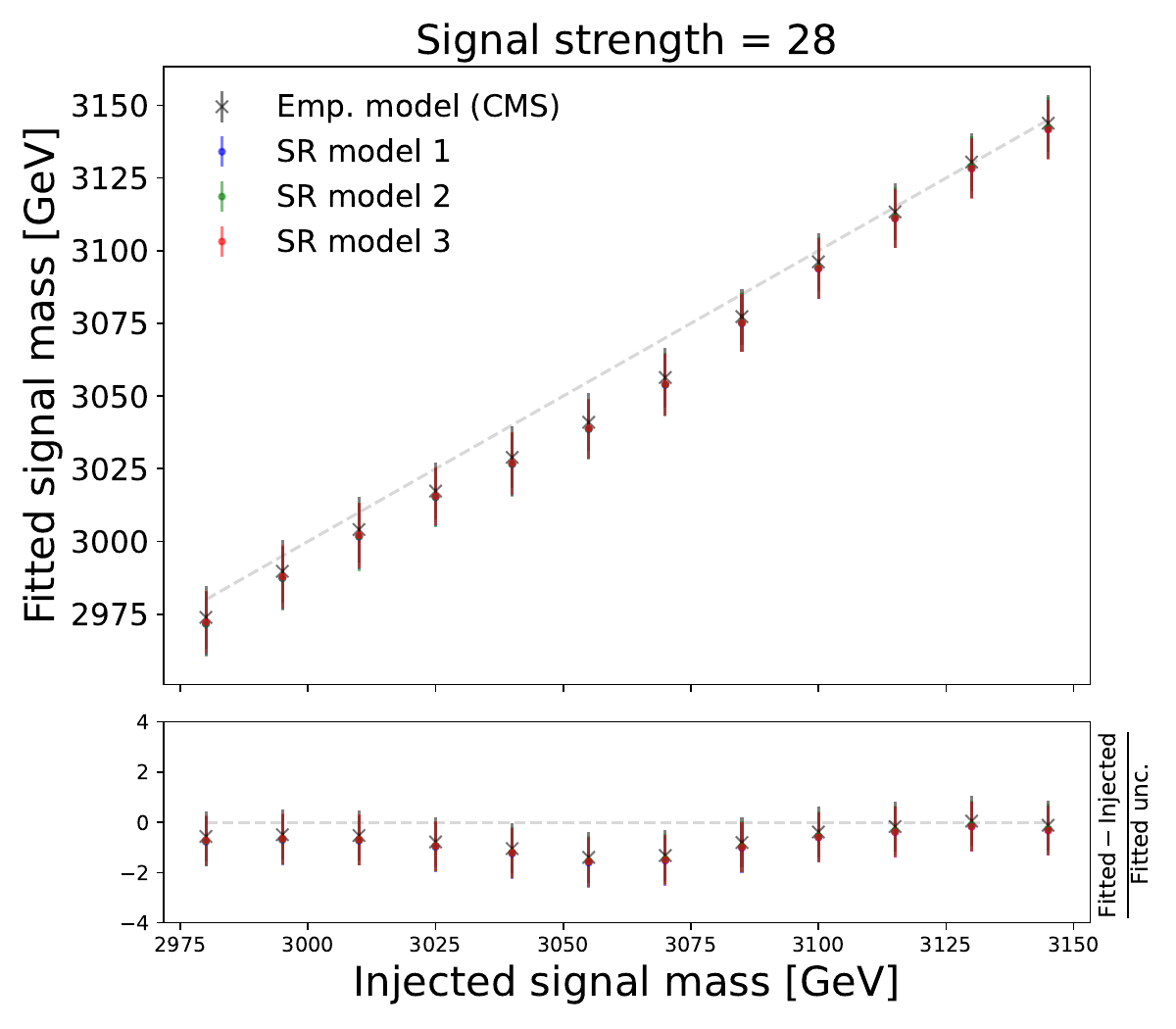}
         \caption{Fitted vs. injected signal mass at a specified signal strength value.}
     \end{subfigure}\vspace{0.5cm}
     \begin{subfigure}[b]{0.6\textwidth}
         \centering
         \includegraphics[width=\textwidth]{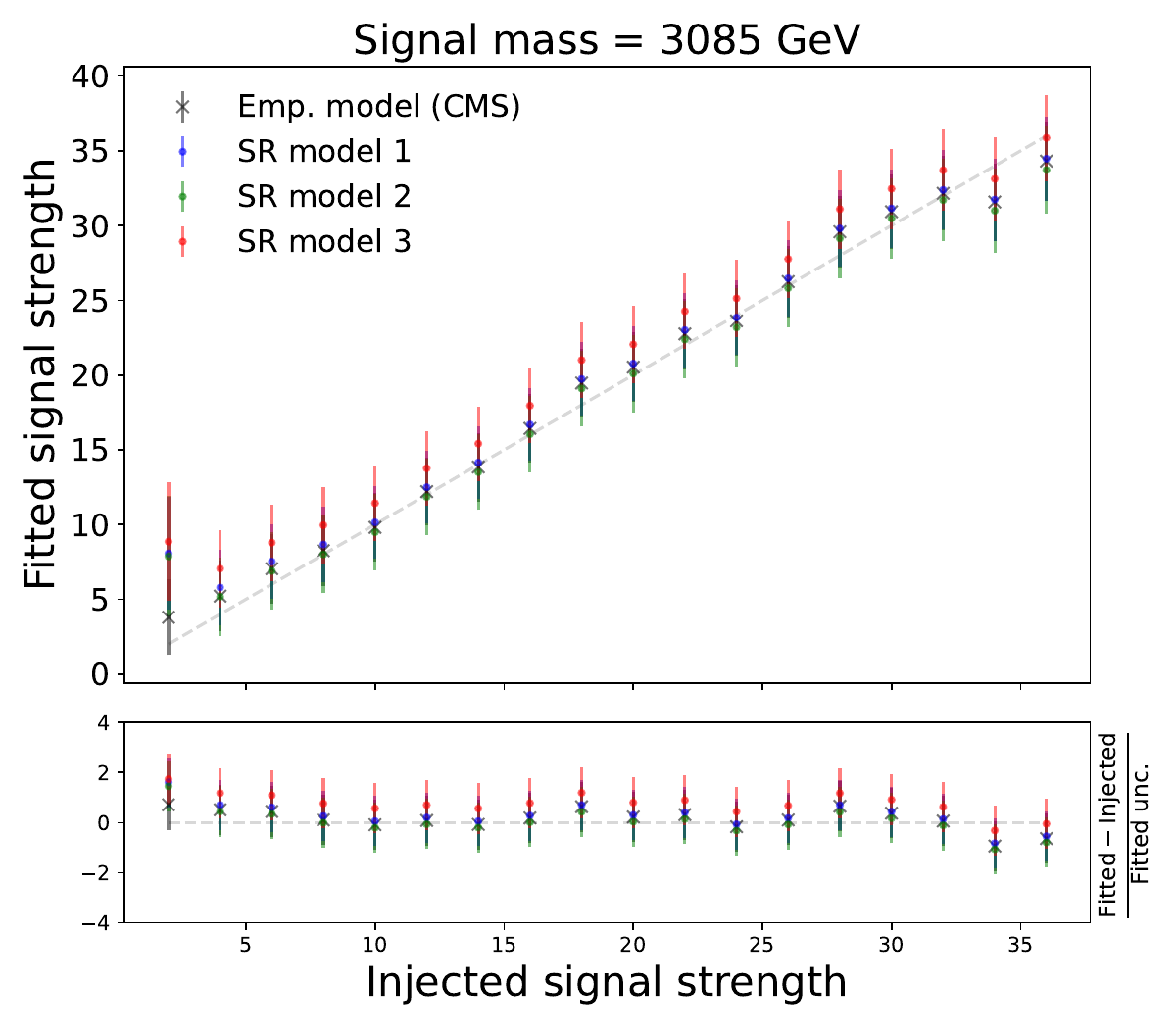}
         \caption{Fitted vs. injected signal strength at a specified signal mass value.}
     \end{subfigure}
     
    \caption{Fitted values vs. the true values of the parameters of the injected signal in the dijet dataset.
    The bottom panels show the residual error in units of the fitted uncertainty.}
    \label{fig:dijet_scan}
\end{figure}

\clearpage

\subsection{Toy dataset 3 (2D) [arbitrary shapes]}
\label{sec:toy-dataset-3}

\begin{figure}[!t]
     \centering
     \begin{subfigure}[b]{0.495\textwidth}
         \centering
         \includegraphics[width=\textwidth]{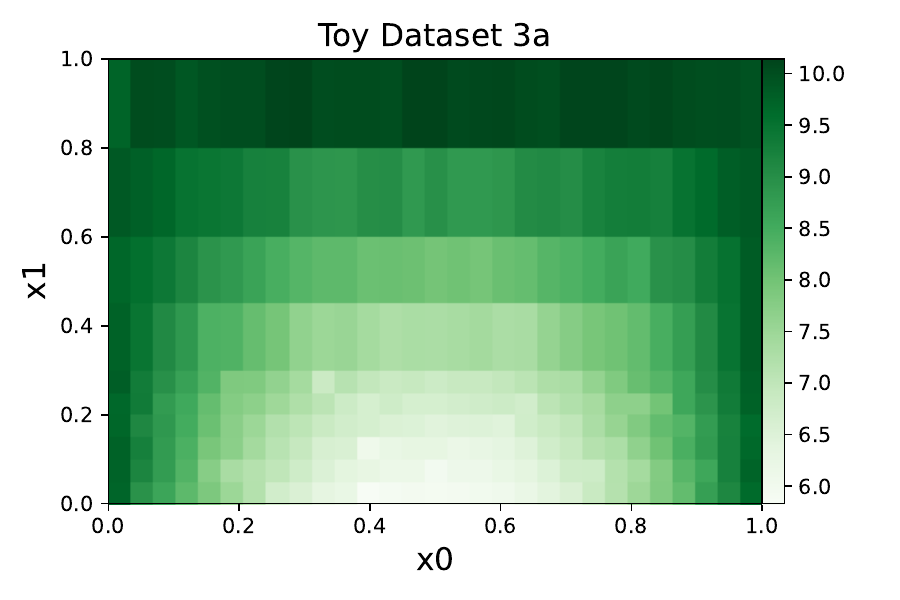}
     \end{subfigure}
     \begin{subfigure}[b]{0.495\textwidth}
         \centering
         \includegraphics[width=\textwidth]{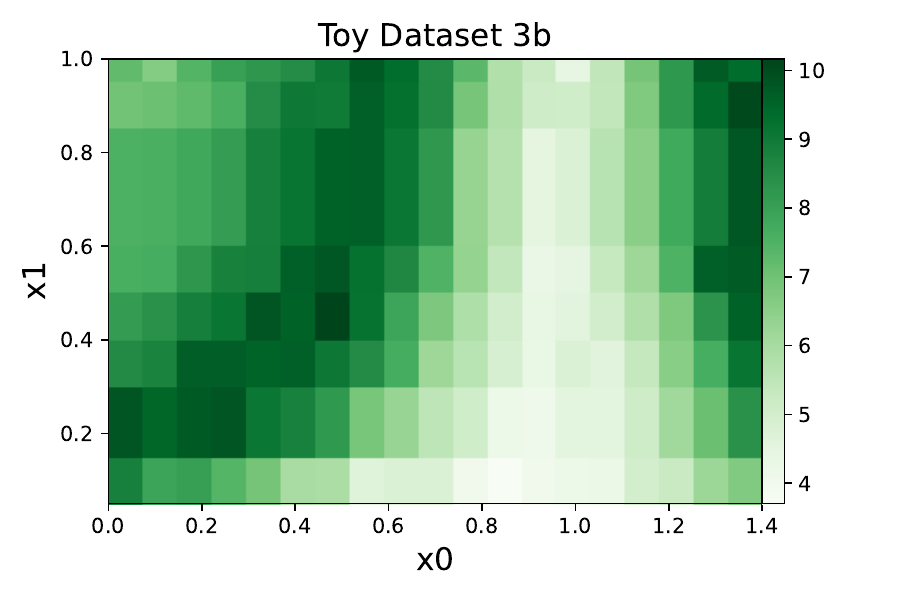}
     \end{subfigure}
     \begin{subfigure}[b]{0.495\textwidth}
         \centering
         \includegraphics[width=\textwidth]{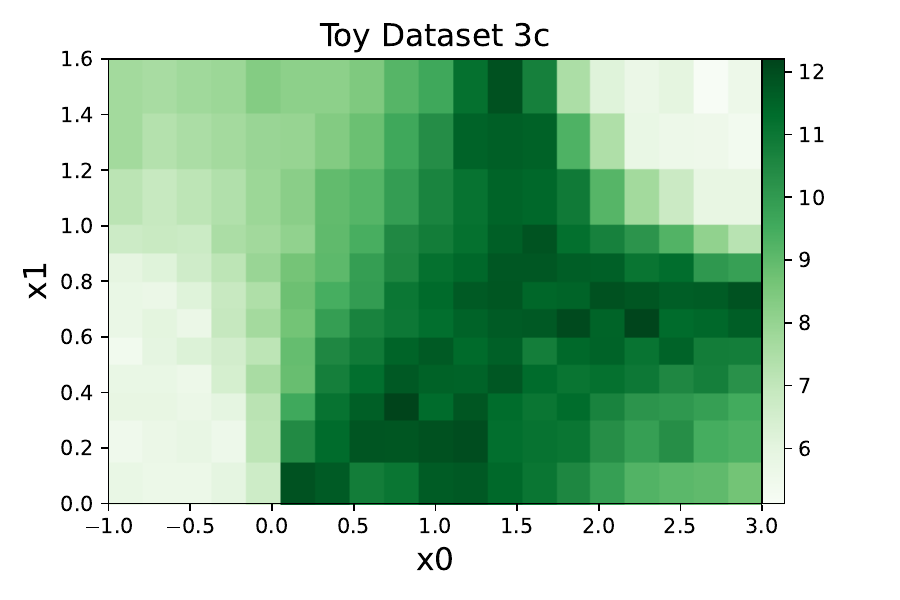}
     \end{subfigure}
    
    \caption{Toy Dataset 3: three 2D binned sub-datasets manually generated without reference to an underlying function.}
    \label{fig:toy_dataset_3}
\end{figure}

In Toy Dataset 3, we consider three 2D binned sub-datasets, labeled 3a, 3b, and 3c, as shown in Fig.~\ref{fig:toy_dataset_3}.
These datasets are manually generated without reference to an underlying function and are used to demonstrate applications such as deriving smooth scale factors from binned data with more than one independent variable.
The framework can be easily extended to datasets with multiple independent variables.

\begin{figure}[!t]
     \centering
     \begin{subfigure}[b]{0.6\textwidth}
         \centering
         \includegraphics[width=\textwidth]{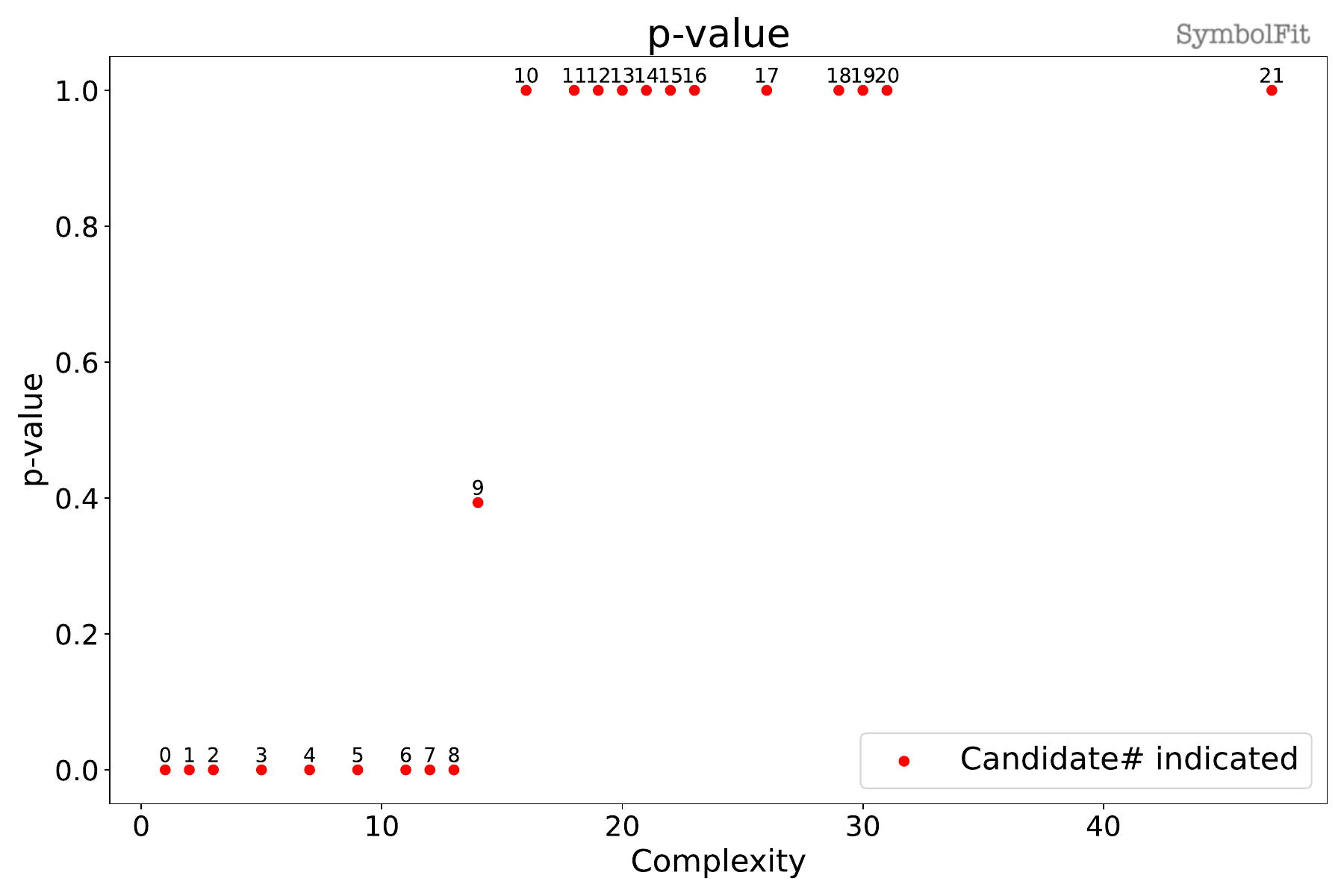}
         \caption{Toy Dataset 3a.}
     \end{subfigure}\\\vspace{0.3cm}
     \begin{subfigure}[b]{0.6\textwidth}
         \centering
         \includegraphics[width=\textwidth]{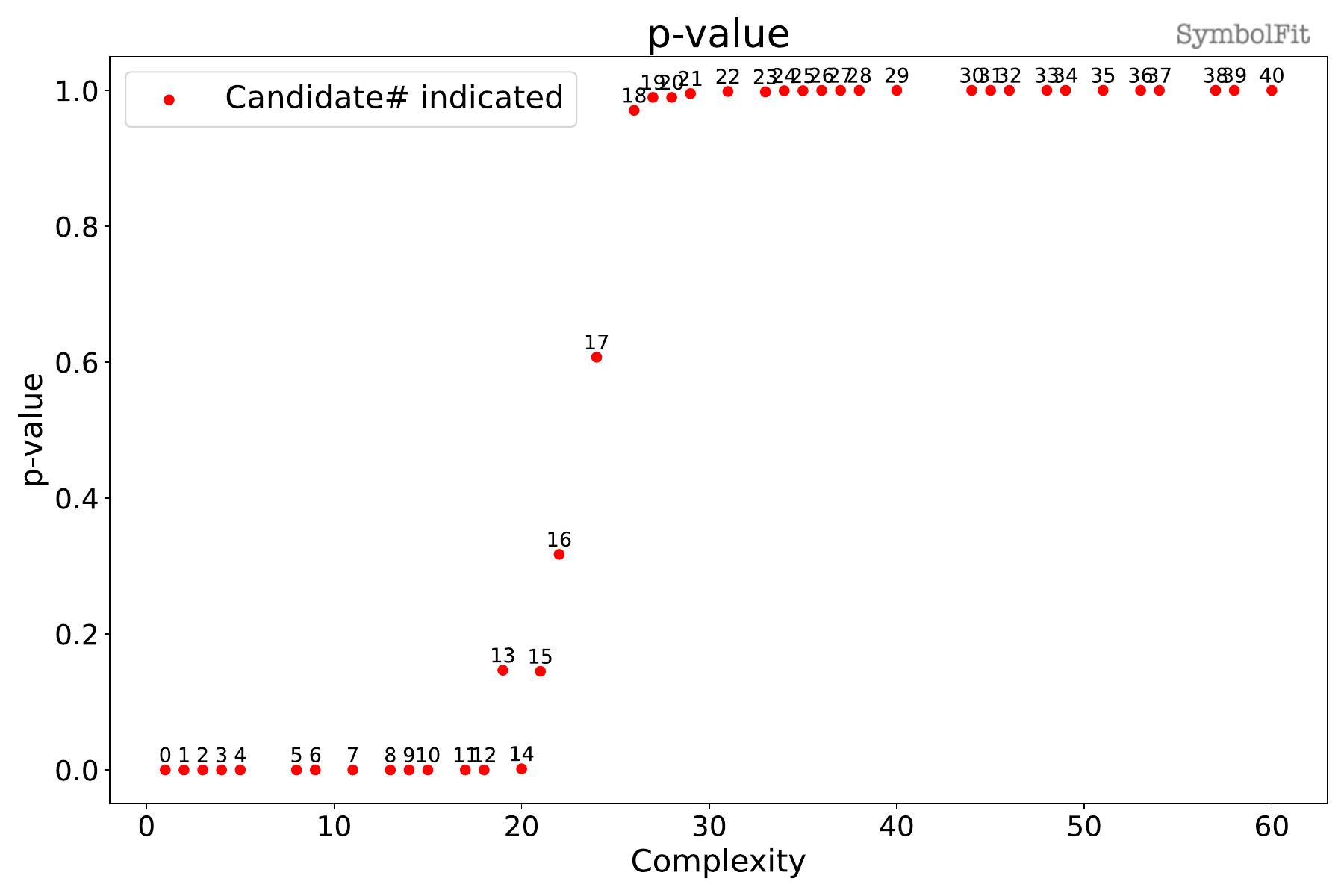}
         \caption{Toy Dataset 3b.}
     \end{subfigure}\\\vspace{0.3cm}
     \begin{subfigure}[b]{0.6\textwidth}
         \centering
         \includegraphics[width=\textwidth]{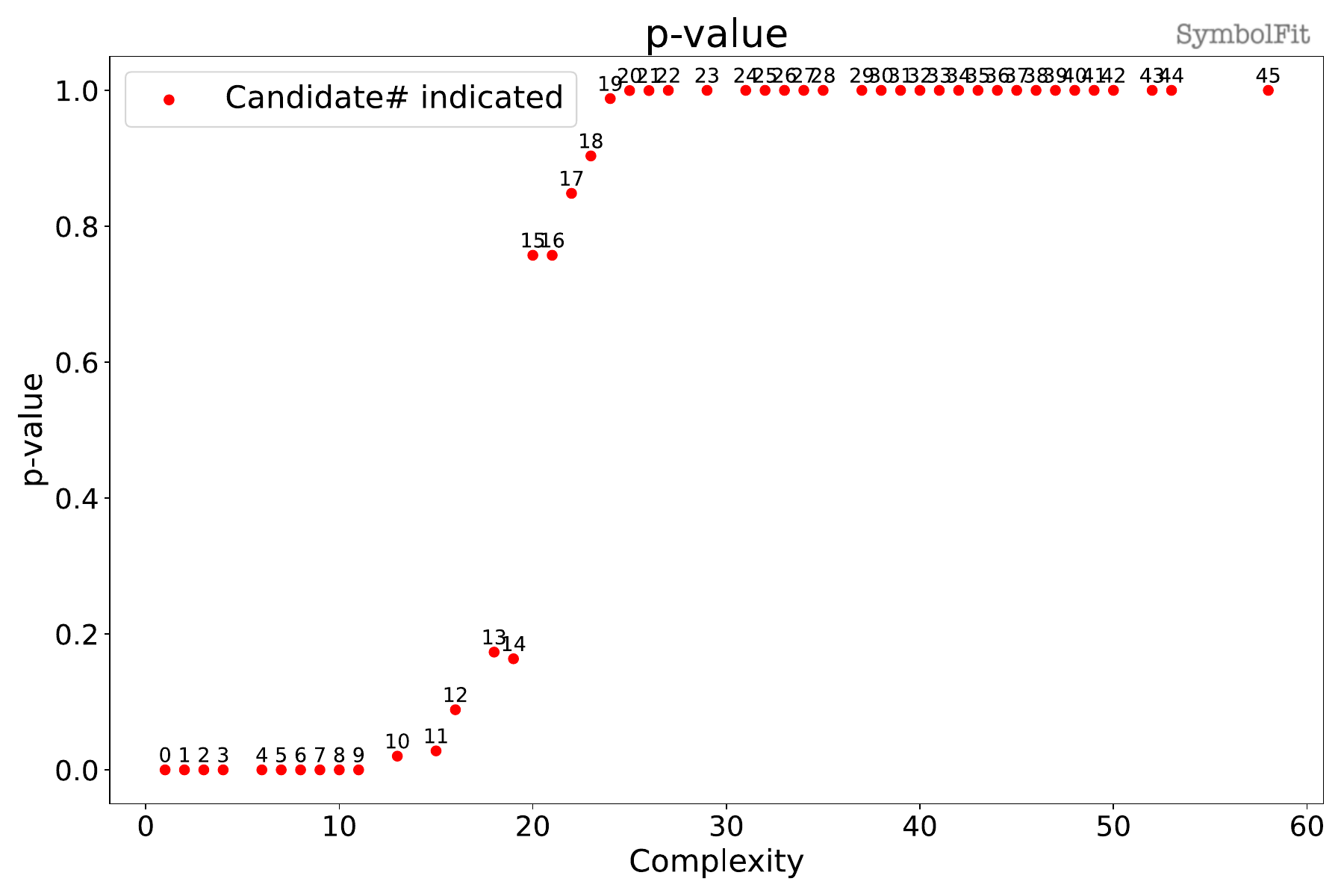}
         \caption{Toy Dataset 3c.}
     \end{subfigure}\vspace{0.3cm}
    
    \caption{p-value vs. function complexity.
    A total of 22, 41, and 46 candidate functions (labeled \#0--\#21, \#0--\#40, and \#0--\#45) were obtained from a single fit on Toy Dataset 3a, 3b, and 3c, respectively.}
    \label{fig:toy_dataset_3_chi2}
\end{figure}

We use the same $\tt{PySR}$ configuration applied to Toy Dataset 1, as shown in List.~\ref{config-toy1}, to fit these 2D binned datasets.
For each sub-dataset, a single run of $\tt{SymbolFit}$ is performed to generate a batch of candidate functions.
Fig.~\ref{fig:toy_dataset_3_chi2} shows the p-value plotted against function complexity.
Several candidate functions for each sub-dataset are selected and listed in Tab.~\ref{tab:toy3_candidates}.

\begin{figure}[!t]
     \centering
     \begin{subfigure}[b]{1\textwidth}
         \centering
         \includegraphics[width=\textwidth]{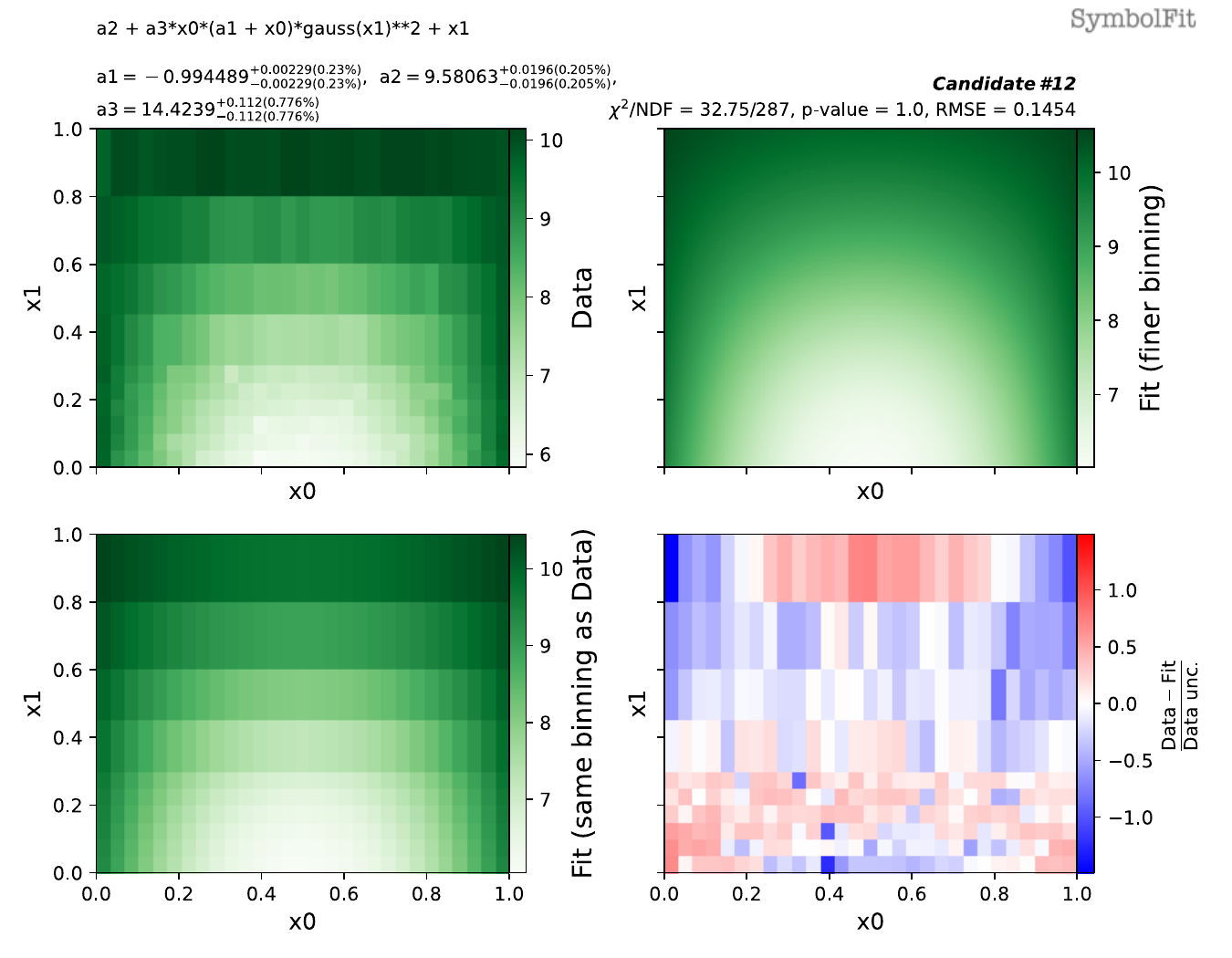}
     \end{subfigure}
    
    \caption{Candidate function \#12 for Toy Dataset 3a (see Tab.~\ref{tab:toy3_candidates}).
        The parameterized form of this function is shown at the top of the figure, along with the best-fit values and associated uncertainties.
        Upper left: the binned data being fitted.
        Lower left: the candidate function plotted with the same binning as the fitting data.
        Upper right: the candidate function plotted with a finer binning.
        Lower right: the residual error, $\frac{\text{Data}-\text{Fit}}{\text{Uncertainty}}$, in units of the data uncertainty.}
    \label{fig:toy_dataset_3a_candidates}
\end{figure}

\begin{figure}[!t]
     \centering
     \begin{subfigure}[b]{1\textwidth}
         \centering
         \includegraphics[width=\textwidth]{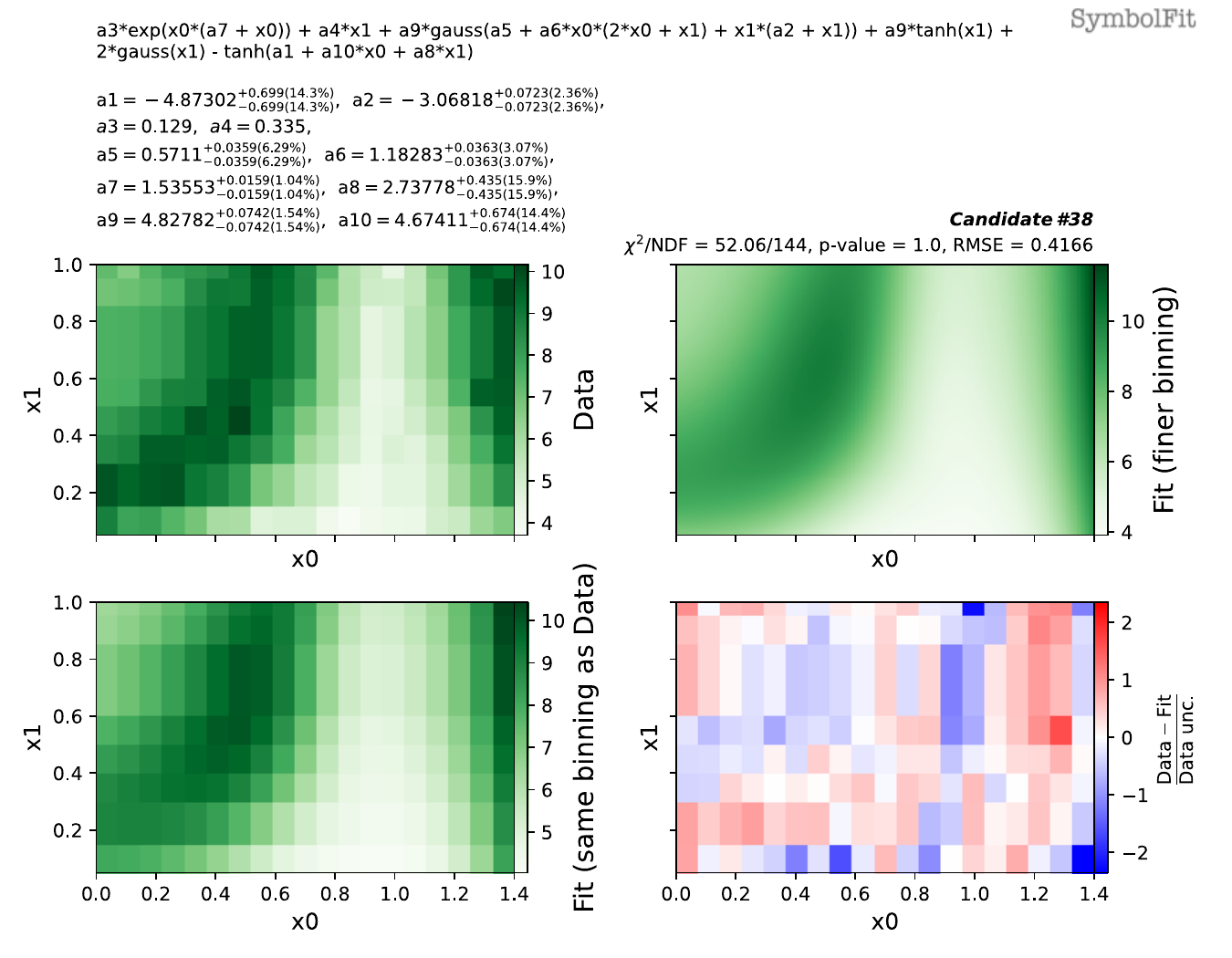}
     \end{subfigure}
    
    \caption{Candidate function \#38 for Toy Dataset 3b (see Tab.~\ref{tab:toy3_candidates}).
        The parameterized form of this function is shown at the top of the figure, along with the best-fit values and associated uncertainties.
        Upper left: the binned data being fitted.
        Lower left: the candidate function plotted with the same binning as the fitting data.
        Upper right: the candidate function plotted with a finer binning.
        Lower right: the residual error, $\frac{\text{Data}-\text{Fit}}{\text{Uncertainty}}$, in units of the data uncertainty.}
    \label{fig:toy_dataset_3b_candidates}
\end{figure}

\begin{figure}[!t]
     \centering
     \begin{subfigure}[b]{1\textwidth}
         \centering
         \includegraphics[width=\textwidth]{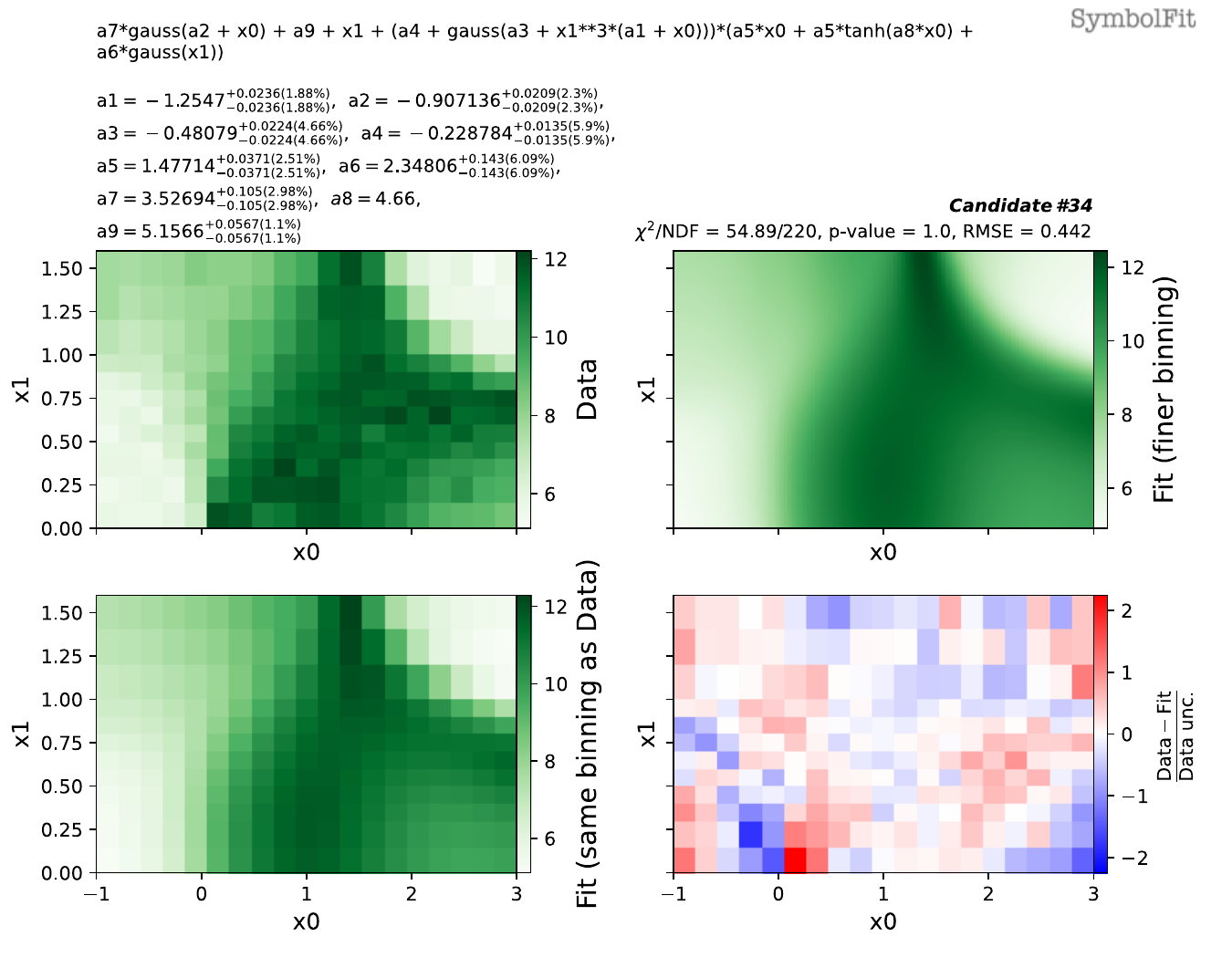}
     \end{subfigure}
    
    \caption{Candidate function \#34 for Toy Dataset 3c (see Tab.~\ref{tab:toy3_candidates}).
        The parameterized form of this function is shown at the top of the figure, along with the best-fit values and associated uncertainties.
        Upper left: the binned data being fitted.
        Lower left: the candidate function plotted with the same binning as the fitting data.
        Upper right: the candidate function plotted with a finer binning.
        Lower right: the residual error, $\frac{\text{Data}-\text{Fit}}{\text{Uncertainty}}$, in units of the data uncertainty.}
    \label{fig:toy_dataset_3c_candidates}
\end{figure}

A candidate function is shown for each of the three sub-datasets: \#26 for sub-dataset 3a in Fig.~\ref{fig:toy_dataset_3a_candidates}, \#39 for sub-dataset 3b in Fig.~\ref{fig:toy_dataset_3b_candidates}, and \#37 for sub-dataset 3c in Fig.~\ref{fig:toy_dataset_3c_candidates}.

\clearpage

\begin{table*}[!t]
\caption{Examples candidate functions for Toy Dataset 3 are listed.
The example candidate functions--\#12 for Toy Dataset 3a, \#38 for 3b, and \#34 for 3c--are plotted in Fig.~\ref{fig:toy_dataset_3a_candidates}, Fig.~\ref{fig:toy_dataset_3b_candidates}, and Fig.~\ref{fig:toy_dataset_3c_candidates}, respectively.
Numerical values are rounded to three significant figures for display purposes.}
\label{tab:toy3_candidates}
\centering
\resizebox{\textwidth}{!}{
\begin{tabular}{c|l|c|c|c|c}\hline
    \multicolumn{6}{c}{\bf{Toy Dataset 3a}} \\ \hline

    \textbf{Complexity} & \textbf{Candidate function} & \textbf{\# param.} & \textbf{$\chi^2/\text{NDF}$} & \textbf{$\chi^2/\text{NDF}$} & \textbf{p-value} \\
    & (after ROF) & & (before ROF) & (after ROF) & (after ROF) \\ \hline
    & & & & & \\
    14 (\#9) & $8.08\gauss(x_0(-2.76 + x_1)) + 9.46x_0 + x_1$ & 3 & 292.9 / 287 = 1.02 & 292.9 / 287 = 1.02 & 0.393 \\
    & & & & & \\ \hline
    & & & & & \\
    19 \textbf{(\#12)} & $9.58 + 14.4x_0(-0.994 + x_0)(\gauss(x_1))^2 + x_1$ & 3 & 41.06 / 287 = 0.1431 & 32.75 / 287 = 0.1141 & 1.0 \\
    & & & & & \\ \hline
    & & & & & \\
    20 (\#13) & $9.92 + 16.0x_0(-1 + x_0)(-0.541x_1 + \gauss(x_1))$ & 4 & 15.8 / 286 = 0.0552 & 13.22 / 286 = 0.0462 & 1.0 \\
    & & & & & \\ \hline

    \multicolumn{6}{c}{\bf{Toy Dataset 3b}} \\\hline
    
    \textbf{Complexity} & \textbf{Candidate function} & \textbf{\# param.} & \textbf{$\chi^2/\text{NDF}$} & \textbf{$\chi^2/\text{NDF}$} & \textbf{p-value} \\
    & (after ROF) & & (before ROF) & (after ROF) & (after ROF) \\ \hline
    & & & & & \\
    22 (\#16) & $1.38\exp(x_0^2) + 7.08\gauss(-0.736x_1 + x_0^2 + x_0) + \tanh(x_1)$ & 3 & 156.7 / 149 = 1.052 & 156.7 / 149 = 1.052 & 0.3172 \\
    & & & & & \\ \hline
    & & & & & \\
    24 (\#17) & $1.3\exp(x_0^2) + 6.83\gauss(-0.726x_1 + x_0^2 + x_0) + \tanh(2.48x_1)$ & 3 & 144.4 / 149 = 0.969 & 143.7 / 149 = 0.964 & 0.607 \\
    & & & & & \\ \hline
    & & & & & \\
    26 (\#18) & $0.345 + 5.73x_1\gauss(x_1) + 5.73\gauss(-0.809x_1 + 2.62x_0^2) +$ & 3 & 118.4 / 149 = 0.7945 & 118.1 / 149 = 0.7928 & 0.971 \\
    & $\exp(x_0^2)$ & & & & \\
    & & & & & \\\hline
    & & & & & \\
    57 \textbf{(\#38)} & $4.83\gauss(0.571 + 1.18x_0(2x_0 + x_1) + x_1(-3.07 + x_1)) +$ & 8 & 52.8 / 144 = 0.3667 & 52.06 / 144 = 0.3615 & 1.0 \\
    & $+ 2\gauss(x_1) - \tanh(-4.87 + 4.67x_0 + 2.74x_1) +$ & & & & \\
    & $0.129\exp(x_0(1.54 + x_0))+4.83\tanh(x_1)+0.335x_1$ & & & & \\
    & & & & & \\ \hline

    \multicolumn{6}{c}{\bf{Toy Dataset 3c}} \\\hline
    
    \textbf{Complexity} & \textbf{Candidate function} & \textbf{\# param.} & \textbf{$\chi^2/\text{NDF}$} & \textbf{$\chi^2/\text{NDF}$} & \textbf{p-value} \\
    & (after ROF) & & (before ROF) & (after ROF) & (after ROF) \\ \hline
    & & & & & \\
    20 (\#15) & $-0.886x_0(\gauss(x_0x_1) + \exp(x_1)) + 6.74\tanh(x_0) + 8.24$ & 3 & 214.2 / 225 = 0.9521 & 209.9 / 225 = 0.9328 & 0.7573 \\
    & & & & & \\ \hline
    & & & & & \\
    23 (\#18) & $1.9x_0(-0.296 + \gauss(x_1^2)) + 4.38\gauss(-1.14 + x_0) + 6.56$ & 5 & 198.1 / 223 = 0.8881 & 196 / 223 = 0.8789 & 0.9035 \\
    & & & & & \\ \hline
    & & & & & \\
    25 (\#20) & $4.33\gauss(-1.14 + x_0) + 6.57 + x_0(-0.238 + \gauss(x_1^2))\times$ & 5 & 160.3 / 223 = 0.7188 & 157.3 / 223 = 0.7054 & 0.9997 \\
    & $(1.33 + x_1)$ & & & & \\
    & & & & & \\\hline
    & & & & & \\
    42 \textbf{(\#34)} & $5.16 + x_1 + (-0.229 + \gauss(-0.481 + x_1^3(-1.25 + x_0)))\times$ & 8 & 59.38 / 220 = 0.2699 & 54.89 / 220 = 0.2495 & 1.0 \\
    & $(1.48x_0 + 1.48\tanh(4.66x_0) + 2.35\gauss(x_1)) +$ & & & & \\
    & $3.53\gauss(-0.907 + x_0)$ & & & & \\
    & & & & & \\ \hline
    
    \end{tabular}%
     }
\end{table*}

\clearpage

\section{Summary}
\label{sec:summary}

We have developed a framework called $\tt{SymbolFit}$ that automates parametric modeling without the need for a priori specification of a functional form to fit data.
The framework utilizes symbolic regression to machine-search for suitable functional forms and incorporates a re-optimization step to improve the candidate functions and provide uncertainty estimates.
Due to the nature of genetic programming, each symbolic regression fit generates a batch of candidate functions with a variety of forms that can potentially model the data well.
This offers flexibility and allows users to select the most suitable candidates for their downstream tasks.

Our primary focus is on applications in high-energy physics data analysis, specifically in signal and background modeling, as well as the derivation of smooth scale factors from binned data.
There is no reason it cannot be applied to other fields where parametric modeling is needed.
We have demonstrated our framework using five real proton-proton collision datasets from new physics searches at the CERN LHC, as well as several toy datasets, including two-dimensional binned data.
Our framework has been shown to easily generate a variety of suitable candidate functions for nontrivial distributions using a very simple fit configuration, even without prior knowledge of the final functional forms.
The candidate functions obtained from our framework are readily comparable to the empirical functions derived from traditional methods, which would have required extensive manual and iterative effort.
This suggests that our method is preferable to traditional methods since the labor-intensive process of finding adequate functional forms is now automated using machine learning.

Since the fit outputs in this approach are parametric closed-form functions, the resulting model representation is identical to that from traditional parametric modeling methods.
This allows seamless integration with established downstream statistical tools used in LHC experiments such as $\tt{Combine}$ and $\tt{pyhf}$ for hypothesis testing.
Furthermore, the ease of generating a wide range of well-fitted functions within this framework facilitates flexible modeling, as the choice of functions can be treated as a source of systematic uncertainty using well-established techniques, such as the discrete profiling method.

We have developed an API for the framework, designed for easy use by the high-energy physics community.
This API automates and streamlines the process of finding suitable functions with uncertainty estimates for modeling binned data, significantly reducing manual effort.
Our goal is to transform the approach to parametric modeling in high-energy physics experiments, moving away from traditional fitting methods that rely on manually determining empirical functions on a case-by-case basis, which is both time-consuming and prone to bias.

\section*{Acknowledgements}
DR is supported by the U.S. Department of Energy (DOE), Office of Science, Office of High Energy Physics Early Career Research program under Award No. DE-SC0025324.
SD is supported by the U.S. DOE, Office of Science, Office of High Energy Physics, under Award No. DE-SC0017647.
JD is supported by the Research Corporation for Science Advancement (RCSA) under grant \#CS-CSA-2023-109, Alfred P. Sloan Foundation under grant \#FG-2023-20452, U.S. DOE, Office of Science, Office of High Energy Physics Early Career Research program under Award No. DE-SC0021187
JD, PH, and VL are supported by the U.S. National Science Foundation (NSF) Harnessing the Data Revolution (HDR) Institute for Accelerating AI Algorithms for Data Driven Discovery (A3D3) under Cooperative Agreement PHY-2117997.
PH is also supported by the Institute for Artificial Intelligence and Fundamental Interactions (IAIFI), under the NSF grant \#PHY-2019786.
EL is supported by the U.S. DOE, Office of Science, Office of High Energy Physics, under Award No. DE-SC0007901.

\section*{References}
\bibliographystyle{naturemag}
\bibliography{references}

\appendix
\addtocontents{toc}{\fixappendix}
\section{More examples}
\label{apd:more-examples}

\subsection{Toy dataset 2 (1D) [arbitrary shapes]}
\label{sec:toy-dataset-2}

\begin{figure}[!t]
     \centering
     \begin{subfigure}[b]{0.495\textwidth}
         \centering
         \includegraphics[width=\textwidth]{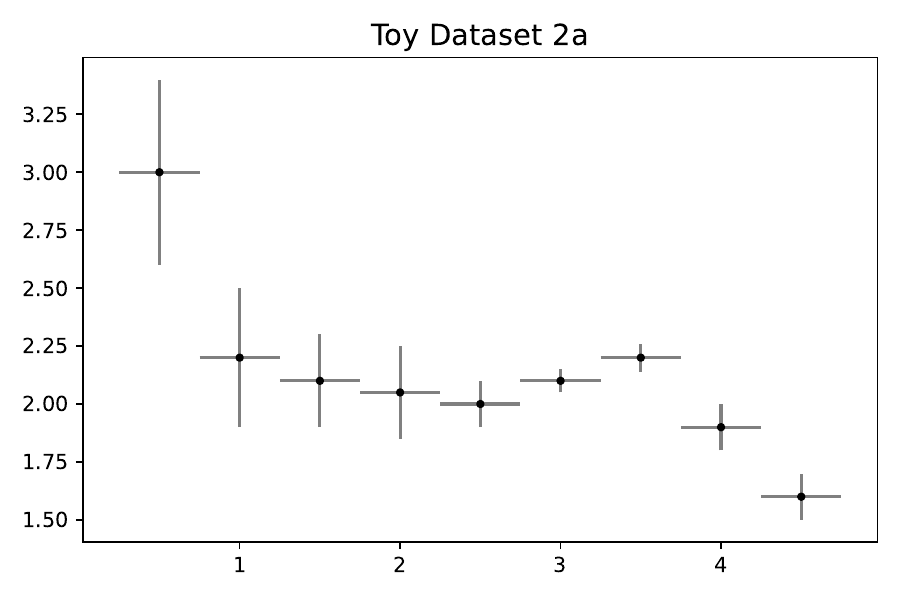}
     \end{subfigure}
     \begin{subfigure}[b]{0.495\textwidth}
         \centering
         \includegraphics[width=\textwidth]{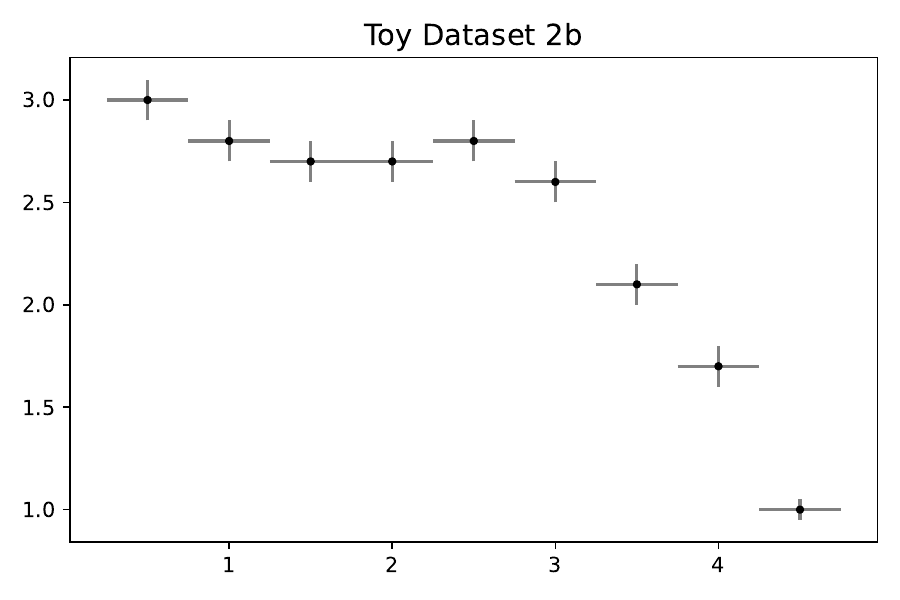}
     \end{subfigure}
     \begin{subfigure}[b]{0.495\textwidth}
         \centering
         \includegraphics[width=\textwidth]{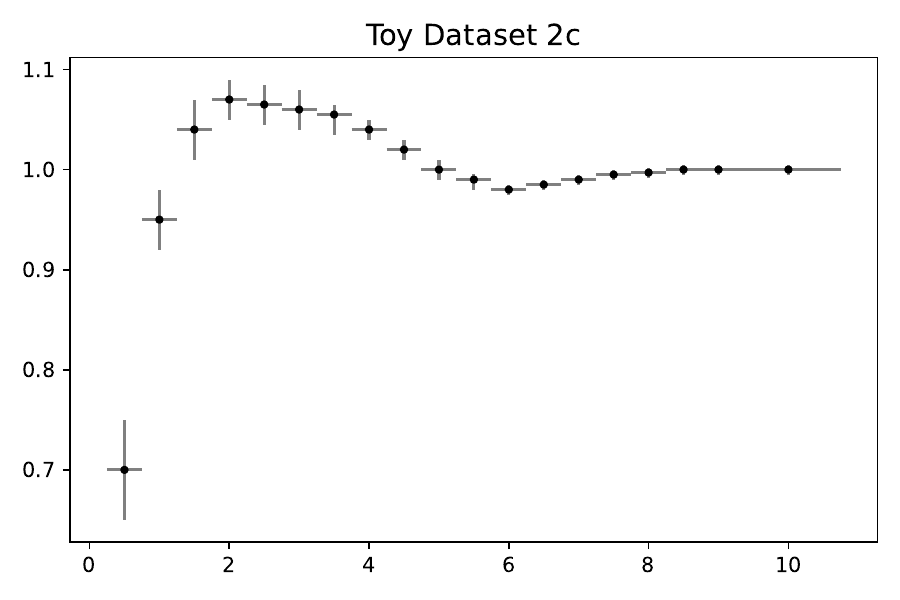}
     \end{subfigure}
     
    \caption{Toy Dataset 2: three 1D binned sub-datasets manually generated without reference to an underlying function.}
    \label{fig:toy_dataset_2}
\end{figure}

In Toy Dataset 2, we consider three 1D binned sub-datasets, labeled 2a, 2b, and 2c, as shown in Fig.~\ref{fig:toy_dataset_2}.
These datasets are manually generated without reference to an underlying function and are used to demonstrate applications such as deriving smooth scale factors from binned data.

\begin{figure}[!t]
     \centering
     \begin{subfigure}[b]{0.6\textwidth}
         \centering
         \includegraphics[width=\textwidth]{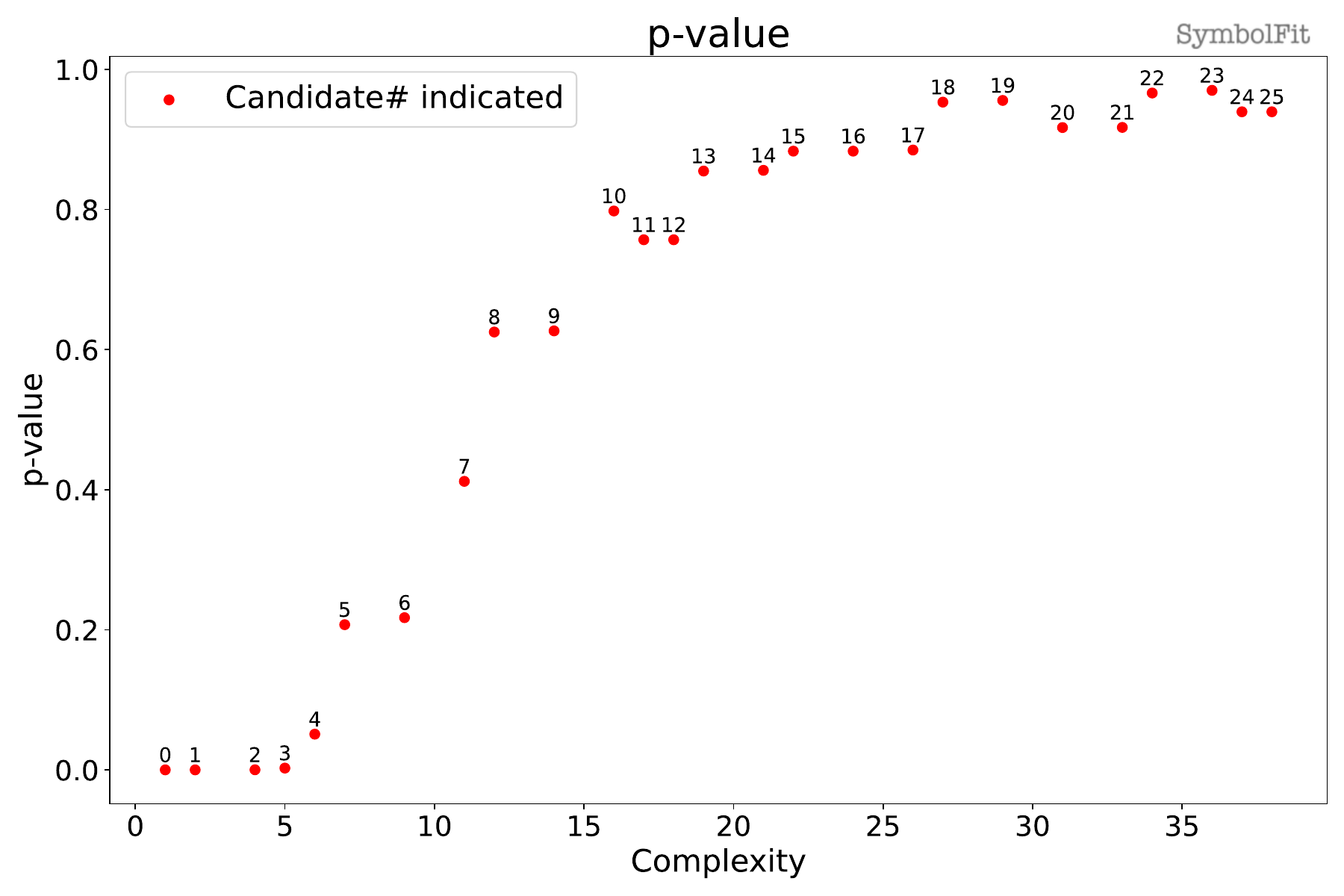}
         \caption{Toy Dataset 2a.}
     \end{subfigure}\\\vspace{0.3cm}
     \begin{subfigure}[b]{0.6\textwidth}
         \centering
         \includegraphics[width=\textwidth]{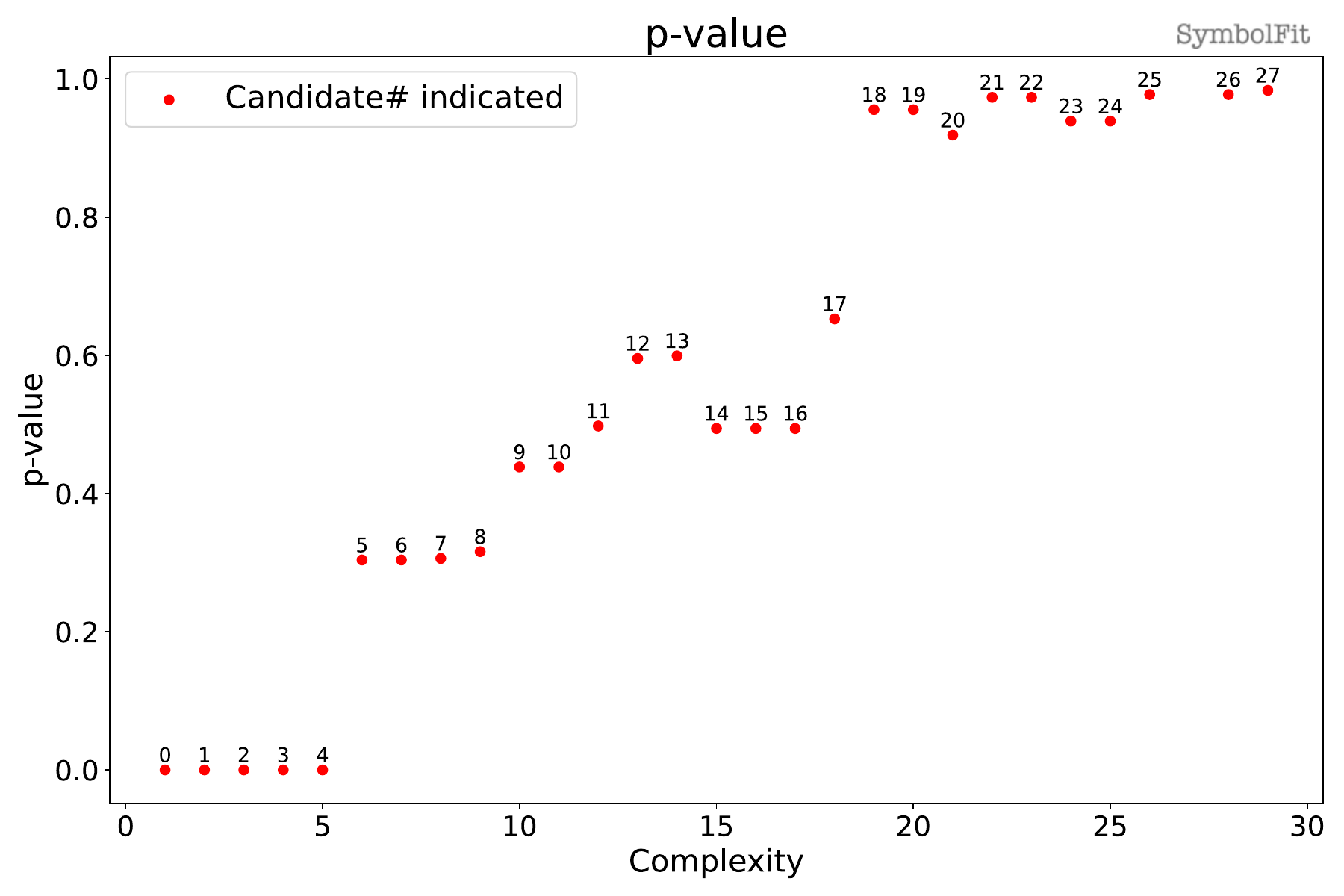}
         \caption{Toy Dataset 2b.}
     \end{subfigure}\\\vspace{0.3cm}
     \begin{subfigure}[b]{0.6\textwidth}
         \centering
         \includegraphics[width=\textwidth]{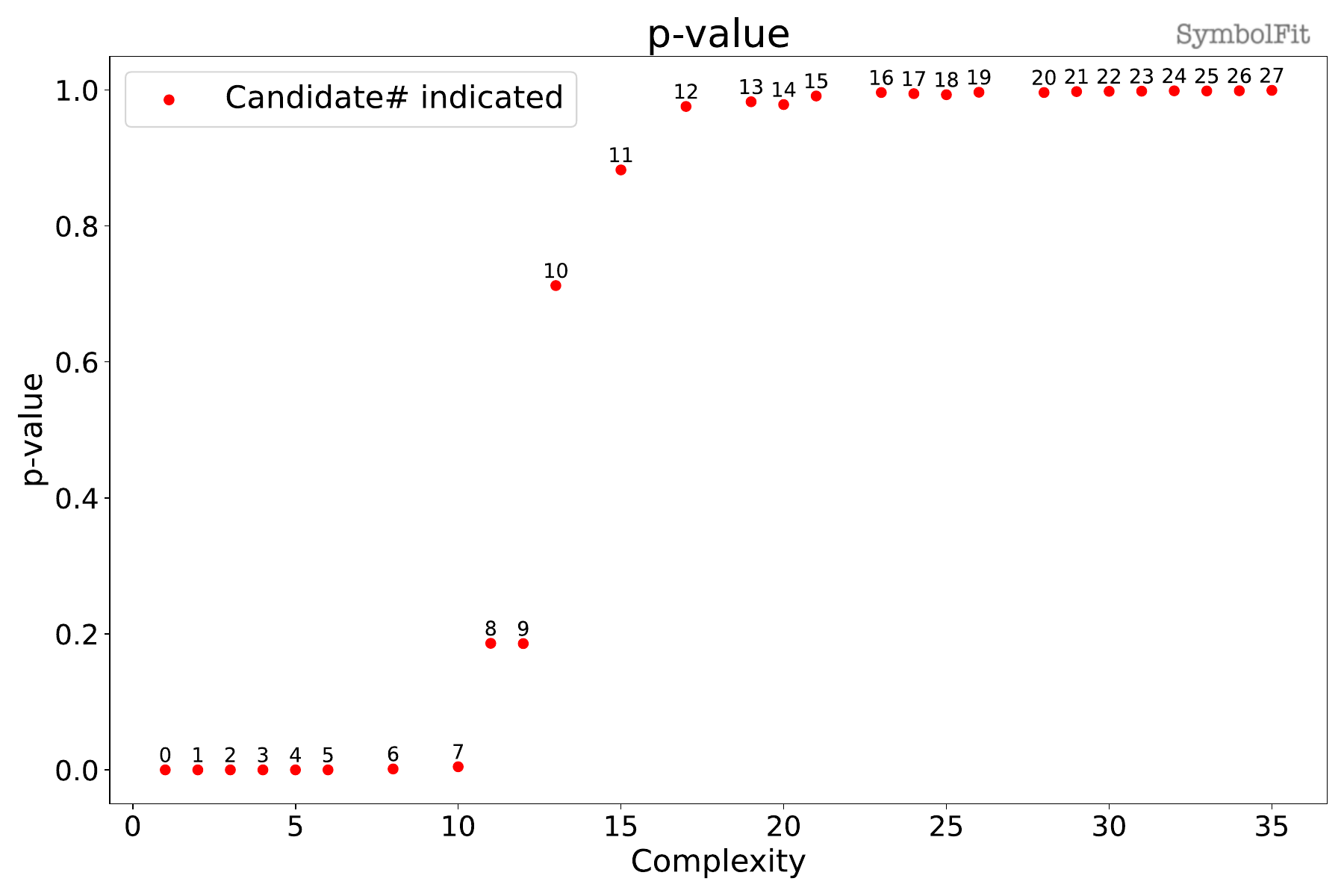}
         \caption{Toy Dataset 2c.}
     \end{subfigure}\\\vspace{0.3cm}
     
    \caption{p-value vs. function complexity.
    A total of 26, 28, and 28 candidate functions (labeled \#0--\#25, \#0--\#27, and \#0--\#27) were obtained from a single fit on Toy Dataset 2a, 2b, and 2c, respectively.}
    \label{fig:toy_dataset_2_chi2}
\end{figure}

We use the same $\tt{PySR}$ configuration applied to the five LHC datasets, as shown in List.~\ref{config-lhc}, to fit these 1D binned datasets.
For each sub-dataset, a single run of $\tt{SymbolFit}$ is performed to generate a batch of candidate functions.
Fig.~\ref{fig:toy_dataset_2_chi2} shows the p-value plotted against function complexity.
Several candidate functions for each sub-datasets are selected and listed in Tab.~\ref{tab:toy2_candidates}.

\begin{figure}[!t]
     \centering
     \begin{subfigure}[b]{0.495\textwidth}
         \centering
         \includegraphics[width=\textwidth]{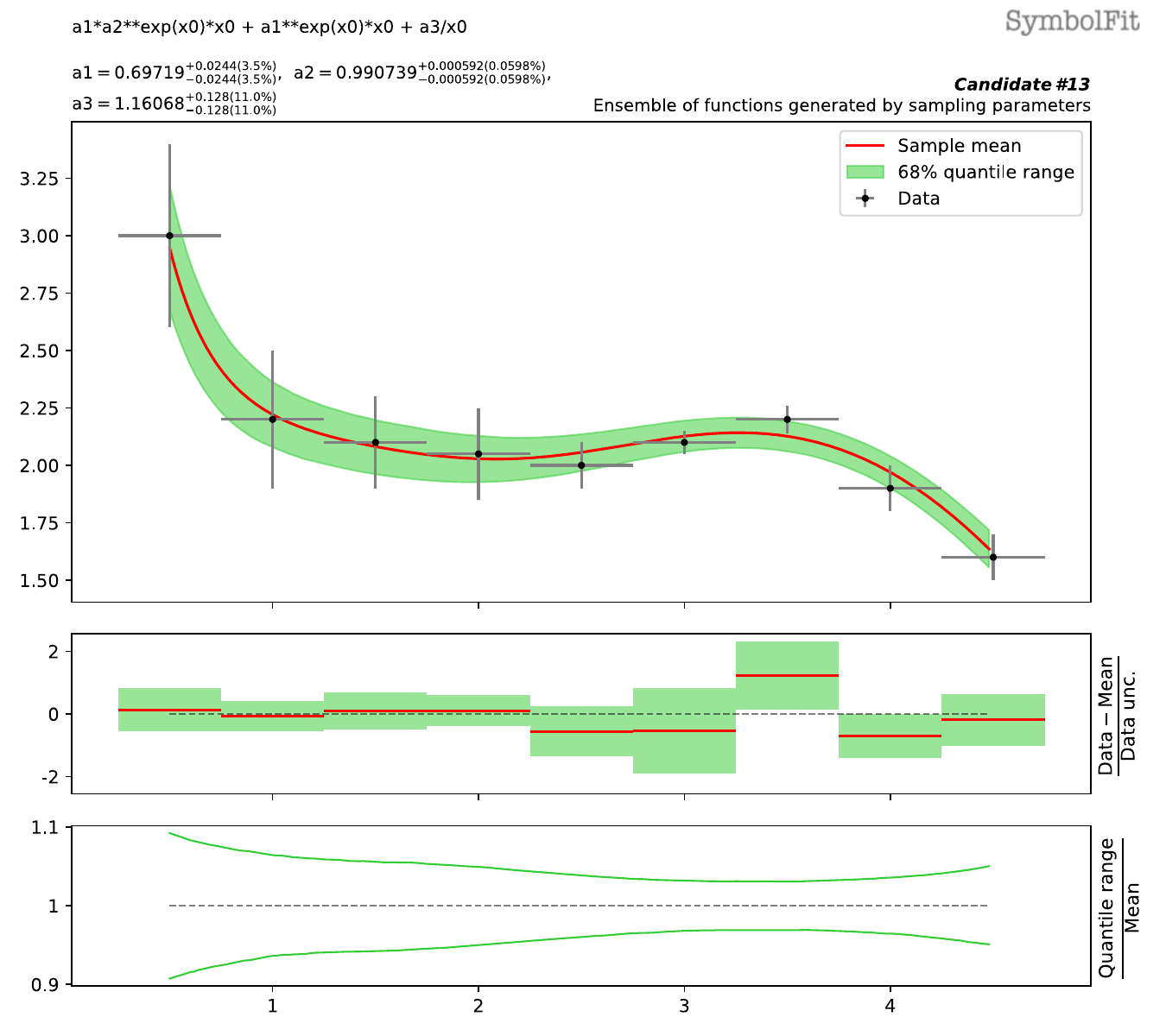}
         \caption{Candidate function \#13 for Toy Dataset 2a.}
     \end{subfigure}
     \begin{subfigure}[b]{0.495\textwidth}
         \centering
         \includegraphics[width=\textwidth]{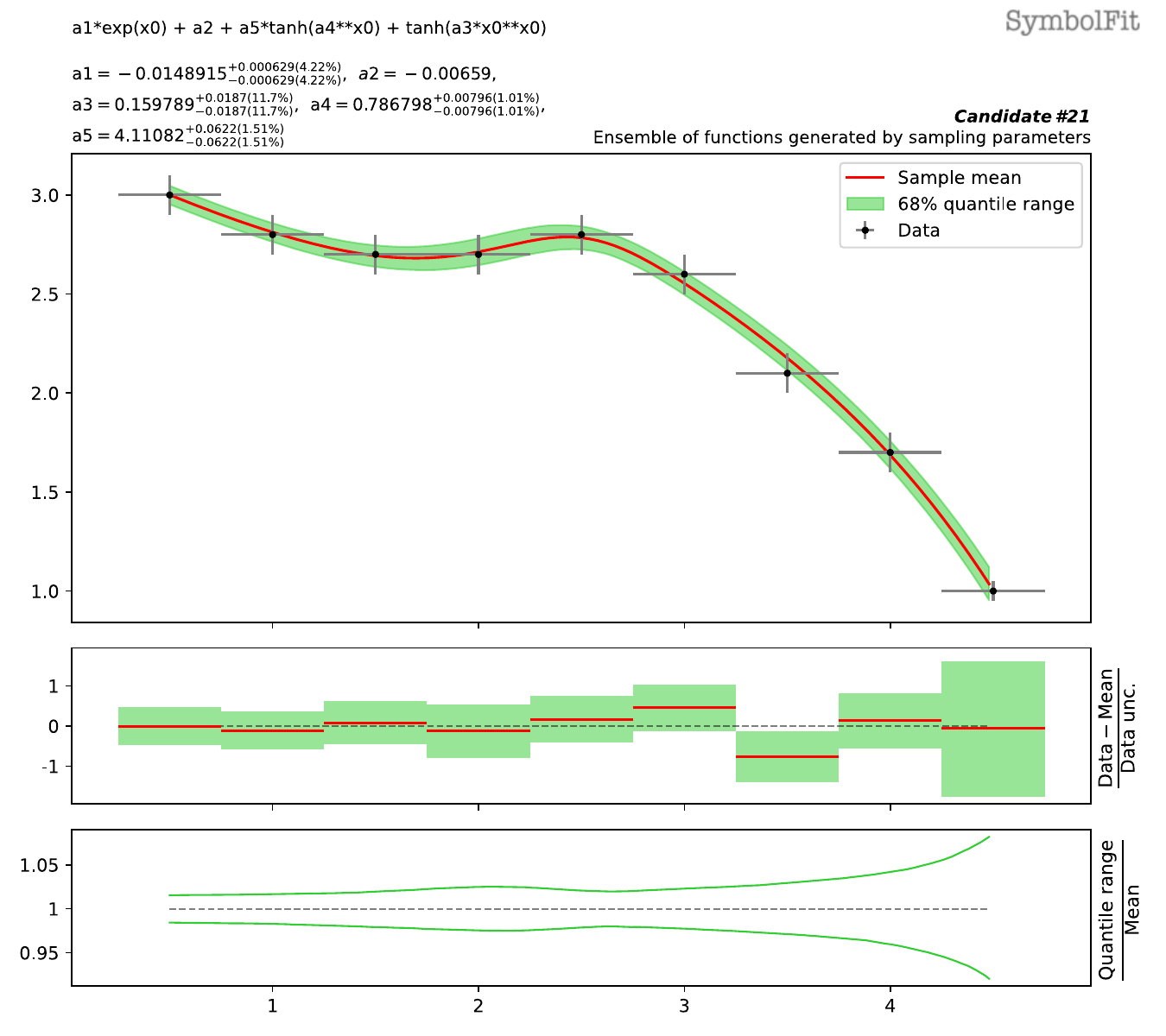}
         \caption{Candidate function \#21 for Toy Dataset 2b.}
     \end{subfigure}\\\vspace{0.5cm}
     \begin{subfigure}[b]{0.495\textwidth}
         \centering
         \includegraphics[width=\textwidth]{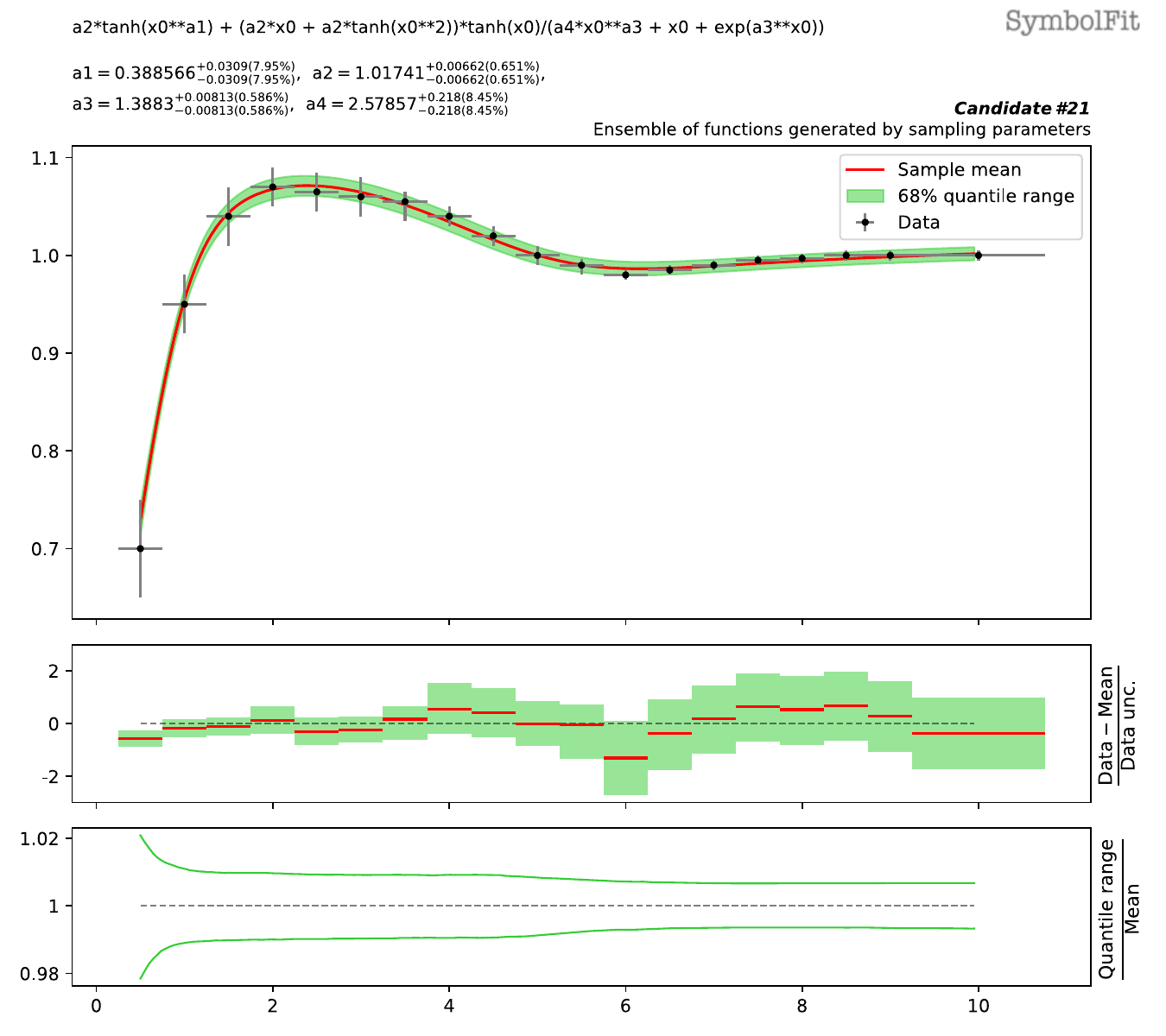}
         \caption{Candidate function \#21 for Toy Dataset 2c.}
     \end{subfigure}\vspace{0.5cm}
    
    \caption{Example candidate functions for Toy Dataset 2 (see Tab.~\ref{tab:toy2_candidates}).
    To visualize the total uncertainty coverage of each candidate function, the green band in each subfigure represents the 68\% quantile range of functions obtained by sampling parameters, taking into account the best-fit values and the covariance matrix within a multidimensional normal distribution.
    The red line denotes the mean of the function ensemble.
    At the top of each subfigure, the candidate function and the fitted parameters are shown.
    The middle panel shows the weighted residual error: $\frac{\text{Data}-\text{Mean}}{\text{Data unc.}}$.
    The bottom panel shows the ratio of the 68\% quantile range to the mean.}
    \label{fig:toy_dataset_2_candidate}
\end{figure}

Fig.~\ref{fig:toy_dataset_2_candidate} shows candidate functions \#13, \#21, and \#21 with uncertainty coverage, for sub-datasets 2a, 2b, and 2c, respectively.

\begin{table*}[!t]
\caption{Examples candidate functions for Toy Dataset 2 are listed.
The example candidate functions--\#13 for Toy Dataset 2a, \#21 for 2b, and \#21 for 2c--are plotted in Fig.~\ref{fig:toy_dataset_2_candidate}.
Numerical values are rounded to three significant figures for display purposes.}
\label{tab:toy2_candidates}
\centering
\resizebox{\textwidth}{!}{
\begin{tabular}{c|l|c|c|c|c}\hline
    \multicolumn{6}{c}{\bf{Toy Dataset 2a}} \\\hline
    
    \textbf{Complexity} & \textbf{Candidate function} & \textbf{\# param.} & \textbf{$\chi^2/\text{NDF}$} & \textbf{$\chi^2/\text{NDF}$} & \textbf{p-value} \\
    & (after ROF) & & (before ROF) & (after ROF) & (after ROF) \\ \hline
    &  & & & & \\
    12 (\#8) & $0.667\times 0.991^{\exp(x)}x + 1.44/x$ & 3 & 5.273 / 6 = 0.8789 & 4.382 / 6 = 0.7303 & 0.6251 \\
    & & & & & \\ \hline
    &  & & & & \\
    19 \textbf{(\#13)} & $0.697\times 0.991^{\exp(x)}x + 0.697^{\exp(x)}x + 1.16/x$ & 3 & 2.863 / 6 = 0.4772 & 2.619 / 6 = 0.4365 & 0.8549 \\
    & & & & & \\ \hline
    &  & & & & \\
    38 (\#25) & $0.907^{0.0993\exp(x)}x(0.0609 + \tanh(0.596^{0.632/x})) +$ & 3 & 1.776 / 6 = 0.2961 & 1.771 / 6 = 0.2951 & 0.9395 \\
    & $0.948/x + 0.948^{\exp(1.63x)}-0.115 $ & & & & \\
    &  & & & & \\ \hline

    \multicolumn{6}{c}{\bf{Toy Dataset 2b}} \\\hline
    
    \textbf{Complexity} & \textbf{Candidate function} & \textbf{\# param.} & \textbf{$\chi^2/\text{NDF}$} & \textbf{$\chi^2/\text{NDF}$} & \textbf{p-value} \\
    & (after ROF) & & (before ROF) & (after ROF) & (after ROF) \\ \hline
    &  & & & & \\
    14 (\#13) & $-0.0219\exp(x) + 1.96 + \tanh(x)^{-0.586 + x}$ & 2 & 5.613 / 7 = 0.8018 & 5.501 / 7 = 0.7858 & 0.5991 \\
    & & & & & \\ \hline
    &  & & & & \\
    19 (\#18) & $-0.0161\exp(x) + 0.177 + 3.02\times 0.825^{x} +$ & 4 & 2.092 / 5 = 0.4184 & 1.084 / 5 = 0.2169 & 0.9555 \\
    & $\tanh(0.177x^x)$ & & & & \\
    &  & & & & \\ \hline
    &  & & & & \\
    22 \textbf{(\#21)} & $-0.0149\exp(x) -0.00659 + 4.11\tanh(0.787^x) +$ & 4 & 0.857 / 5 = 0.1714 & 0.8561 / 5 = 0.1712 & 0.9733 \\
    & $\tanh(0.16x^x)$ & & & & \\
    &  & & & & \\ \hline

    \multicolumn{6}{c}{\bf{Toy Dataset 2c}} \\\hline
    
    \textbf{Complexity} & \textbf{Candidate function} & \textbf{\# param.} & \textbf{$\chi^2/\text{NDF}$} & \textbf{$\chi^2/\text{NDF}$} & \textbf{p-value} \\
    & (after ROF) & & (before ROF) & (after ROF) & (after ROF) \\ \hline
    &  & & & & \\
    13 (\#10) & $0.626x\exp(-1.38^x) + \tanh(x^{0.502})$ & 3 & 16.75 / 16 = 1.047 & 12.45 / 16 = 0.7784 & 0.7122 \\
    & & & & & \\ \hline
    &  & & & & \\
    15 (\#11) & $x/(2x + \exp(1.41^x)) + \tanh(x^{0.505})$ & 2 & 10.67 / 17 = 0.6276 & 10.48 / 17 = 0.6165 & 0.8823 \\
    & & & & & \\ \hline
    &  & & & & \\
    29 \textbf{(\#21)} & $(1.02x + 1.02\tanh(x^2))\tanh(x)/(2.58x^{1.39} + x +$ & 4 & 7.767 / 15 = 0.5178 & 4.079 / 15 = 0.2719 & 0.9975 \\
    & $\exp(1.39^x))+ 1.02\tanh(x^{0.389})$ & & & & \\
    &  & & & & \\ \hline

    \end{tabular}%
     }
\end{table*}

\clearpage

\subsection{CMS high-mass diphoton dataset (1D) [background modeling]}
\label{sec:diphoton}

CMS performed a search for high-mass diphoton resonances using proton-proton collision data at a center-of-mass energy of $\sqrt{s}=13$ TeV and reported no significant deviations from the Standard Model prediction~\cite{diphoton}.
The dataset for the diphoton spectrum is publicly available on HEPDATA at Ref.~\cite{hepdata.diphoton}.
In the analysis, CMS considered four empirical functions to model the background contribution in the distribution of the diphoton invariant mass, $m_{\gamma\gamma}$, and one of them is:
\begin{equation}
    \label{eq:diphoton}
    f(x)=p_0x^{p_1+p_2\log(x)},
\end{equation}
where $x=m_{\gamma\gamma}/\sqrt{s}$ is dimensionless and $p_{\{0,1,2\}}$ are free parameters.

We perform the same experiments conducted on the dijet dataset, as detailed in Sec.~\ref{sec:dijet}.
Starting from the original diphoton spectrum, we generate pseudodata by injecting a perturbed Gaussian signal centered at $m_{\gamma\gamma}=1320$ GeV ($s_1$), with a width of 74 GeV ($2s_2)$ and a signal strength of $s_0=15$.
To model the background, we blind the signal region by masking the $m_{\gamma\gamma}$ bins between 980 and 1400 GeV in the pseudodata and perform the fits.

\begin{figure}[!t]
     \centering
     \begin{subfigure}[b]{0.495\textwidth}
         \centering
         \includegraphics[width=\textwidth]{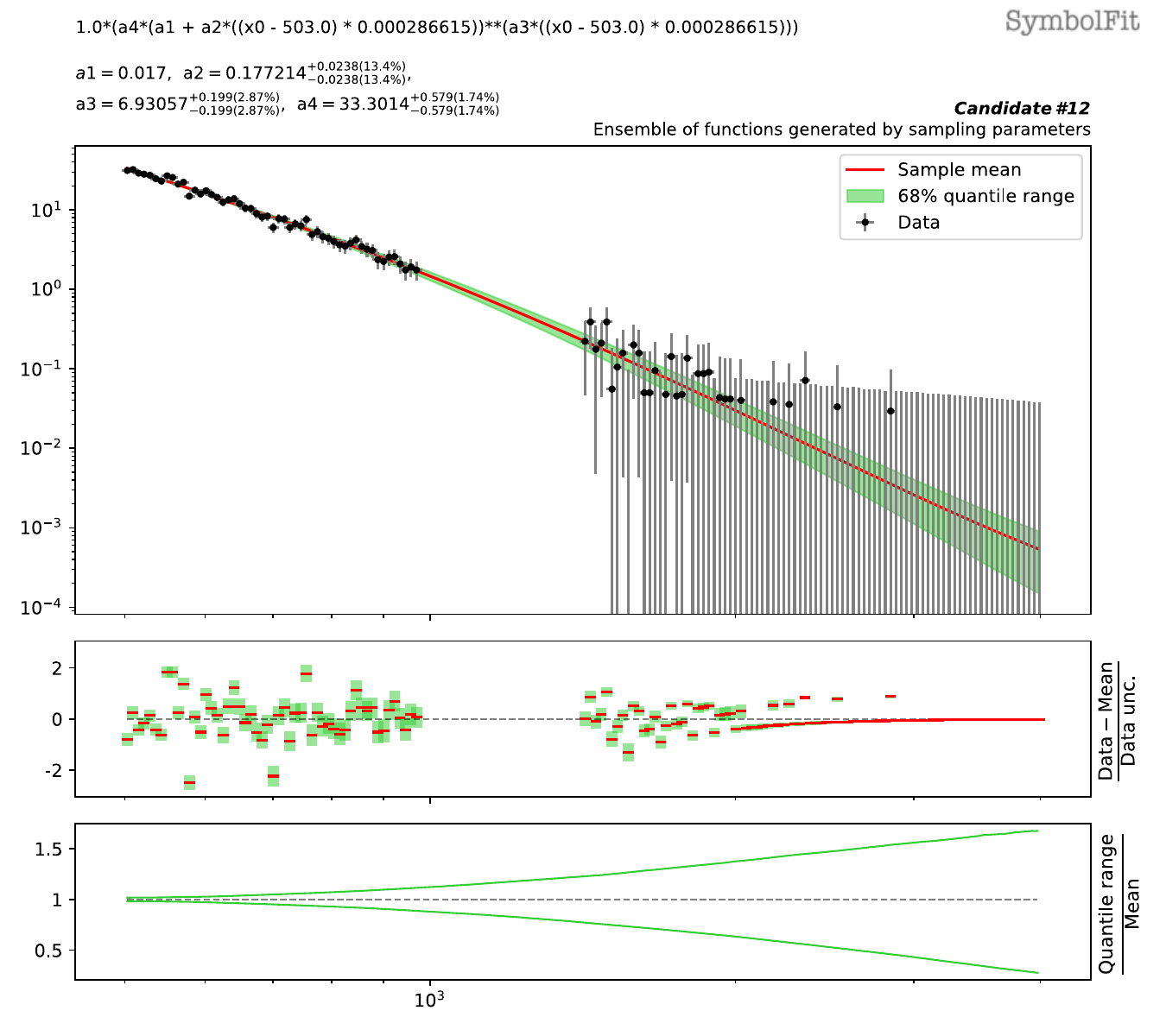}
         \caption{SR model 1.}
     \end{subfigure}
     \begin{subfigure}[b]{0.495\textwidth}
         \centering
         \includegraphics[width=\textwidth]{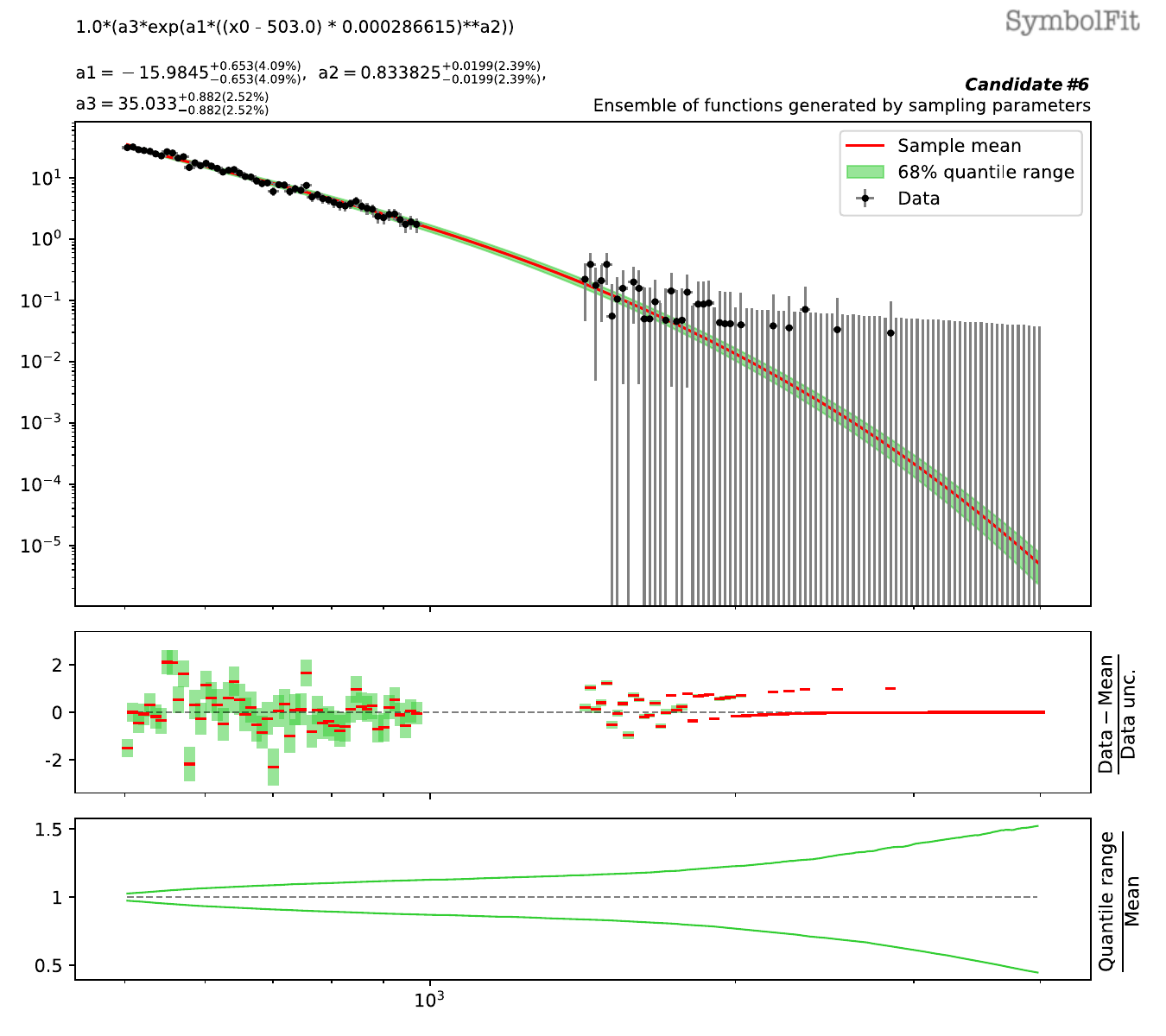}
         \caption{SR model 2.}
     \end{subfigure}\\\vspace{0.5cm}
     \begin{subfigure}[b]{0.495\textwidth}
         \centering
         \includegraphics[width=\textwidth]{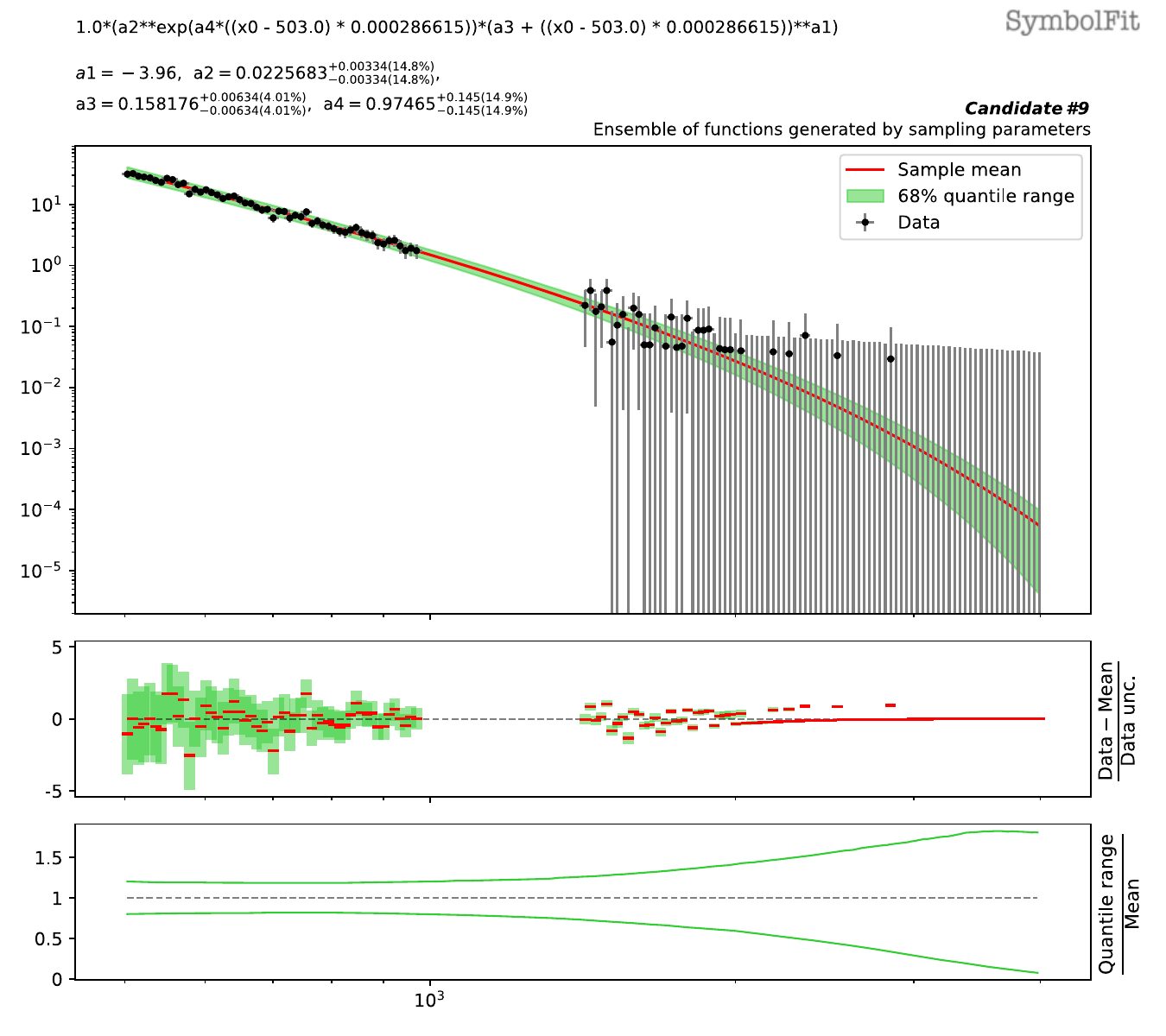}
         \caption{SR model 3.}
     \end{subfigure}\vspace{0.5cm}
     
    \caption{The three SR models fitted to the pseudodata of the diphoton spectrum with the signal region blinded (see Tab.~\ref{tab:diphoton_candidates}).
    To visualize the total uncertainty coverage of each candidate function, the green band in each subfigure represents the 68\% quantile range of functions obtained by sampling parameters, taking into account the best-fit values and the covariance matrix within a multidimensional normal distribution.
    The red line denotes the mean of the function ensemble.
    At the top of each subfigure, the candidate function and the fitted parameters are shown.
    The middle panel shows the weighted residual error: $\frac{\text{Data}-\text{Mean}}{\text{Data unc.}}$.
    The bottom panel shows the ratio of the 68\% quantile range to the mean.}
    \label{fig:diphoton_sampling}
\end{figure}

\begin{figure}[!t]
     \centering
     \begin{subfigure}[b]{1\textwidth}
         \centering
         \includegraphics[width=\textwidth]{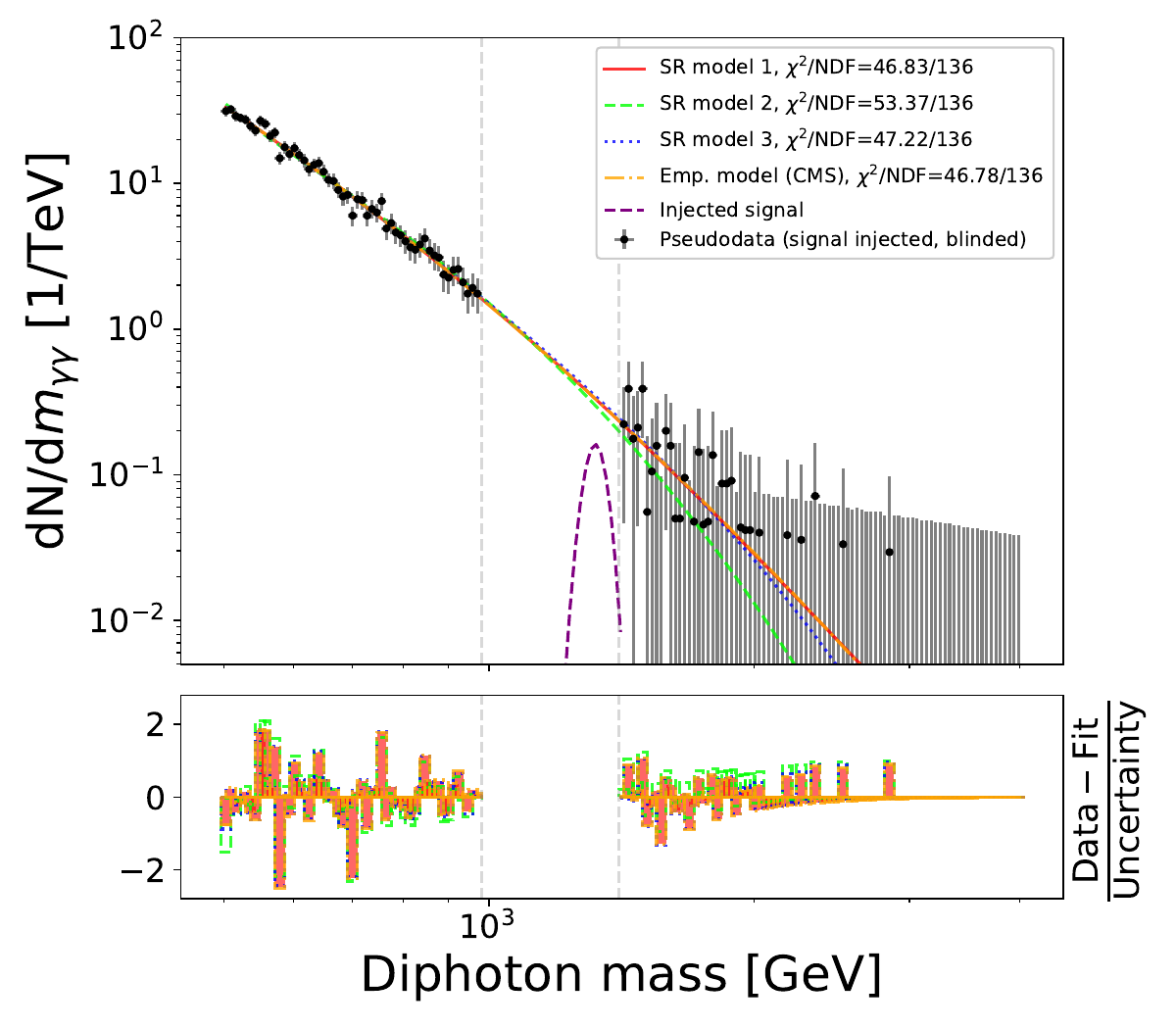}
     \end{subfigure}
     
    \caption{Pseudodata of the diphoton spectrum with the injected signal shown in the blinded signal region. The three SR models (see Tab.~\ref{tab:diphoton_candidates}) are compared against the empirical model used by CMS. The lower panel shows the residual error per bin, measured in units of the data uncertainty. It can be seen that the three SR models, generated easily from three separate fits using the same simple fit configuration with different random seeds, yields results that are readily comparable to the CMS empirical model that would have required extensive manual effort to obtain.}
    \label{fig:diphoton_blinded}
\end{figure}

Three $\tt{SymbolFit}$ runs using different random seeds are carried out, applying the same $\tt{PySR}$ configuration as used for the dijet dataset (see List.~\ref{config-lhc}), except that the maximum complexity is set at 20 instead of 80, since the diphoton distribution shape is less complex.
Tab.~\ref{tab:diphoton_candidates} lists the three SR models, each obtained from a fit initialized with a different random seed.
The $\chi^2/\text{NDF}$ scores improve significantly after the ROF step compared to the original functions returned by $\tt{PySR}$.
The three background models fit the blinded pseudodata well, as shown in Fig.~\ref{fig:diphoton_sampling} for the total uncertainty coverage and Fig.~\ref{fig:diphoton_blinded} for a comparison with the empirical model used by CMS.

\begin{table*}[!t]
\caption{The candidate functions are obtained from three fits using different random seeds, fitted to the pseudodata of the diphoton spectrum with the (injected) signal region blinded.
The fits were performed on a scaled dataset (to enhance fit stability and prevent numerical overflow), and the functions can be transformed back to describe the original spectrum using the transformation: $x\rightarrow 0.000287(x - 503)$.
These functions are plotted and compared with the blinded pseudodata in Fig.~\ref{fig:diphoton_blinded}.
Numerical values are rounded to three significant figures for display purposes.}
\label{tab:diphoton_candidates}
\centering
\resizebox{\textwidth}{!}{
\begin{tabular}{c|l|c|c|c|c}\hline
    & \textbf{Candidate function} & \textbf{\# param.} & \textbf{$\chi^2/\text{NDF}$} & \textbf{$\chi^2/\text{NDF}$} & \textbf{p-value} \\ 

    & (after ROF) & & (before ROF) & (after ROF) & (after ROF) \\ \hline
    & & & & & \\
    SR model 1 & $33.3(0.017 + 0.177x)^{6.93x}$ & 3 & 47.12 / 136 = 0.3465 & 46.83 / 136 = 0.3443 & 1.0 \\
    & &  &  & & \\ \hline
    & & & & & \\
    SR model 2 & $35.0\exp(-16.0x^{0.834})$ & 3 & 63.44 / 136 = 0.4665 & 53.37 / 136 = 0.3925 & 1.0 \\
    & $ $ & & & & \\ \hline
    & & & & & \\
    SR model 3 & $0.0226\exp(0.975x)(0.158 + x)^{-3.96}$ & 3 & 48.94 / 136 = 0.3598 & 47.22 / 136 = 0.3472 & 1.0 \\
    & $ $ & & & & \\ \hline
    
    \end{tabular}%
    }
\end{table*}

\begin{figure}[!t]
     \centering
     \begin{subfigure}[b]{0.495\textwidth}
         \centering
         \includegraphics[width=\textwidth]{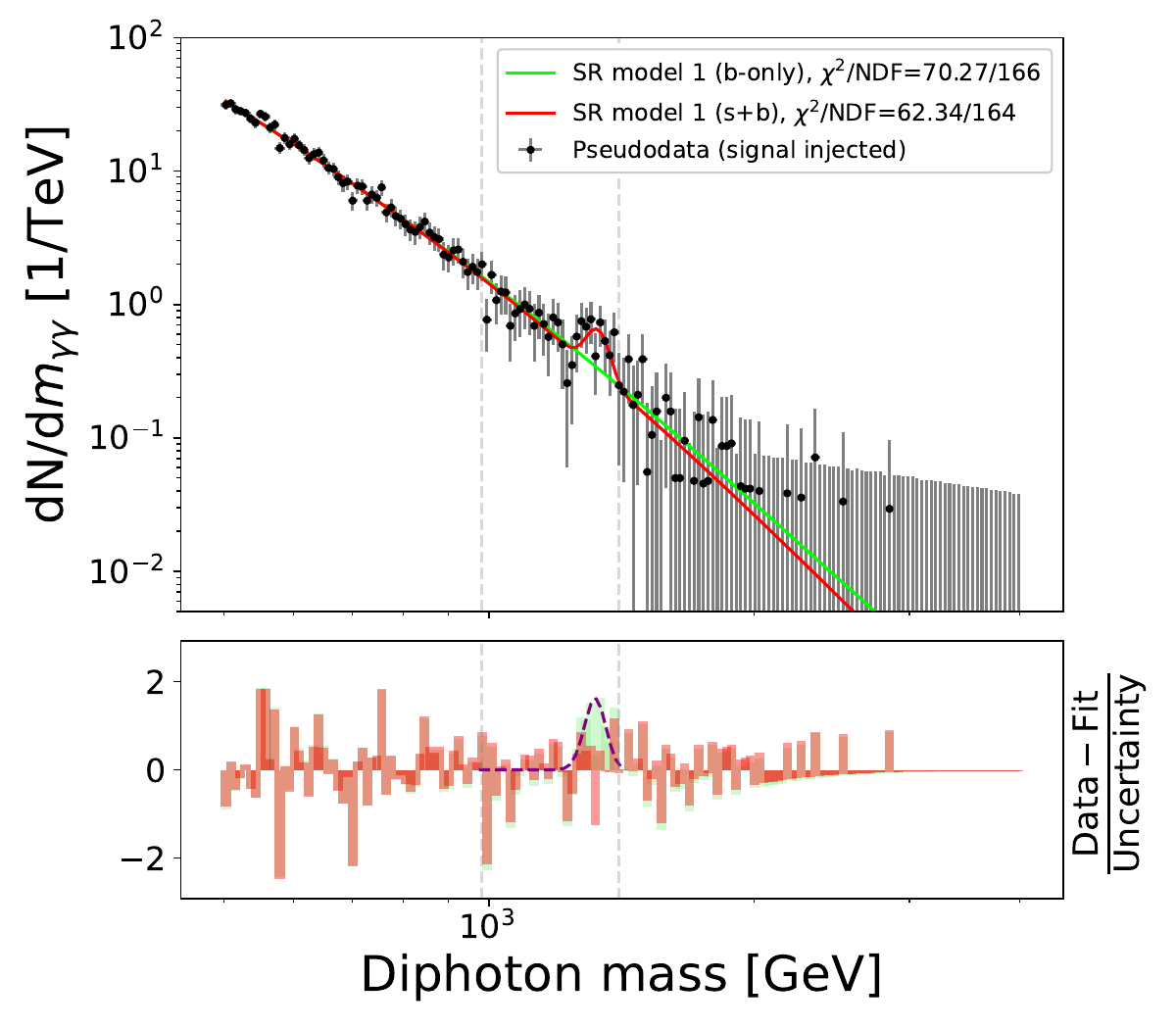}
         \caption{SR model 1.}
     \end{subfigure}
     \begin{subfigure}[b]{0.495\textwidth}
         \centering
         \includegraphics[width=\textwidth]{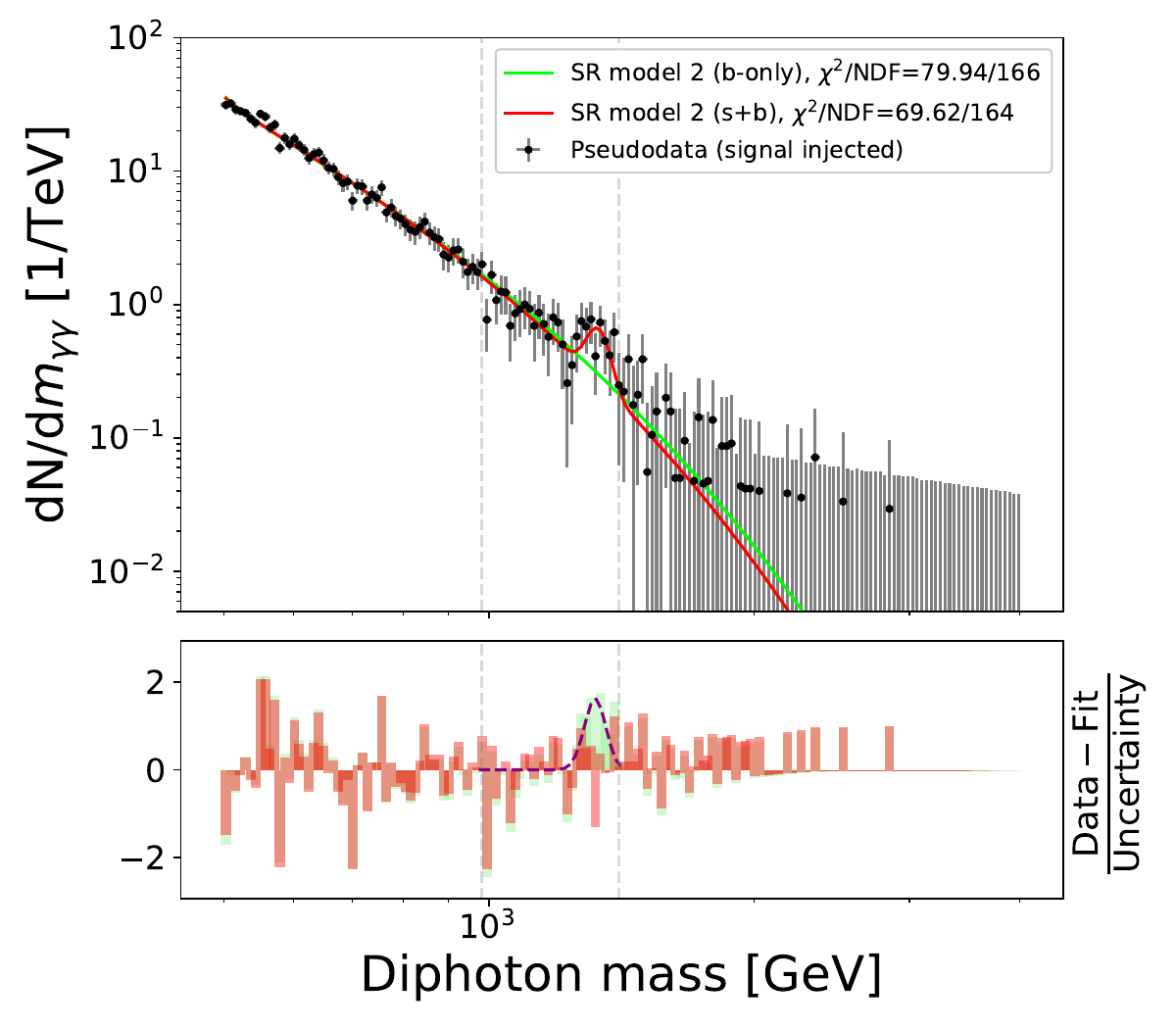}
         \caption{SR model 2.}
     \end{subfigure}\\\vspace{0.5cm}
     \begin{subfigure}[b]{0.495\textwidth}
         \centering
         \includegraphics[width=\textwidth]{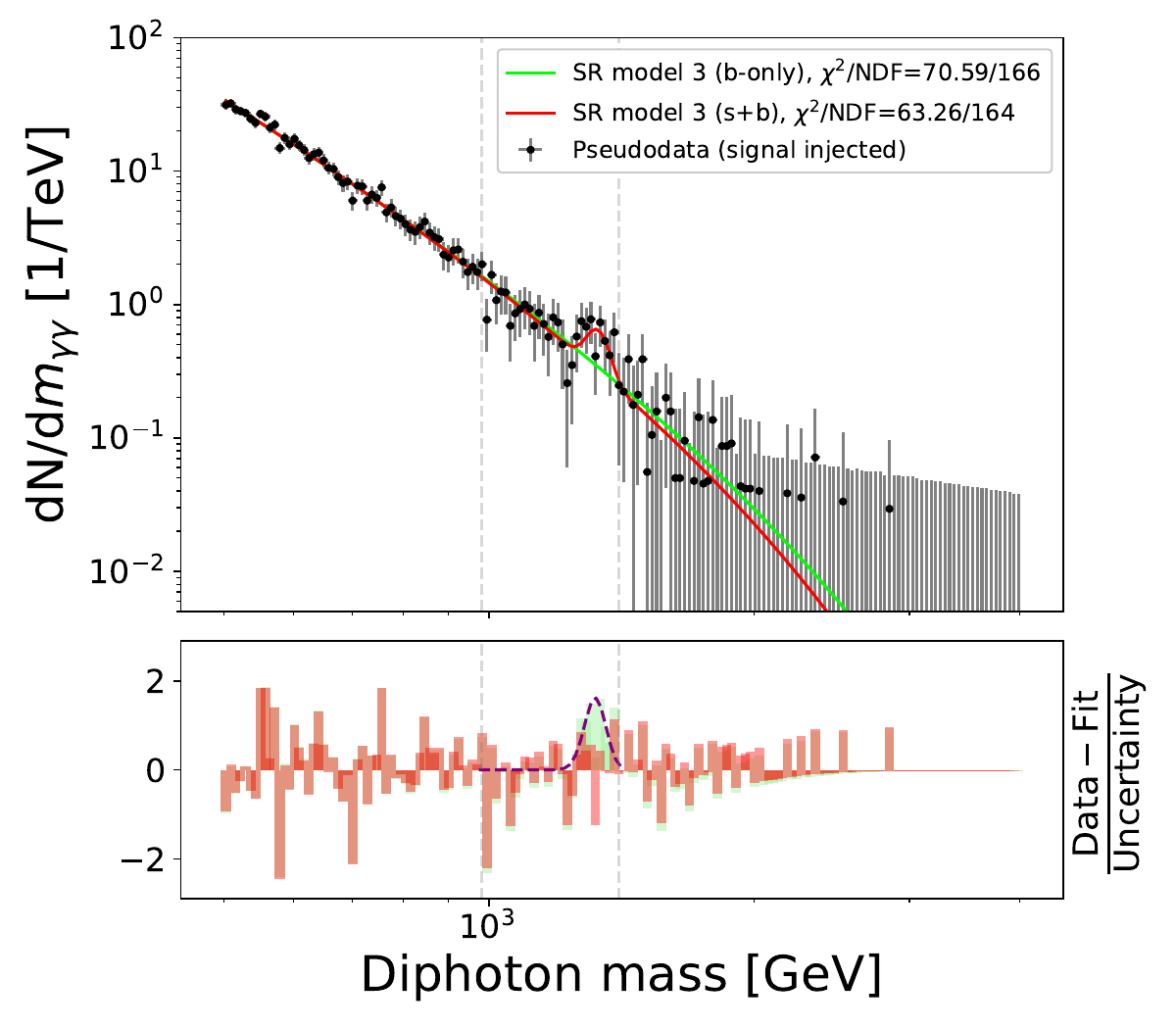}
         \caption{SR model 3.}
     \end{subfigure}
     \begin{subfigure}[b]{0.495\textwidth}
         \centering
         \includegraphics[width=\textwidth]{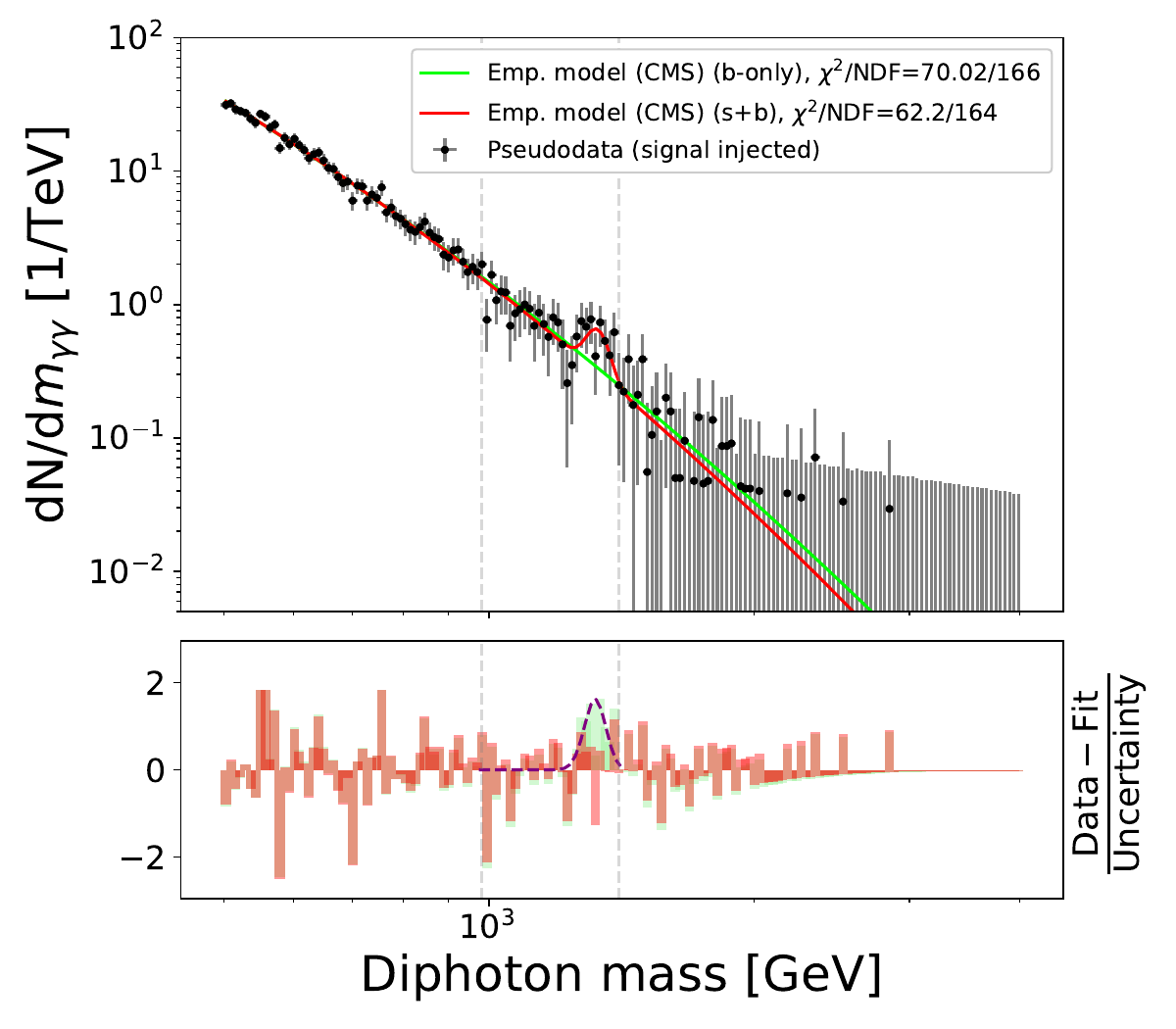}
         \caption{Empirical model (CMS).}
     \end{subfigure}\vspace{0.5cm}
     
    \caption{Comparison of the b-only fits and the s+b fits to the unblinded pseudodata of the diphoton spectrum. The lower panel shows the residual error per bin, measured in units of the data uncertainty. The shape of the injected signal is also shown.}
    \label{fig:diphoton_unblinded}
\end{figure}

Next, we unblind the pseudodata and perform b-only fits and s+b fits on the full pseudodata spectrum.
These results are shown in Fig.\ref{fig:diphoton_unblinded}.
In all three SR models, as well as the CMS empirical model, the excess of events over the background around the injected signal location observed in the b-only fits is reduced in the s+b fits, demonstrating that the models are sensitive to the injected signal.
Tab.~\ref{tab:diphoton_chi2} lists the $\chi^2/\text{NDF}$ scores for each model, showing the fit performance in response to the presence of the injected signal.

\begin{figure}[!t]
     \centering
     \begin{subfigure}[b]{0.6\textwidth}
         \centering
         \includegraphics[width=\textwidth]{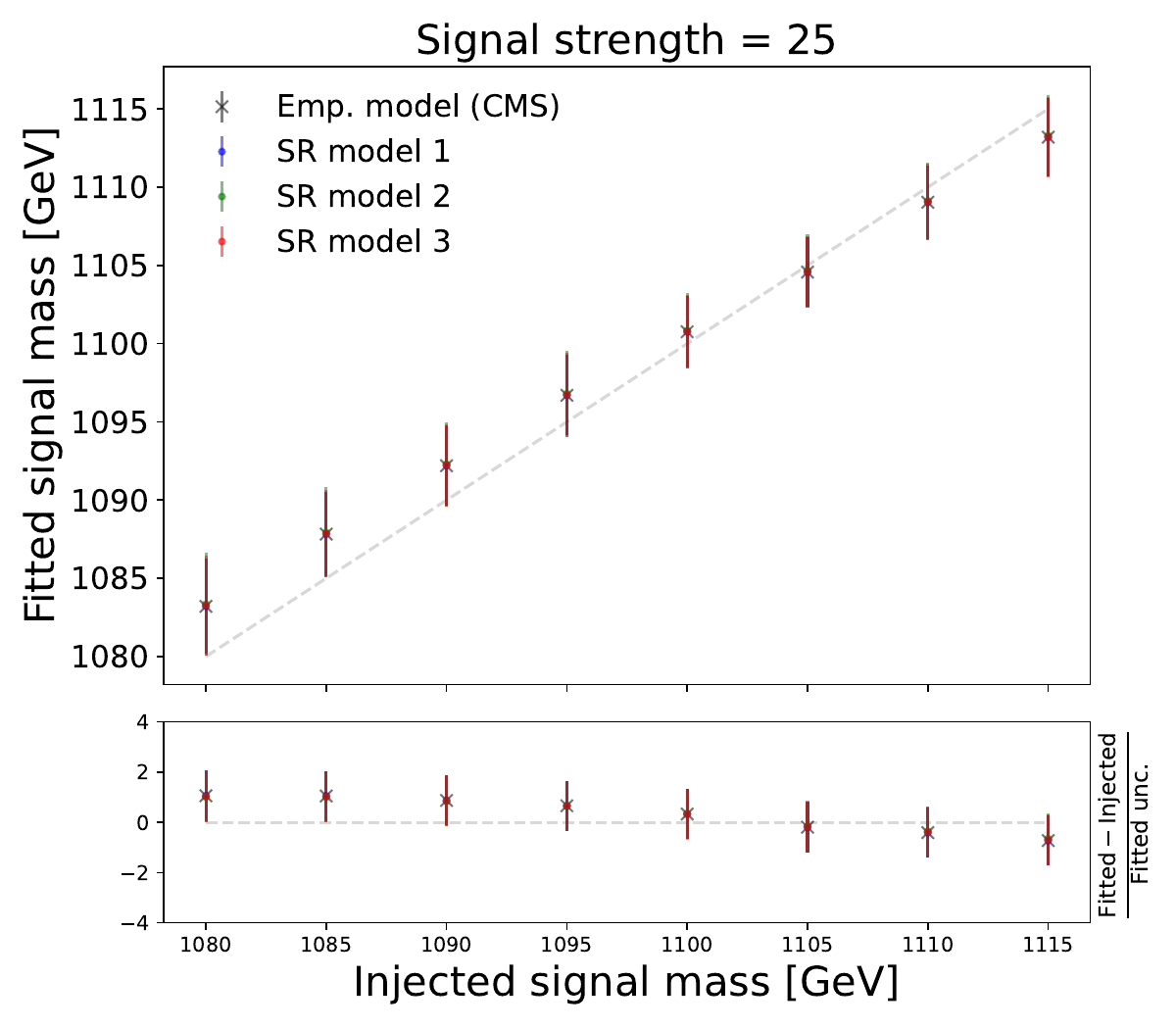}
         \caption{Fitted vs. injected signal mass at a specified signal strength value.}
     \end{subfigure}\\\vspace{0.5cm}
     \begin{subfigure}[b]{0.6\textwidth}
         \centering
         \includegraphics[width=\textwidth]{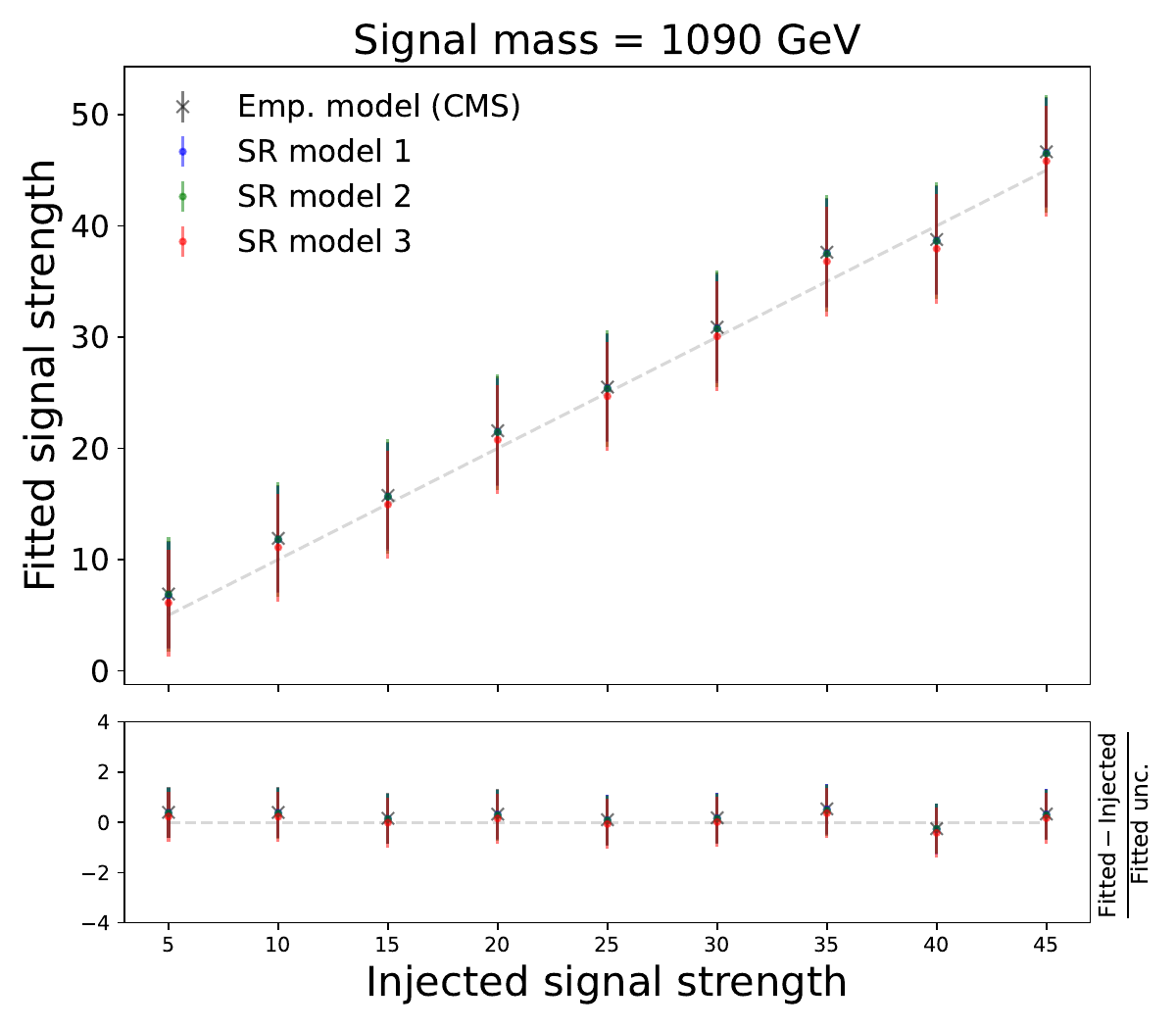}
         \caption{Fitted vs. injected signal strength at a specified signal mass value.}
     \end{subfigure}
     
    \caption{Fitted values vs. the true values of parameters of the injected signal in the diphoton dataset.
    The bottom panels show the residual error in units of the fitted uncertainty.}
    \label{fig:diphoton_scan}
\end{figure}

To assess whether the SR models can accurately extract the injected signals, we generate multiple sets of pseudodata by injecting Gaussian signals with different mean values ranging from 1080 to 1120 GeV and varying signal strengths between 5 and 50.
We then perform the s+b fits to extract the corresponding signal parameters.
Fig.~\ref{fig:diphoton_scan} shows the extracted signal parameters plotted against their injected values.
All three SR models are capable of extracting the correct signal parameter values within reasonable uncertainties and are comparable to the empirical model used by CMS.

\begin{table*}[!t]
\caption{Comparison of the $\chi^2/\text{NDF}$ scores from three types of fits to the diphoton dataset: the b-only fits to the blinded pseudodata, b-only fits to the unblinded pseudodata, and s+b fits to the unblinded pseudodata. The background models used for the fits are listed in Tab.~\ref{tab:diphoton_candidates}, and the fits are shown in Fig.~\ref{fig:diphoton_blinded} (blinded) and Fig.~\ref{fig:diphoton_unblinded} (unblinded).}
\label{tab:diphoton_chi2}
\centering
\resizebox{\textwidth}{!}{
\begin{tabular}{l|c|c|c}\hline
    & \textbf{$\chi^2/\text{NDF}$ (b-only, blinded)} & \textbf{$\chi^2/\text{NDF}$ (b-only, unblinded)} & \textbf{$\chi^2/\text{NDF}$ (s+b, unblinded)} \\ \hline
    
    SR model 1 & 46.83 / 136 = 0.3443 & 70.27 / 166 = 0.4233 & 62.34 / 164 = 0.3801 \\
    & & & \\ \hline

    SR model 2 & 53.37 / 136  = 0.3924 & 79.94 / 166 = 0.4816 & 69.62 / 164 = 0.4245 \\
    & & & \\ \hline

    SR model 3 & 47.22 / 136 = 0.3472 & 70.59 / 166 = 0.4252 & 63.26 / 164 = 0.3857 \\
    & & & \\ \hline

    Emp. model (CMS) & 46.78 / 136 = 0.344 & 70.02 / 166 = 0.4218 & 62.2 / 164 = 0.3793 \\
    & & & \\ \hline
    
    \end{tabular}%
    }
\end{table*}

\clearpage

\subsection{CMS trijet dataset (1D) [background modeling]}
\label{sec:trijet}

CMS performed a search for high-mass trijet resonances using proton-proton collision data at a center-of-mass energy of $\sqrt{s}=13$ TeV and reported no significant deviations from the Standard Model prediction~\cite{trijet}.
The dataset for the trijet spectrum is publicly available on HEPDATA at Ref.~\cite{hepdata.trijet}.
In the analysis, CMS considered four empirical functions to model the background contribution in the distribution of the trijet invariant mass, $m_{\text{jjj}}$, and one of them is:
\begin{equation}
    \label{eq:trijet}
    f(x)=\frac{p_0(1-x)^{p_1}}{x^{p_2+p_3\log(x)}},
\end{equation}
where $x=m_{\text{jjj}}/\sqrt{s}$ is dimensionless and $p_{\{0,1,2,3\}}$ are free parameters.
Eq.~\ref{eq:trijet} corresponds to Eq.~\ref{eq:trijet-n} with $N=3$ determined by an F-test.

We perform the same experiments conducted on the dijet dataset, as detailed in Sec.~\ref{sec:dijet}.
Starting from the original trijet spectrum, we generate pseudodata by injecting a perturbed Gaussian signal centered at $m_{\text{jjj}}=4000$ GeV ($s_1$) with a width of 400 GeV ($2s_2$) and a signal strength of $s_0=50000$.
To model the background, we blind the signal region by masking the $m_{\text{jjj}}$ bins between 3000 and 5000 GeV in the pseudodata and perform the fits.

\begin{figure}[!t]
     \centering
     \begin{subfigure}[b]{0.495\textwidth}
         \centering
         \includegraphics[width=\textwidth]{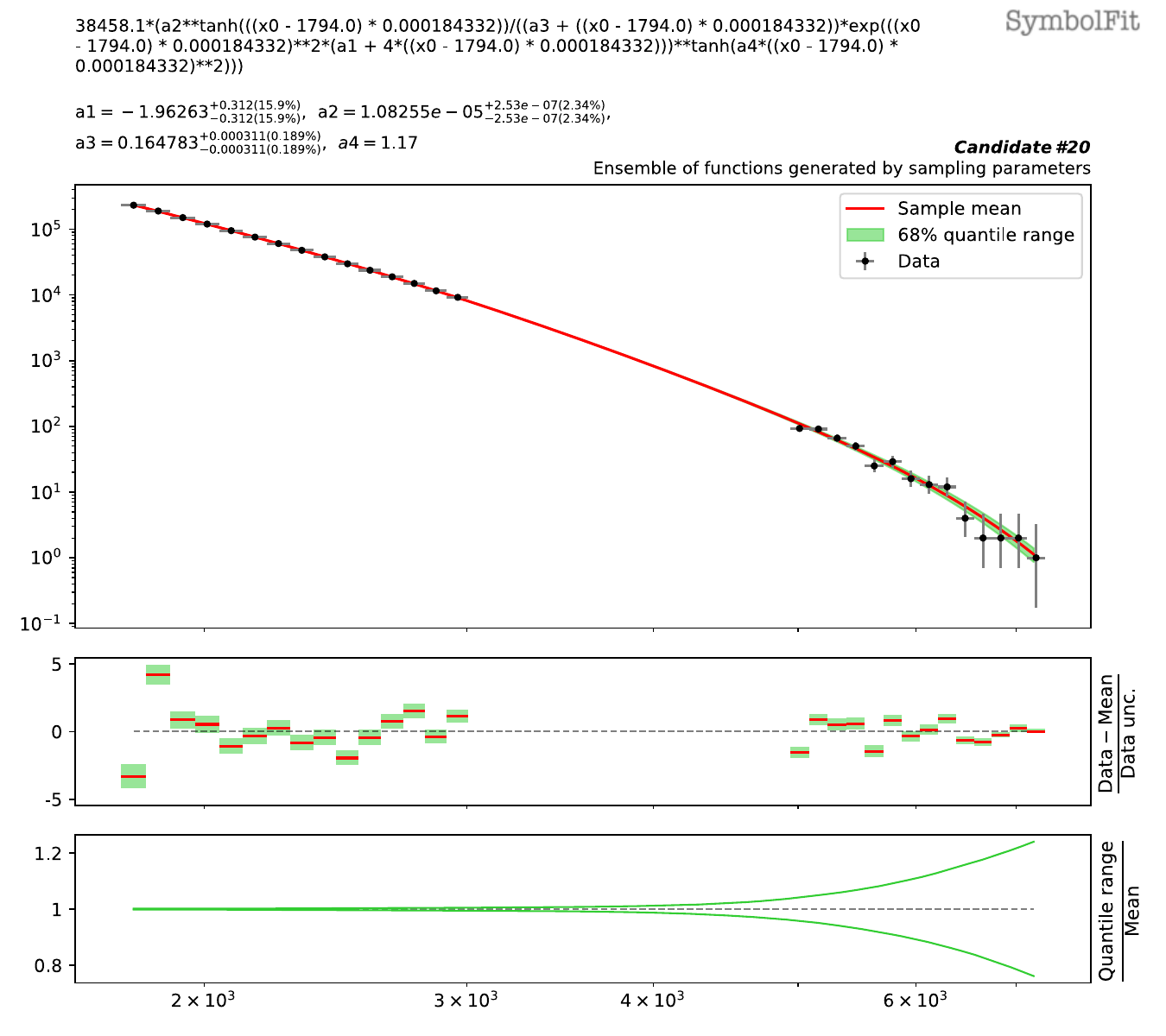}
         \caption{SR model 1.}
     \end{subfigure}
     \begin{subfigure}[b]{0.495\textwidth}
         \centering
         \includegraphics[width=\textwidth]{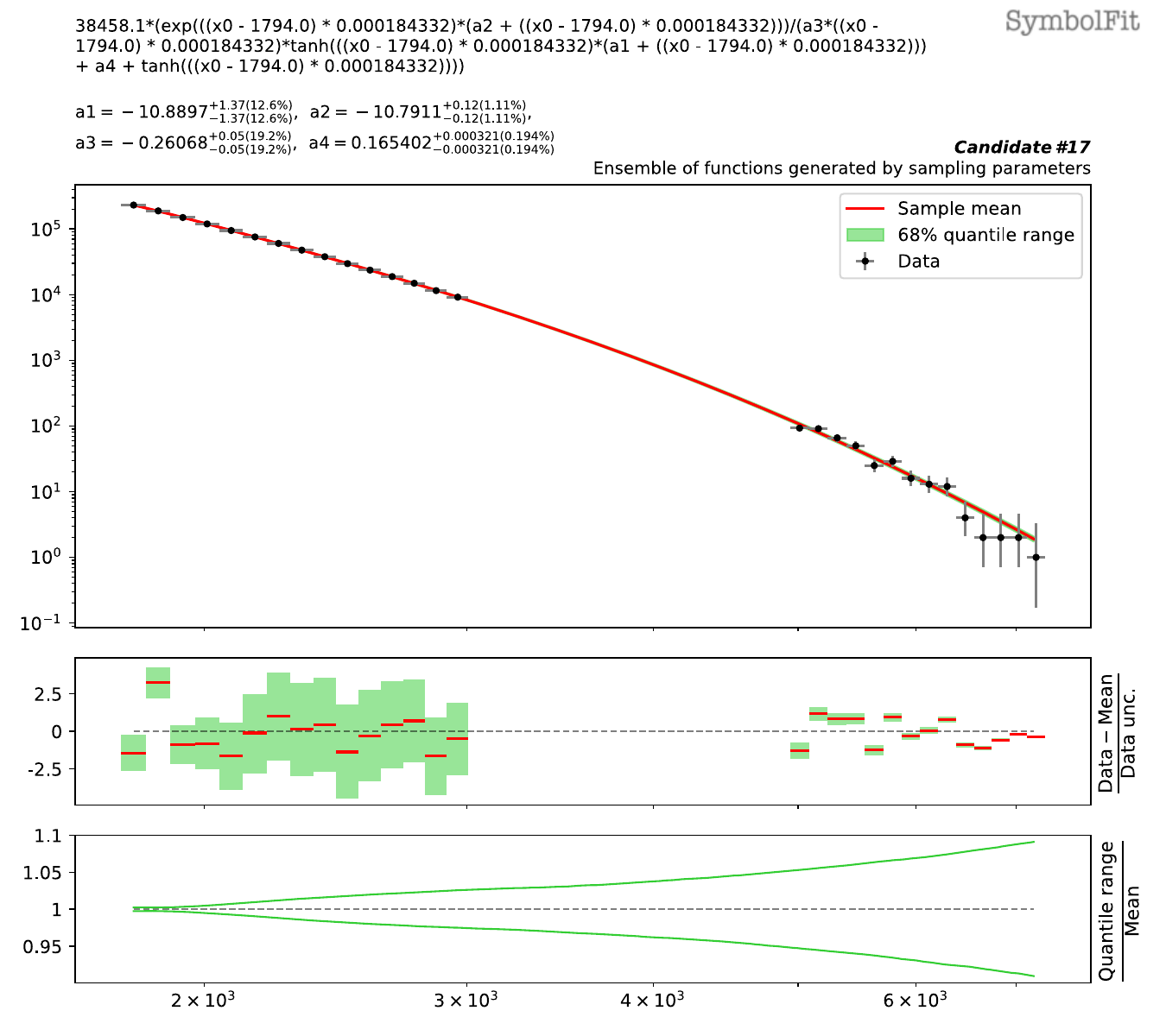}
         \caption{SR model 2.}
     \end{subfigure}\\\vspace{0.5cm}
     \begin{subfigure}[b]{0.495\textwidth}
         \centering
         \includegraphics[width=\textwidth]{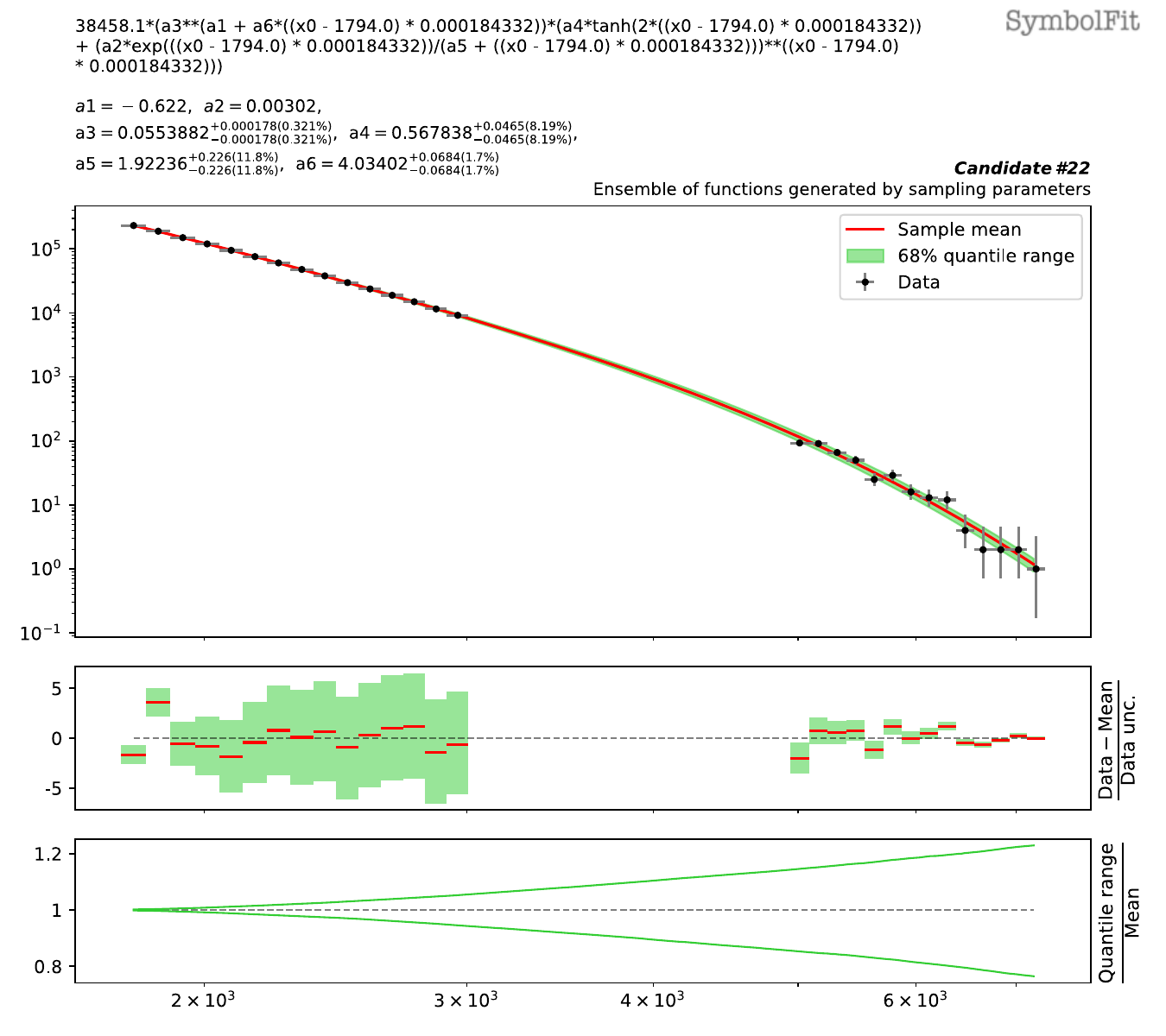}
         \caption{SR model 3.}
     \end{subfigure}\vspace{0.5cm}
     
    \caption{The three SR models fitted to the pseudodata of the trijet spectrum with the signal region blinded (see Tab.~\ref{tab:trijet_candidates}).
    To visualize the total uncertainty coverage of each candidate function, the green band in each subfigure represents the 68\% quantile range of functions obtained by sampling parameters, taking into account the best-fit values and the covariance matrix within a multidimensional normal distribution.
    The red line denotes the mean of the function ensemble.
    At the top of each subfigure, the candidate function and the fitted parameters are shown.
    The middle panel shows the weighted residual error: $\frac{\text{Data}-\text{Mean}}{\text{Data unc.}}$.
    The bottom panel shows the ratio of the 68\% quantile range to the mean.}
    \label{fig:trijet_sampling}
\end{figure}

\begin{figure}[!t]
     \centering
     \begin{subfigure}[b]{1\textwidth}
         \centering
         \includegraphics[width=\textwidth]{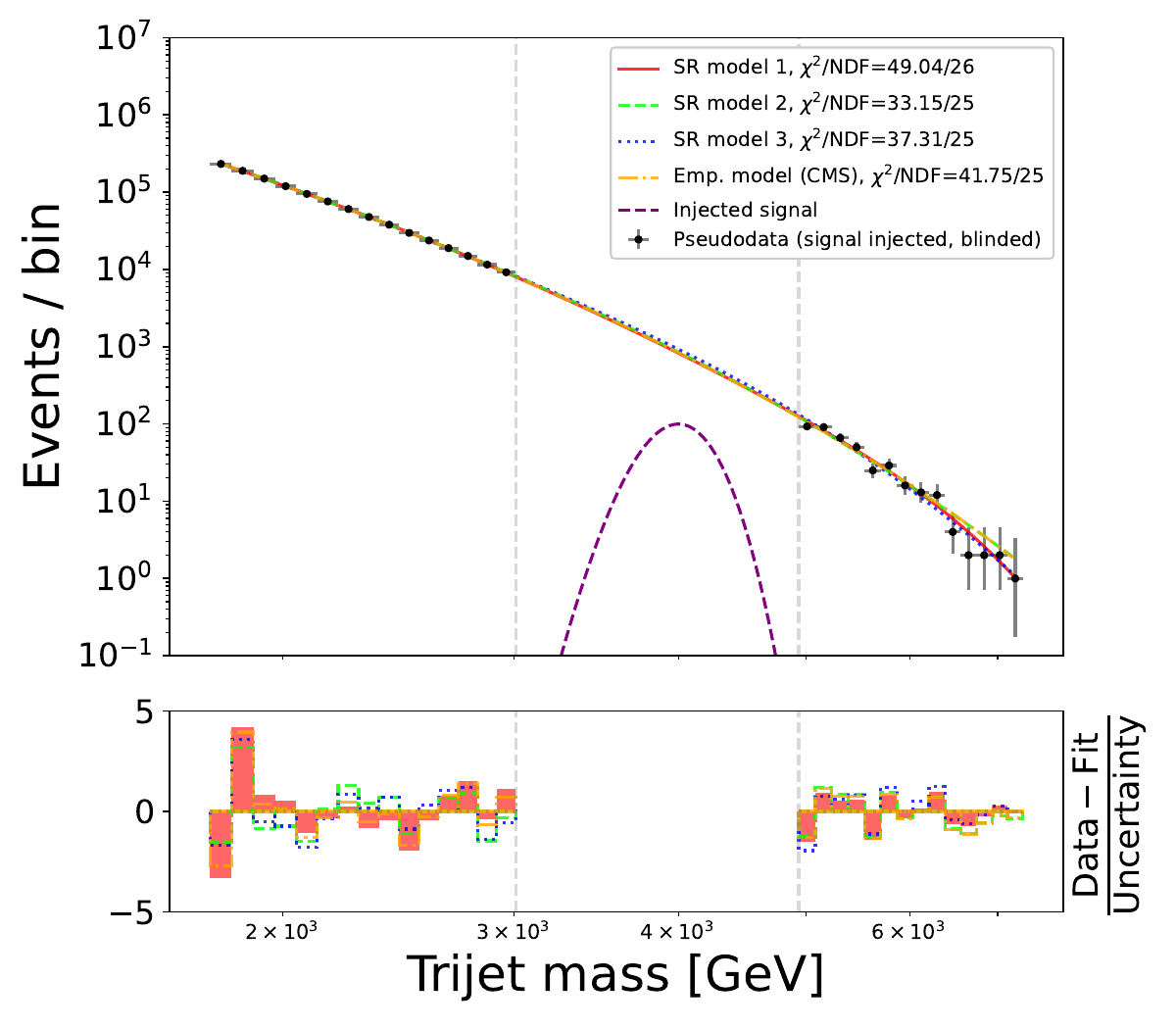}
     \end{subfigure}
     
    \caption{Pseudodata of the trijet spectrum with the injected signal shown in the blinded signal region. The three SR models (see Tab.~\ref{tab:trijet_candidates}) are compared against the empirical model used by CMS. The lower panel shows the residual error per bin, measured in units of the data uncertainty. It can be seen that the three SR models, generated easily from three separate fits using the same simple fit configuration with different random seeds, yields results that are readily comparable to the CMS empirical model that would have required extensive manual effort to obtain.}
    \label{fig:trijet_blinded}
\end{figure}

Three $\tt{SymbolFit}$ runs using different random seeds are carried out, applying the same $\tt{PySR}$ configuration as used for the dijet dataset (see List.~\ref{config-lhc}).
Tab.~\ref{tab:trijet_candidates} lists the three SR models, each obtained from a fit initialized with a different random seed.
The $\chi^2/\text{NDF}$ scores improve significantly after the ROF step compared to the original functions returned by $\tt{PySR}$.
The three background models fit the blinded pseudodata well, as shown in Fig.~\ref{fig:trijet_sampling} for the total uncertainty coverage and Fig.~\ref{fig:trijet_blinded} for a comparison with the empirical model used by CMS.

\begin{table*}[!t]
\caption{The candidate functions are obtained from three fits using different random seeds, fitted to the pseudodata of the trijet spectrum with the (injected) signal region blinded.
The fits were performed on a scaled dataset (to enhance fit stability and prevent numerical overflow), and the functions can be transformed back to describe the original spectrum using the transformation: $f(x)\rightarrow 38458\times f(0.000184(x - 1790))$.
These functions are plotted and compared with the blinded pseudodata in Fig.~\ref{fig:trijet_blinded}.
Numerical values are rounded to three significant figures for display purposes.}
\label{tab:trijet_candidates}
\centering
\resizebox{\textwidth}{!}{
\begin{tabular}{c|l|c|c|c|c}\hline
    & \textbf{Candidate function} & \textbf{\# param.} & \textbf{$\chi^2/\text{NDF}$} & \textbf{$\chi^2/\text{NDF}$} & \textbf{p-value} \\ 

    & (after ROF) & & (before ROF) & (after ROF) & (after ROF) \\ \hline
    & & & & & \\
    SR model 1 & $(1.08\times 10^{-5})^{\tanh(x)}/((0.165 + x)\times$ & 3 & 50.46 / 26 = 1.941 & 49.04 / 26 = 1.886 & 0.00408 \\
    & $\exp(x^2(-1.96 + 4x))^{\tanh(1.17x^2)})$ & & & & \\
    & & & & & \\\hline
    & & & & & \\
    SR model 2 & $\exp(x(-10.8 + x))/(-0.261x\tanh(x\times$ & 4 & 39.49 / 25 = 1.58 & 33.15 / 25 = 1.326 & 0.1273 \\
    & $(-10.9 + x)) + 0.165 + \tanh(x))$ & & & & \\
    & & & & & \\ \hline
    & & & & & \\
    SR model 3 & $0.0554^{-0.622 + 4.03x}(0.568\tanh(2x) + $ & 4 & 38.4 / 25 = 1.536 & 37.31 / 25 = 1.492 & 0.05395 \\
    & $(0.00302\exp(x)/(1.92 + x))^x)$ & & & & \\
    & & & & & \\ \hline
    
    \end{tabular}%
    }
\end{table*}

\begin{figure}[!t]
     \centering
     \begin{subfigure}[b]{0.495\textwidth}
         \centering
         \includegraphics[width=\textwidth]{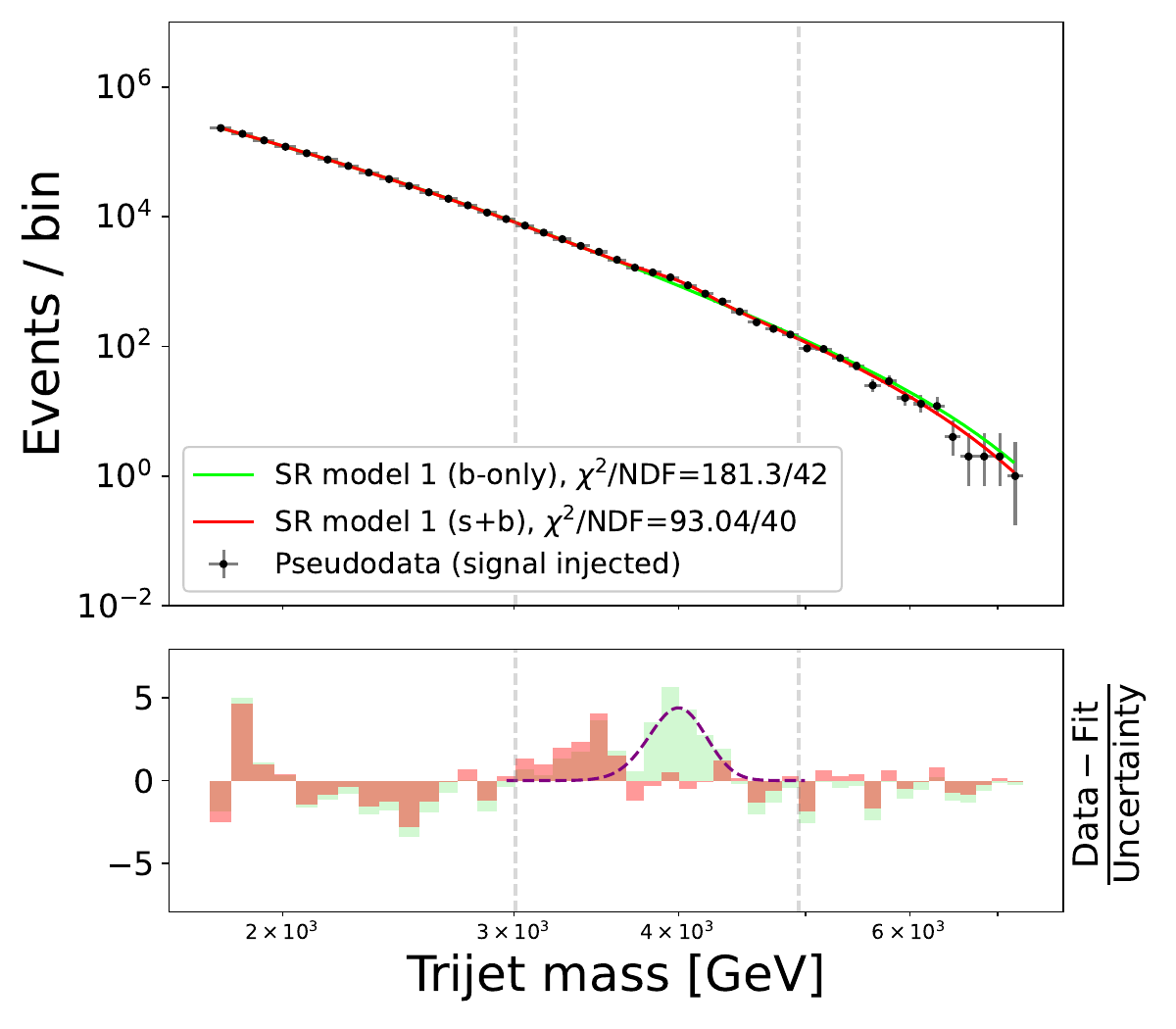}
         \caption{SR model 1.}
     \end{subfigure}
     \begin{subfigure}[b]{0.495\textwidth}
         \centering
         \includegraphics[width=\textwidth]{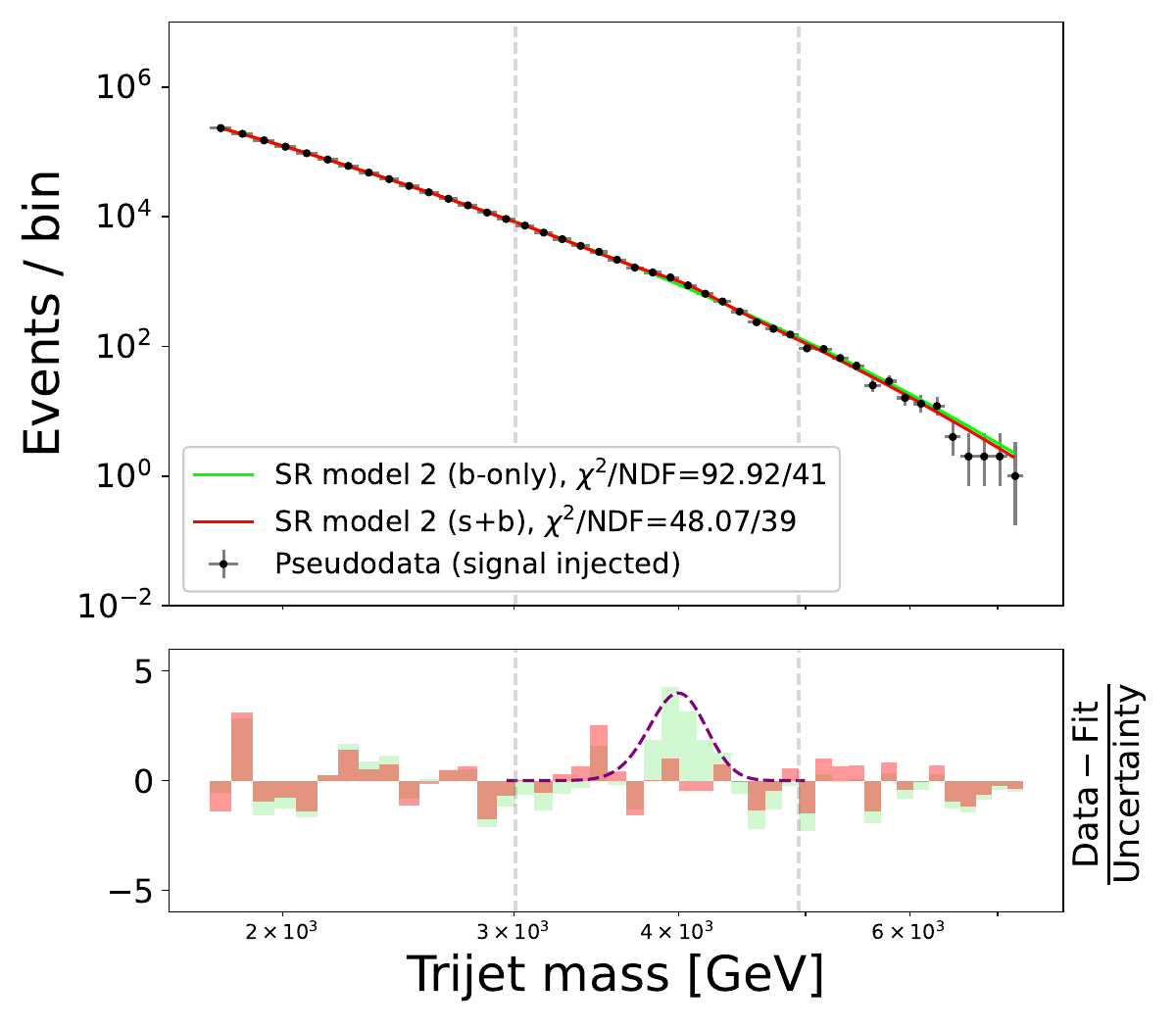}
         \caption{SR model 2.}
     \end{subfigure}\\\vspace{0.5cm}
     \begin{subfigure}[b]{0.495\textwidth}
         \centering
         \includegraphics[width=\textwidth]{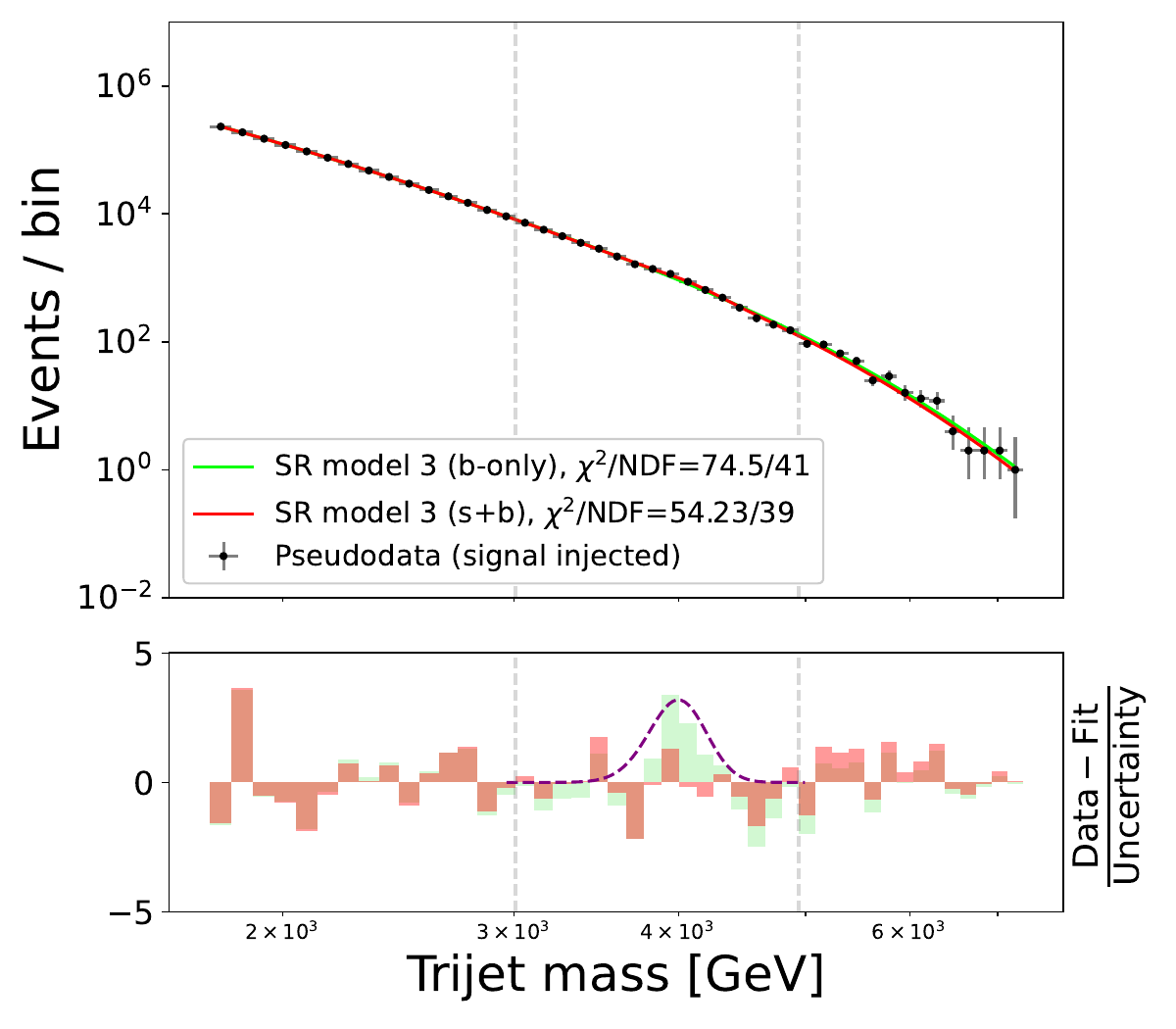}
         \caption{SR model 3.}
     \end{subfigure}
     \begin{subfigure}[b]{0.495\textwidth}
         \centering
         \includegraphics[width=\textwidth]{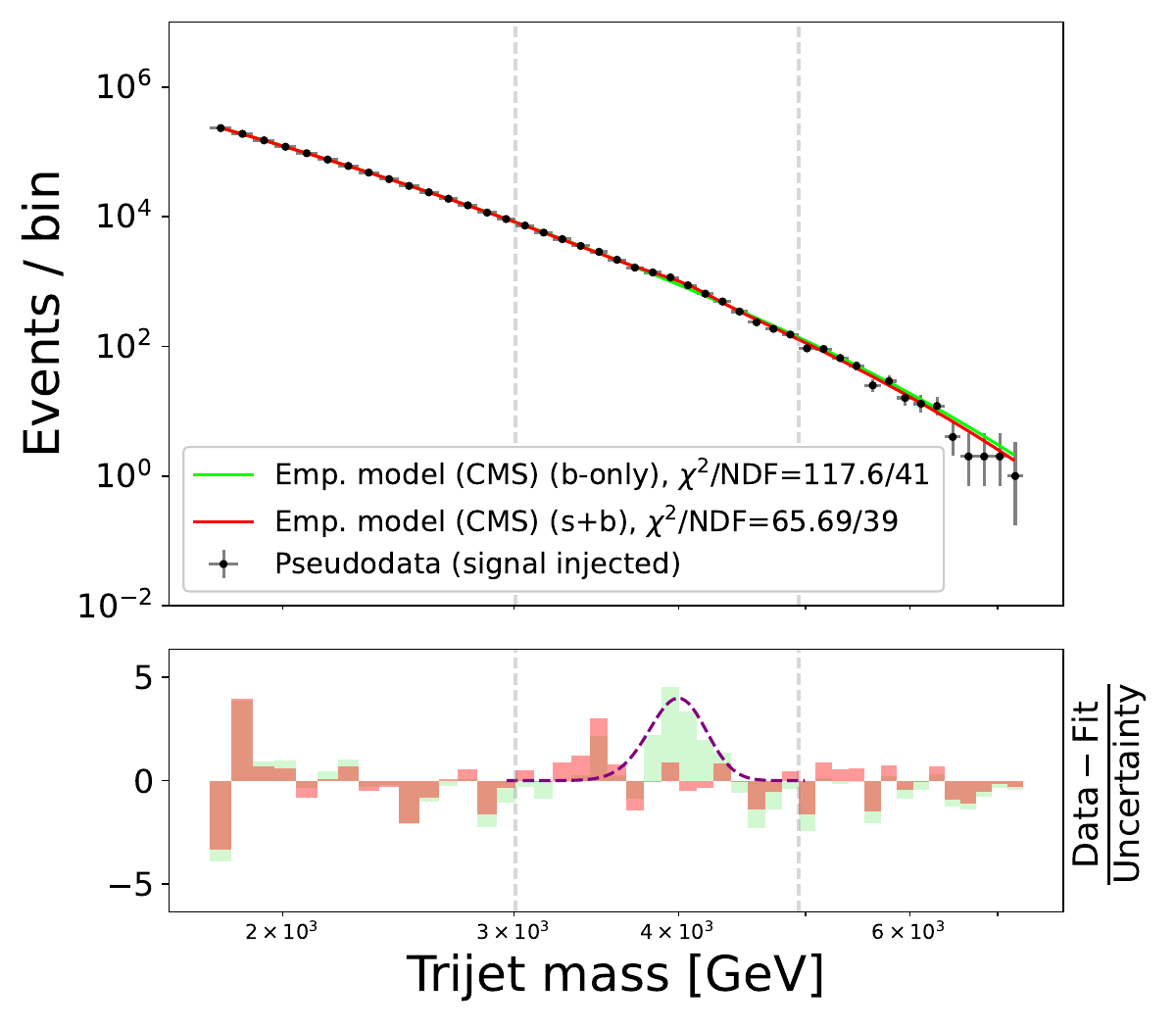}
         \caption{Empirical model (CMS).}
     \end{subfigure}\vspace{0.5cm}
     
    \caption{Comparison of the b-only fits and the s+b fits to the unblinded pseudodata. The lower panel shows the residual error per bin, measured in units of the data uncertainty. The shape of the injected signal is also shown.}
    \label{fig:trijet_unblinded}
\end{figure}

Next, we unblind the pseudodata and perform b-only fits and s+b fits on the full pseudodata spectrum.
These results are shown in Fig.\ref{fig:trijet_unblinded}.
In all three SR models, as well as the CMS empirical model, the excess of events over the background around the injected signal location observed in the b-only fits is reduced in the s+b fits, demonstrating that the models are sensitive to the injected signal.
Tab.~\ref{tab:trijet_chi2} lists the $\chi^2/\text{NDF}$ scores for each model, showing the fit performance in response to the presence of the injected signal.

\begin{figure}[!t]
     \centering
     \begin{subfigure}[b]{0.6\textwidth}
         \centering
         \includegraphics[width=\textwidth]{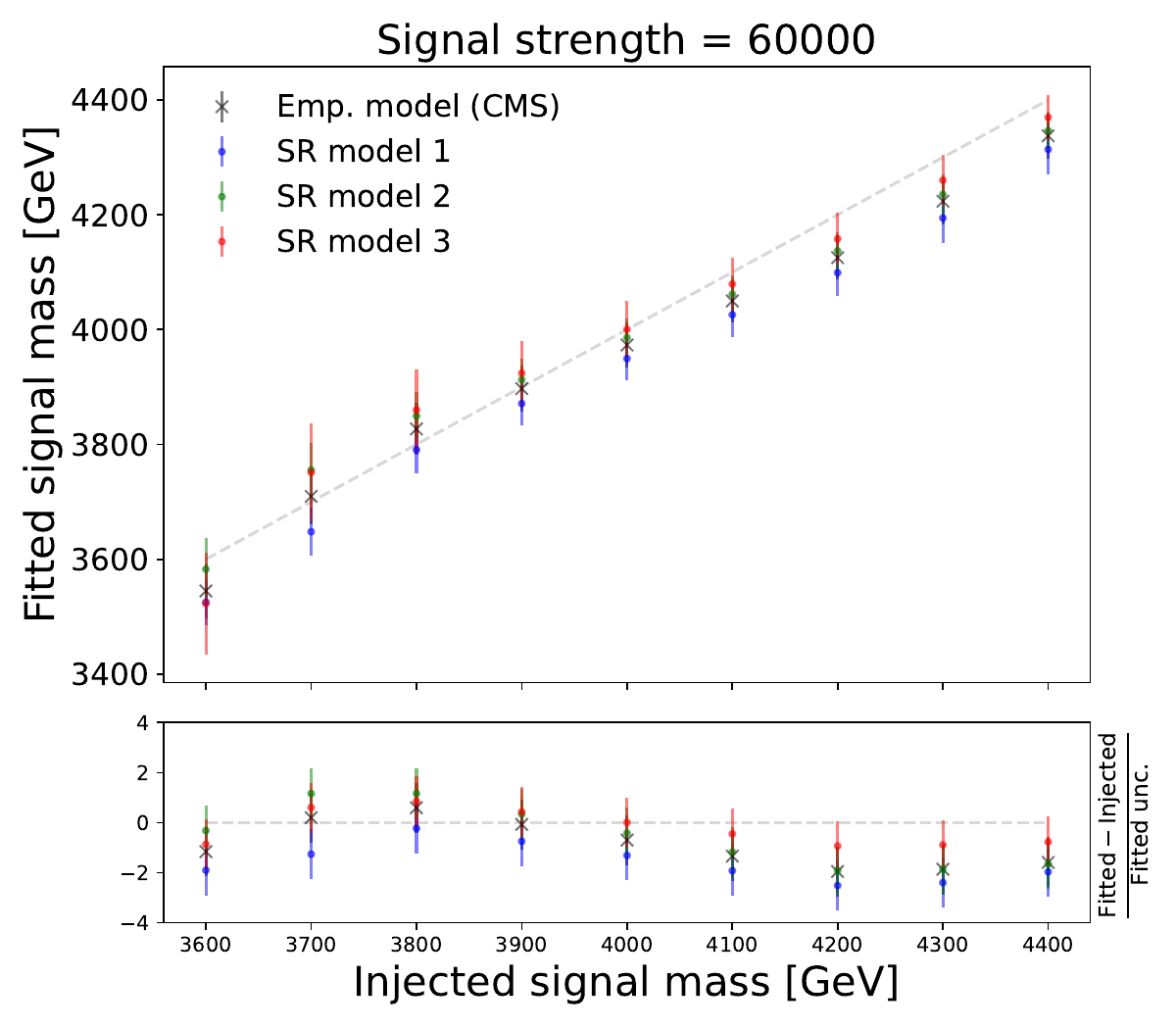}
         \caption{Fitted vs. injected signal mass at a specified signal strength value.}
     \end{subfigure}\\\vspace{0.5cm}
     \begin{subfigure}[b]{0.6\textwidth}
         \centering
         \includegraphics[width=\textwidth]{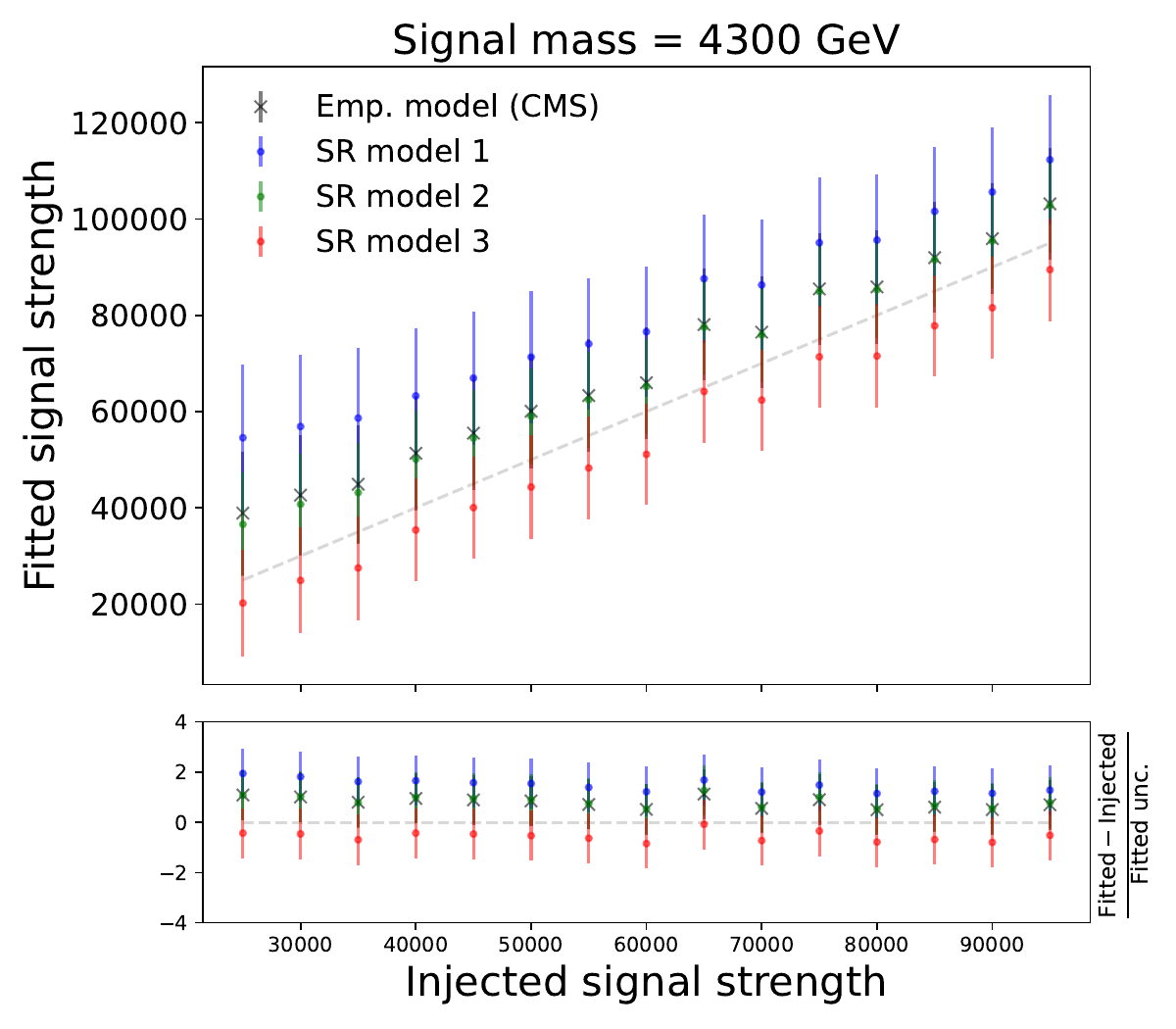}
         \caption{Fitted vs. injected signal strength at a specified signal mass value.}
     \end{subfigure}
     
    \caption{Fitted values vs. the true values of parameters of the injected signal in the trijet dataset.
    The bottom panels show the residual error in units of the fitted uncertainty.}
    \label{fig:trijet_scan}
\end{figure}

To assess whether the SR models can accurately extract the injected signals, we generate multiple sets of pseudodata by injecting Gaussian signals with different mean values ranging from 3600 to 4500 GeV and varying signal strength between 25000 and 100000.
We then perform the s+b fits to extract the corresponding signal parameters.
Fig.~\ref{fig:trijet_scan} shows the extracted signal parameters plotted against their injected values.
All three SR models are capable of extracting the correct signal parameter values within reasonable uncertainties and are comparable to the empirical model used by CMS.

\begin{table*}[!t]
\caption{Comparison of the $\chi^2/\text{NDF}$ scores from three types of fits to the trijet dataset: the b-only fits to the blinded pseudodata, b-only fits to the unblinded pseudodata, and s+b fits to the unblinded pseudodata. The background models used for the fits are listed in Tab.~\ref{tab:trijet_candidates}, and the fits are shown in Fig.~\ref{fig:trijet_blinded} (blinded) and Fig.~\ref{fig:trijet_unblinded} (unblinded).}
\label{tab:trijet_chi2}
\centering
\resizebox{\textwidth}{!}{
\begin{tabular}{l|c|c|c}\hline
    & \textbf{$\chi^2/\text{NDF}$ (b-only, blinded)} & \textbf{$\chi^2/\text{NDF}$ (b-only, unblinded)} & \textbf{$\chi^2/\text{NDF}$ (s+b, unblinded)} \\ \hline
    
    SR model 1 & 49.04 / 26 = 1.886 & 181.3 / 42 = 4.317 & 93.04 / 40 = 2.326 \\
    & & & \\ \hline

    SR model 2 & 33.15 / 25 = 1.326 & 92.92 / 41 = 2.266 & 48.07 / 39 = 1.233 \\
    & & & \\ \hline

    SR model 3 & 37.31 / 25 = 1.492 & 74.5 / 41 = 1.817 & 54.23 / 39 = 1.391 \\
    & & & \\ \hline

    Emp. model (CMS) & 41.75 / 25 = 1.67 & 117.6 / 41 = 2.868 & 65.69 / 39 = 1.684 \\
    & & & \\ \hline
    
    \end{tabular}%
    }
\end{table*}

\clearpage

\subsection{CMS paired-dijet dataset (1D) [background modeling]}
\label{sec:paired-dijet}

CMS performed a search for high-mass four-jet resonances using proton-proton collision data at a center-of-mass energy of $\sqrt{s}=13$ TeV and reported no significant deviations from the Standard Model prediction~\cite{paired-dijet}.
The dataset for the four-jet spectrum is publicly available on HEPDATA at Ref.~\cite{hepdata.paired-dijet}.
In the analysis, CMS considered four empirical functions to model the background contribution in the distribution of the four-jet invariant mass, $m_{\text{jjjj}}$, and one of them is:
\begin{equation}
    \label{eq:fourjet}
    f(x)=\frac{p_0(1-x^{1/3})^{p_1}}{x^{p_2+p_3\log x+p_4\log^2x}},
\end{equation}
where $x=m_{\text{jjjj}}/\sqrt{s}$ is dimensionless and $p_{\{0,1,2,3,4\}}$ are free parameters.

We perform the same experiments conducted on the dijet dataset, as detailed in Sec.~\ref{sec:dijet}.
Starting from the original four-jet spectrum, we generate pseudodata by injecting a perturbed Gaussian signal centered at $m_{\text{jjjj}}=3500$ GeV ($s_1$) with a width of 400 GeV ($2s_2$) and a signal strength of $s_0=2$.
To model the background, we blind the signal region by masking the $m_{\text{jjjj}}$ bins between 3000 and 4000 GeV in the pseudodata and perform the fits.

\begin{figure}[!t]
     \centering
     \begin{subfigure}[b]{0.495\textwidth}
         \centering
         \includegraphics[width=\textwidth]{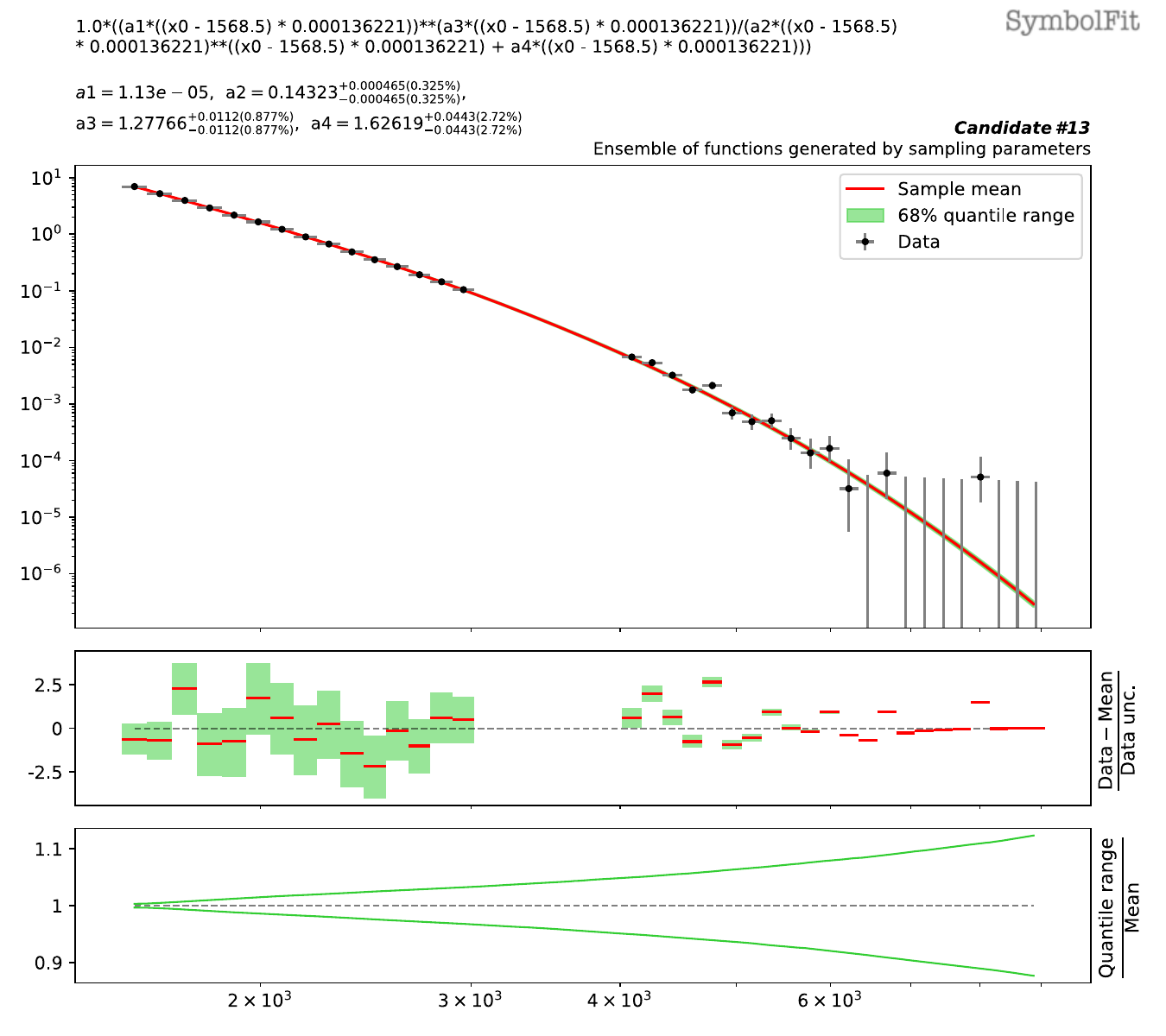}
         \caption{SR model 1.}
     \end{subfigure}
     \begin{subfigure}[b]{0.495\textwidth}
         \centering
         \includegraphics[width=\textwidth]{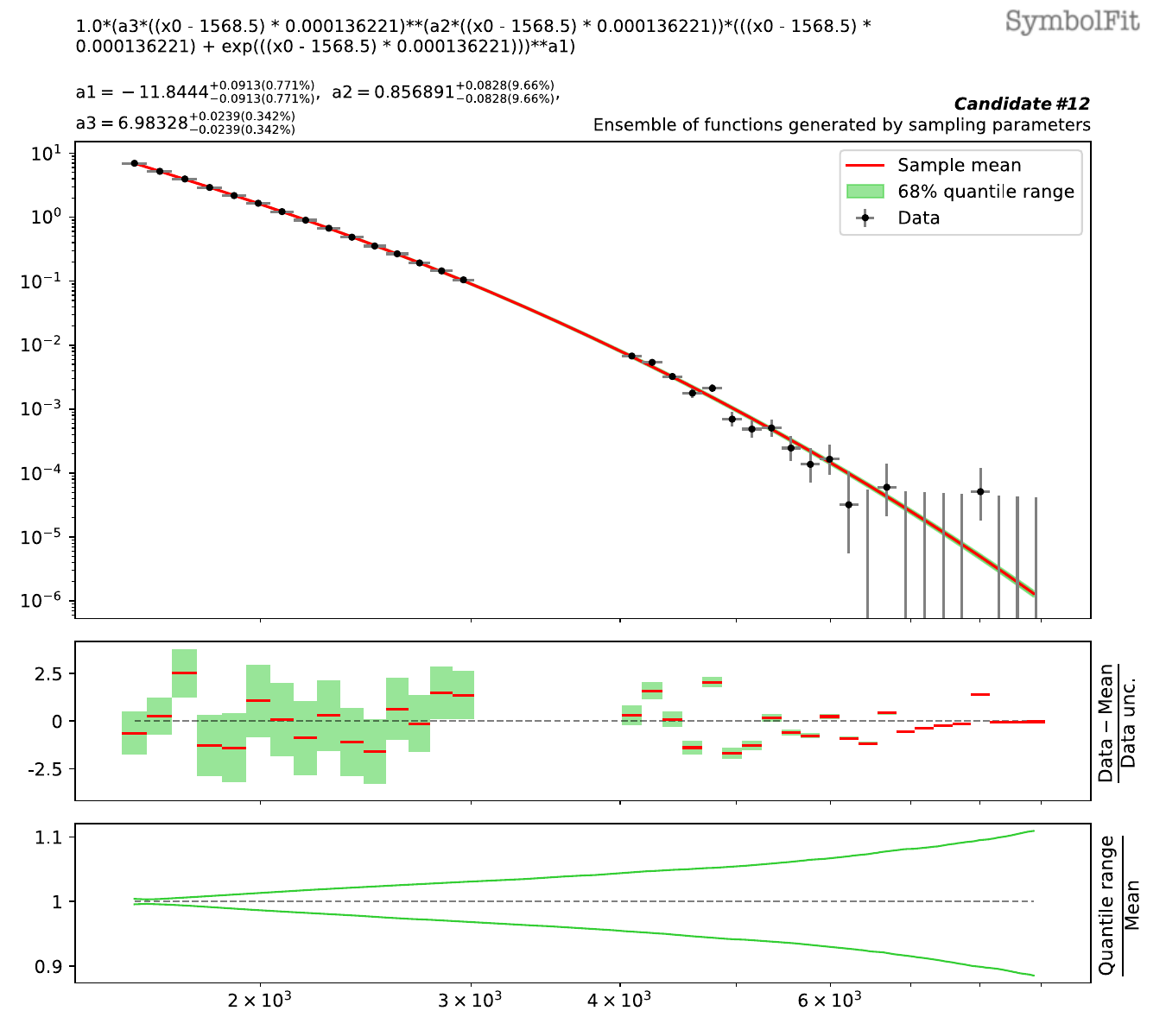}
         \caption{SR model 2.}
     \end{subfigure}\\\vspace{0.5cm}
     \begin{subfigure}[b]{0.495\textwidth}
         \centering
         \includegraphics[width=\textwidth]{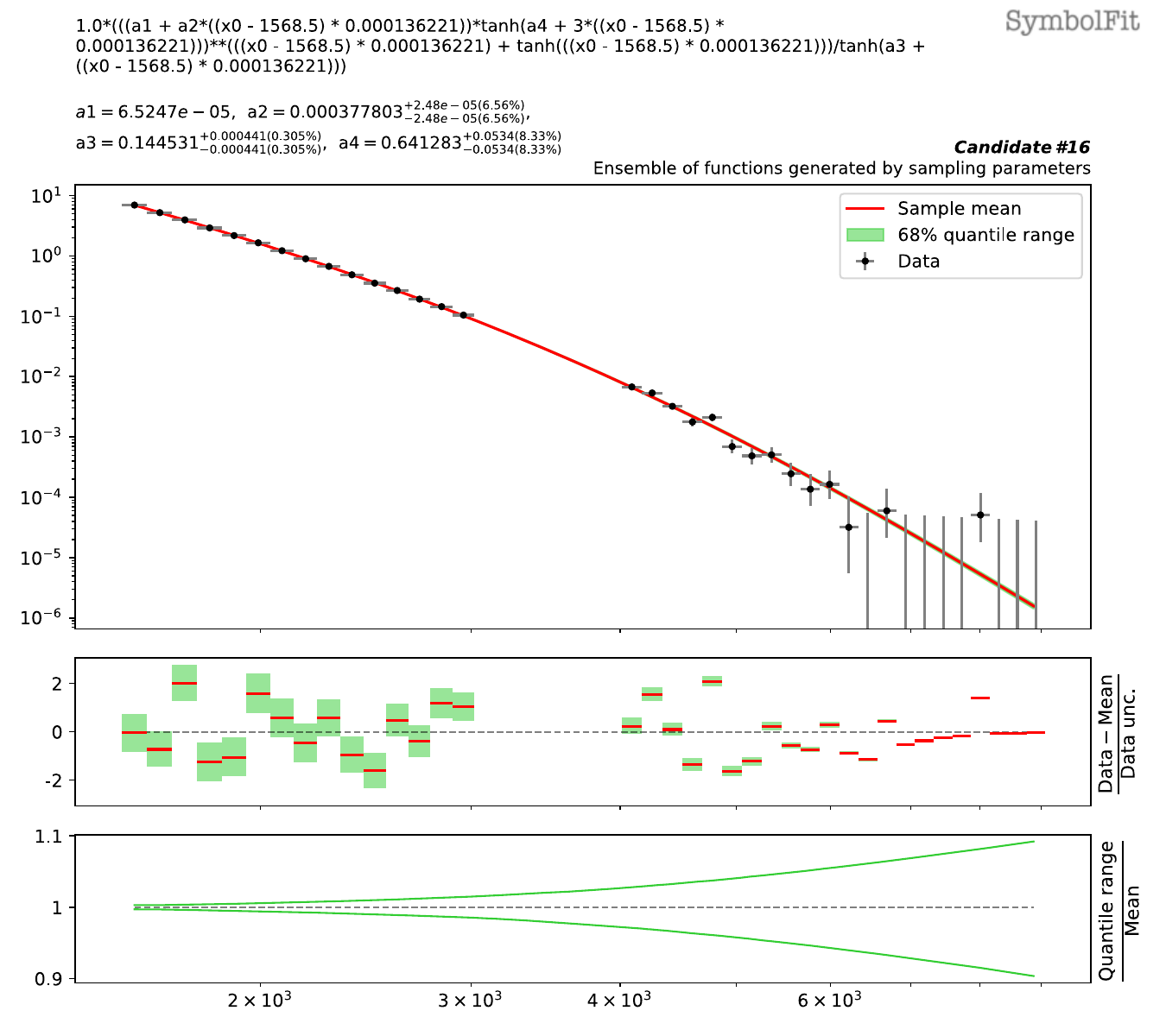}
         \caption{SR model 3.}
     \end{subfigure}\vspace{0.5cm}
     
    \caption{The three SR models fitted to the pseudodata of the paired-dijet spectrum with the signal region blinded (see Tab.~\ref{tab:dijet_pair_candidates}).
    To visualize the total uncertainty coverage of each candidate function, the green band in each subfigure represents the 68\% quantile range of functions obtained by sampling parameters, taking into account the best-fit values and the covariance matrix within a multidimensional normal distribution.
    The red line denotes the mean of the function ensemble.
    At the top of each subfigure, the candidate function and the fitted parameters are shown.
    The middle panel shows the weighted residual error: $\frac{\text{Data}-\text{Mean}}{\text{Data unc.}}$.
    The bottom panel shows the ratio of the 68\% quantile range to the mean.}
    \label{fig:dijet_pair_sampling}
\end{figure}

\begin{figure}[!t]
     \centering
     \begin{subfigure}[b]{1\textwidth}
         \centering
         \includegraphics[width=\textwidth]{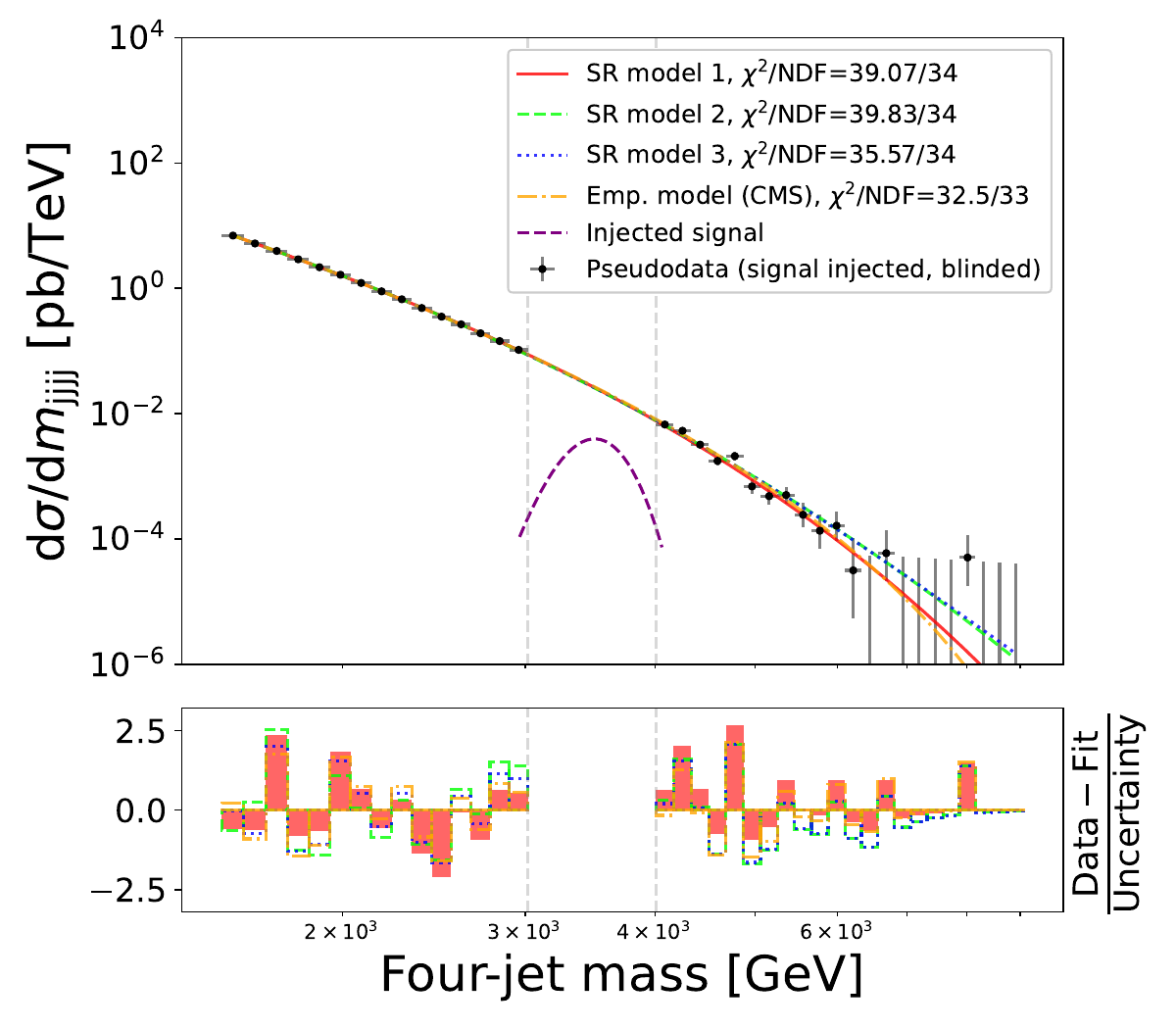}
     \end{subfigure}
     
    \caption{Pseudodata of the four-jet spectrum with the injected signal shown in the blinded signal region. The three SR models (see Tab.~\ref{tab:dijet_pair_candidates}) are compared against the empirical model used by CMS. The lower panel shows the residual error per bin, measured in units of the data uncertainty. It can be seen that the three SR models, generated easily from three separate fits using the same simple fit configuration with different random seeds, yields results that are readily comparable to the CMS empirical model that would have required extensive manual effort to obtain.}
    \label{fig:dijet_pair_blinded}
\end{figure}

Three $\tt{SymbolFit}$ runs using different random seeds are carried out, applying the same $\tt{PySR}$ configuration as used for the dijet dataset (see List.~\ref{config-lhc}).
Tab.~\ref{tab:dijet_pair_candidates} lists the three SR models, each obtained from a fit initialized with a different random seed.
The $\chi^2/\text{NDF}$ scores improve significantly after the ROF step compared to the original functions returned by $\tt{PySR}$.
The three background models fit the blinded pseudodata well, as shown in Fig~\ref{fig:dijet_pair_sampling} for the total uncertainty coverage and Fig.~\ref{fig:dijet_pair_blinded} for a comparison with the empirical model used by CMS.

\begin{table*}[!t]
\caption{The candidate functions are obtained from three fits using different random seeds, fitted to the pseudodata of the four-jet spectrum with the (injected) signal region blinded.
The fits were performed on a scaled dataset (to enhance fit stability and prevent numerical overflow), and the functions can be transformed back to describe the original spectrum using the transformation: $x\rightarrow 0.000136(x - 1568.5)$.
These functions are plotted and compared with the blinded pseudodata in Fig.~\ref{fig:dijet_pair_blinded}.
Numerical values are rounded to three significant figures for display purposes.}
\label{tab:dijet_pair_candidates}
\centering
\resizebox{\textwidth}{!}{
\begin{tabular}{c|l|c|c|c|c}\hline
    & \textbf{Candidate function} & \textbf{\# param.} & \textbf{$\chi^2/\text{NDF}$} & \textbf{$\chi^2/\text{NDF}$} & \textbf{p-value} \\ 

    & (after ROF) & & (before ROF) & (after ROF) & (after ROF) \\ \hline
    &  & & & & \\
    
    SR model 1 & $(1.13\times 10^{-5}x)^{1.28x}/(0.143x^x + 1.63x)$ & 3 & 47.95 / 34 = 1.41 & 39.07 / 34 = 1.149 & 0.2524 \\
    &  & & & & \\ \hline
    &  & & & & \\
    
    SR model 2 & $6.98x^{0.857x}(x + \exp(x))^{-11.8}$ & 3 & 47.36 / 34 = 1.393 & 39.83 / 34 = 1.171 & 0.2267 \\
    &  & & & & \\ \hline
    &  & & & & \\

    SR model 3 & $((6.52\times 10^{-5} + 0.000378x)\tanh(0.641 + $ & 3 & 71.24 / 34 = 2.095 & 35.57 / 34 = 1.046 & 0.3942 \\
    & $3x))^{x + \tanh(x)}/\tanh(0.145 + x)$ & & & & \\
    &  & & & & \\\hline
    
    \end{tabular}%
    }
\end{table*}

\begin{figure}[!t]
     \centering
     \begin{subfigure}[b]{0.495\textwidth}
         \centering
         \includegraphics[width=\textwidth]{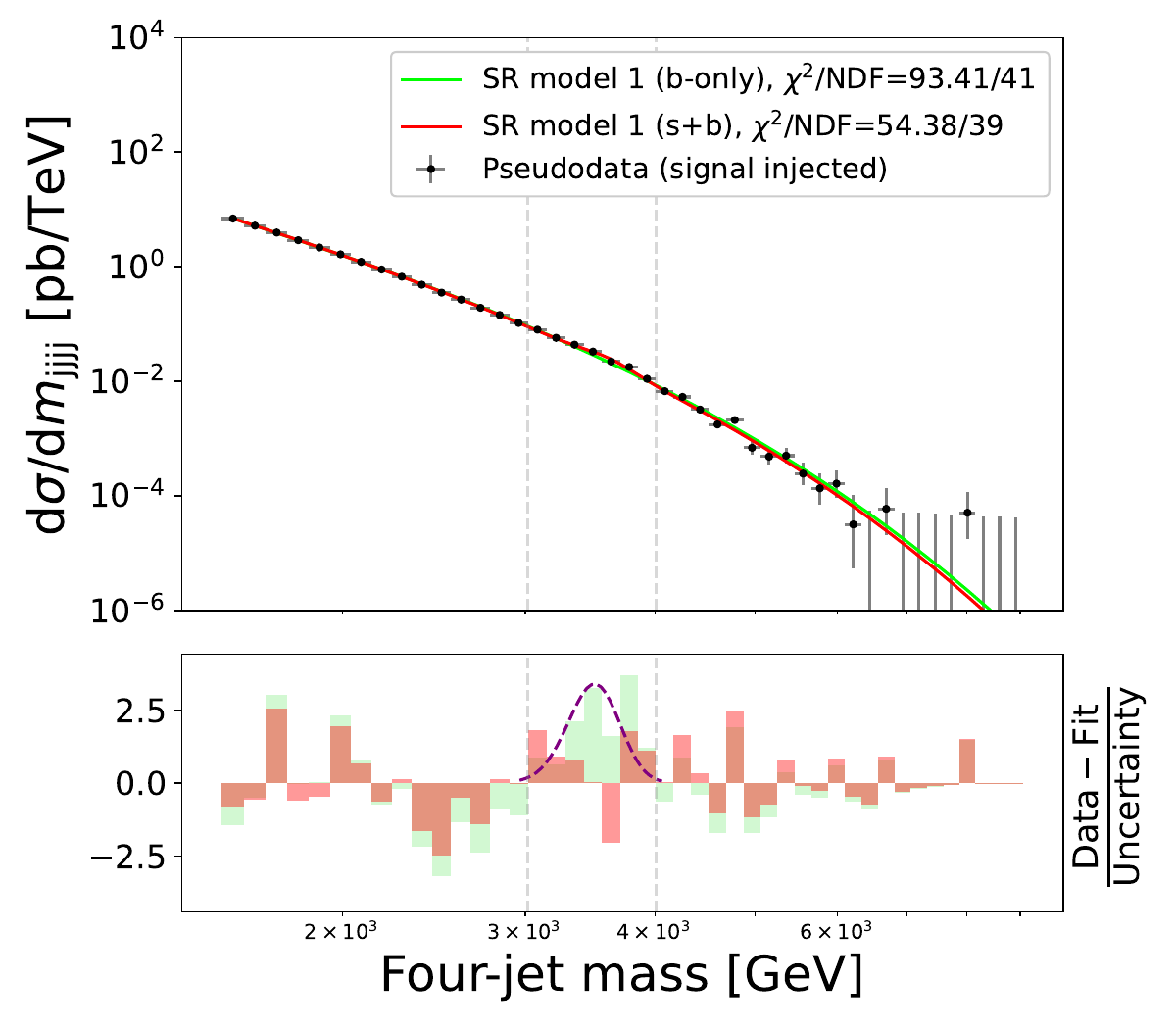}
         \caption{SR model 1.}
     \end{subfigure}
     \begin{subfigure}[b]{0.495\textwidth}
         \centering
         \includegraphics[width=\textwidth]{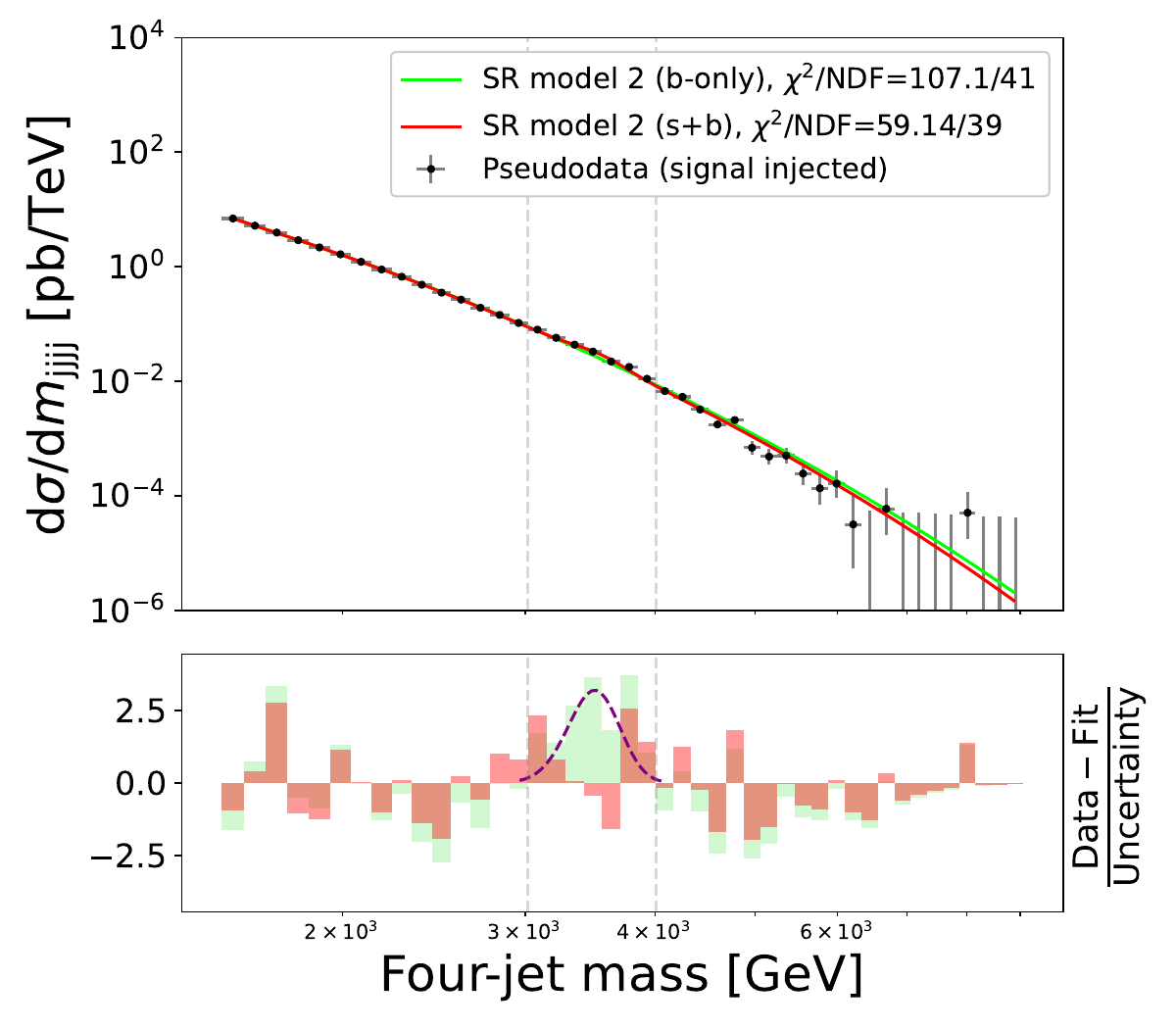}
         \caption{SR model 2.}
     \end{subfigure}\\\vspace{0.5cm}
     \begin{subfigure}[b]{0.495\textwidth}
         \centering
         \includegraphics[width=\textwidth]{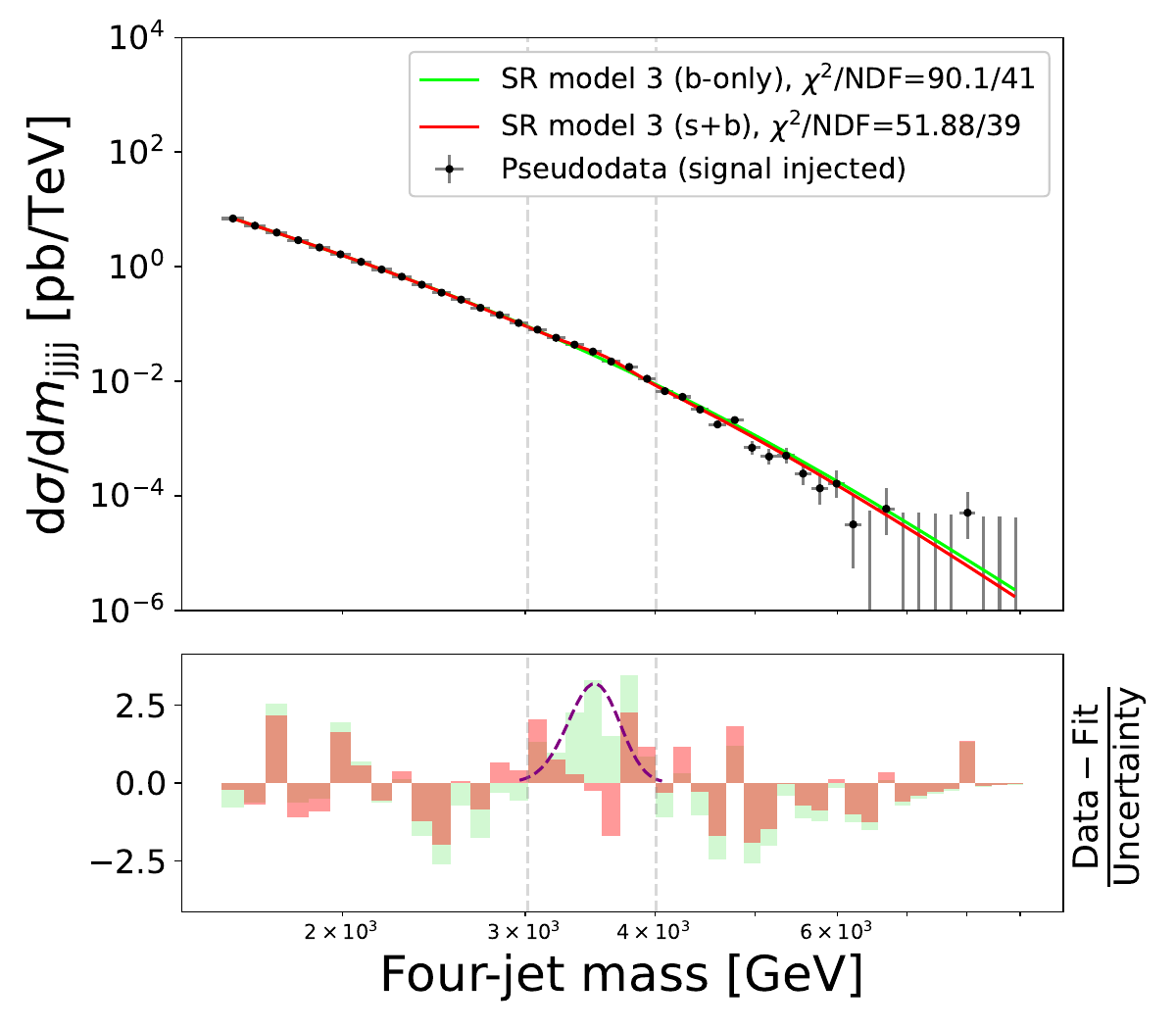}
         \caption{SR model 3.}
     \end{subfigure}
     \begin{subfigure}[b]{0.495\textwidth}
         \centering
         \includegraphics[width=\textwidth]{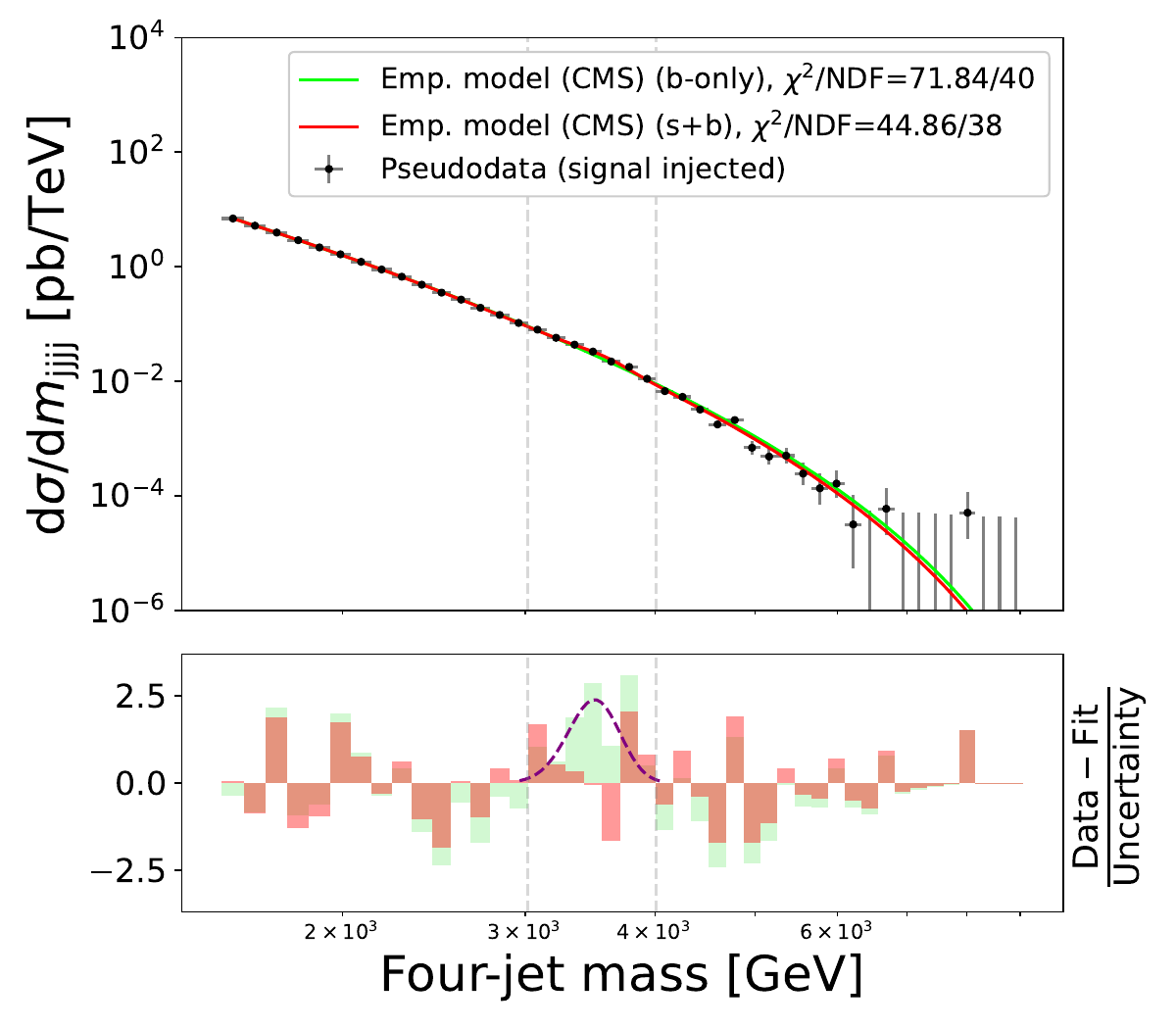}
         \caption{Empirical model (CMS).}
     \end{subfigure}\vspace{0.5cm}
     
    \caption{Comparison of the b-only fits and the s+b fits to the unblinded pseudodata of the four-jet spectrum. The lower panel shows the residual error per bin, measured in units of the data uncertainty. The shape of the injected signal is also shown.}
    \label{fig:dijet_pair_unblinded}
\end{figure}

Next, we unblind the pseudodata and perform b-only fits and s+b fits on the full pseudodata spectrum.
These results are shown in Fig.\ref{fig:dijet_pair_unblinded}.
In all three SR models, as well as the CMS empirical model, the excess of events over the background around the injected signal location observed in the b-only fits is reduced in the s+b fits, demonstrating that the models are sensitive to the injected signal
Tab.~\ref{tab:dijet_pair_chi2} lists the $\chi^2/\text{NDF}$ scores for each model, showing the fit performance in response to the presence of the injected signal.

\begin{figure}[!t]
     \centering
     \begin{subfigure}[b]{0.6\textwidth}
         \centering
         \includegraphics[width=\textwidth]{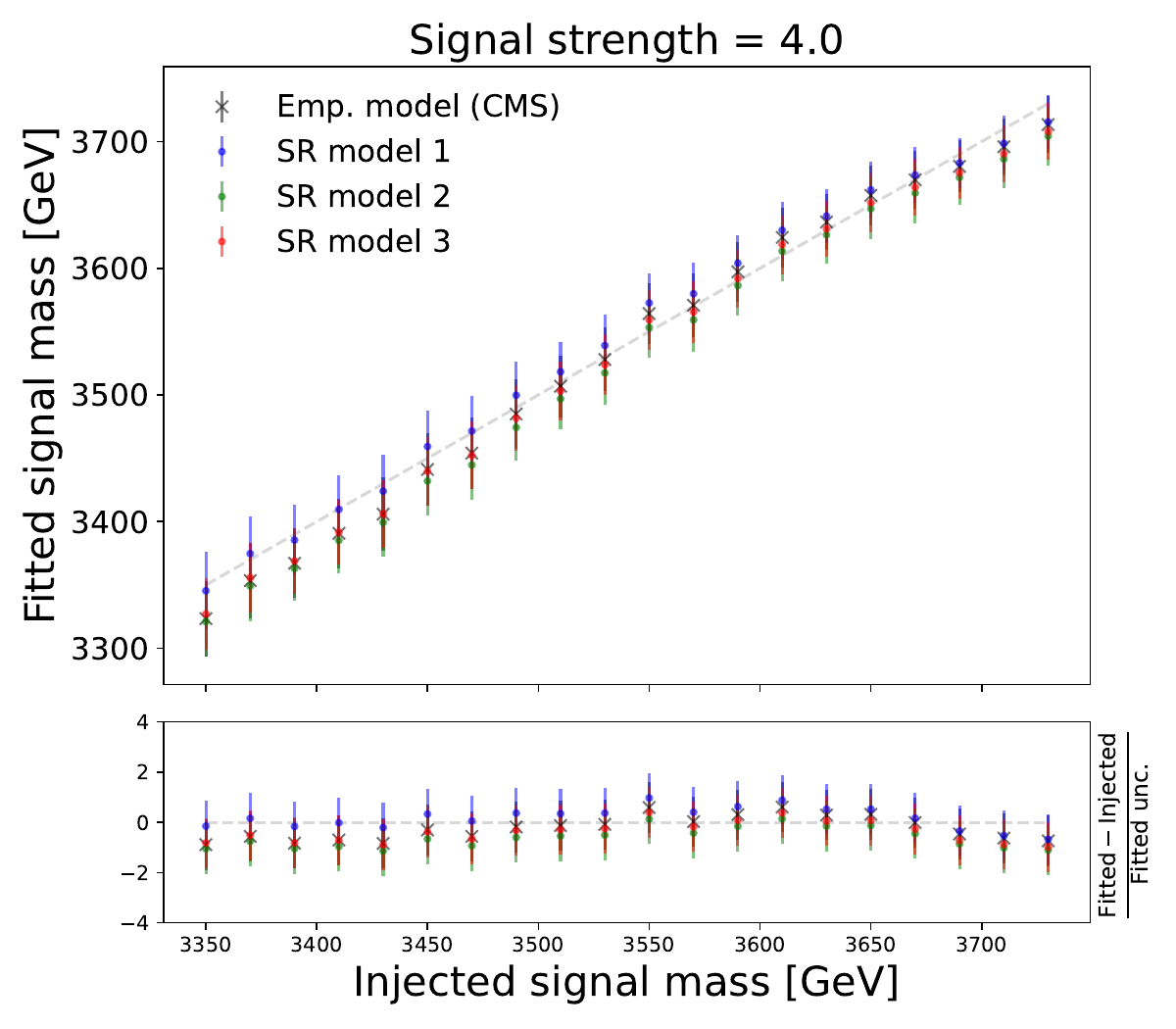}
         \caption{Fitted vs. injected signal mass at a specified signal strength value.}
     \end{subfigure}\\\vspace{0.5cm}
     \begin{subfigure}[b]{0.6\textwidth}
         \centering
         \includegraphics[width=\textwidth]{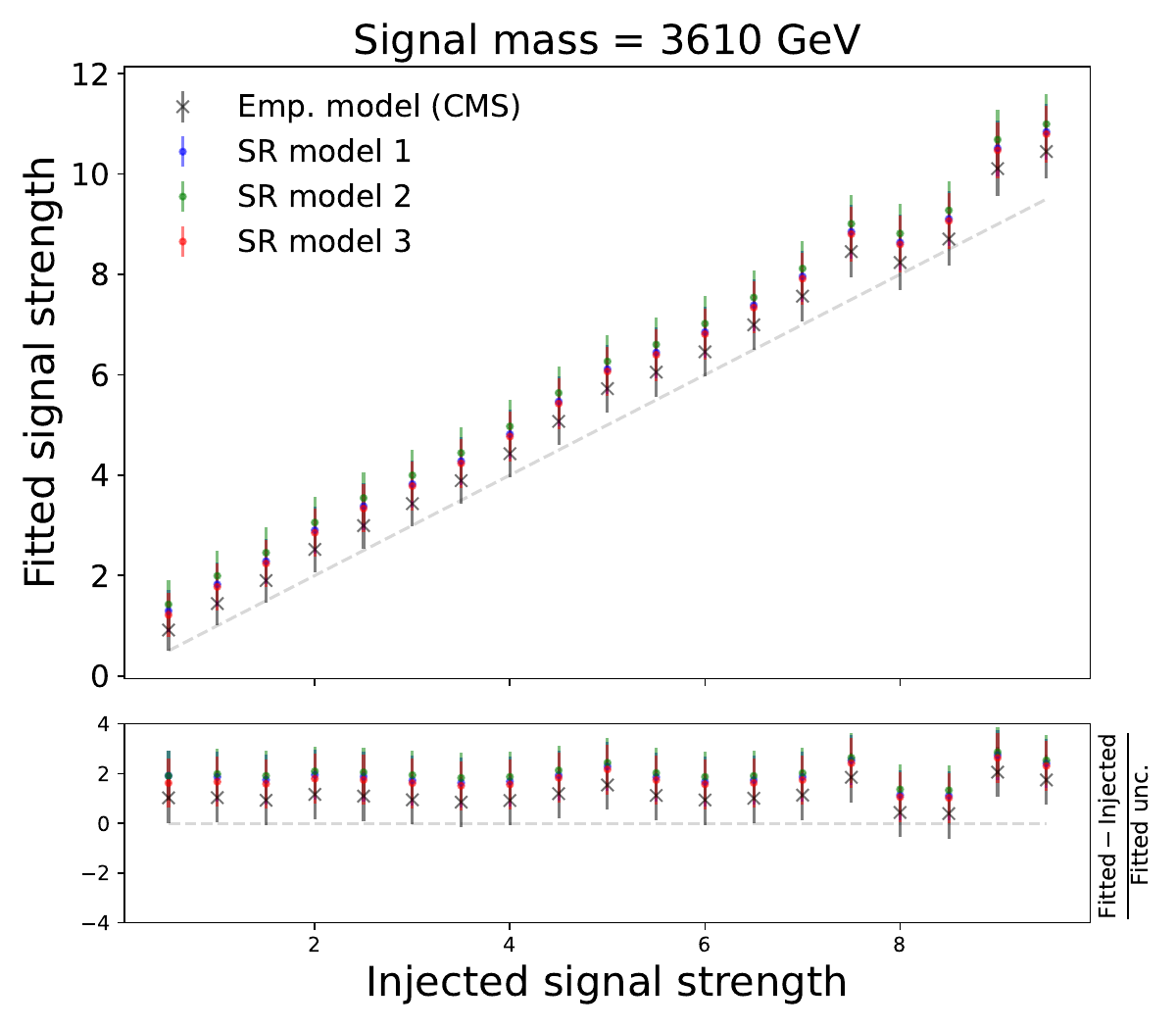}
         \caption{Fitted vs. injected signal strength at a specified signal mass value.}
     \end{subfigure}
     
    \caption{Fitted values vs. the true values of parameters of the injected signal in the paired-dijet dataset. The bottom panels show the residual error in units of the fitted uncertainty.}
    \label{fig:dijet_pair_scan}
\end{figure}

To assess whether the SR models can accurately extract the injected signals, we generate multiple sets of pseudodata by injecting Gaussian signals with different mean values ranging from 3350 to 3750 GeV and varying signal strength between 0.5 and 10.
We then perform the s+b fits to extract the corresponding signal parameters.
Fig.~\ref{fig:dijet_pair_scan} shows the extracted signal parameters plotted against their injected values.
All three SR models are capable of extracting the correct signal parameter values within reasonable uncertainties and are comparable to the empirical model used by CMS.

\begin{table*}[!t]
\caption{Comparison of the $\chi^2/\text{NDF}$ scores from three types of fits to the paired-dijet dataset: the b-only fits to the blinded pseudodata, b-only fits to the unblinded pseudodata, and s+b fits to the unblinded pseudodata. The background models used for the fits are listed in Tab.~\ref{tab:dijet_pair_candidates}, and the fits are shown in 
Fig.~\ref{fig:dijet_pair_blinded} (blinded) and Fig.~\ref{fig:dijet_pair_unblinded} (unblinded).}
\label{tab:dijet_pair_chi2}
\centering
\resizebox{\textwidth}{!}{
\begin{tabular}{l|c|c|c}\hline
    & \textbf{$\chi^2/\text{NDF}$ (b-only, blinded)} & \textbf{$\chi^2/\text{NDF}$ (b-only, unblinded)} & \textbf{$\chi^2/\text{NDF}$ (s+b, unblinded)} \\ \hline
    
    SR model 1 & 39.07 / 34 = 1.149 & 93.41 / 41 = 2.278 & 54.38 / 39 = 1.394 \\
    & & & \\ \hline

    SR model 2 & 39.83 / 34 = 1.171 & 107.1 / 41 = 2.612 & 59.14 / 39 = 1.516 \\
    & & & \\ \hline

    SR model 3 & 35.57 / 34 = 1.046 & 90.1 / 41 = 2.198 & 51.88 / 39 = 1.33 \\
    & & & \\ \hline

    Emp. model (CMS) & 32.5 / 33 = 0.985 & 71.84 / 40 = 1.796 & 44.86 / 38 = 1.181 \\
    & & & \\ \hline
    
    \end{tabular}%
    }
\end{table*}

\clearpage

\subsection{CMS high-mass dimuon dataset (1D) [background modeling]}
\label{sec:dimuon}

CMS performed a search for high-mass dimuon resonances using proton-proton collision data at a center-of-mass energy of $\sqrt{s}=13$ TeV and reported no significant deviations from the Standard Model prediction~\cite{dimuon}.
The dataset for the dimuon spectrum is publicly available on HEPDATA at Ref.~\cite{hepdata.dimuon}.
In the analysis, CMS considered three different functions to model the background contribution in the distribution of the dimuon invariant mass, $m_{\mu\mu}$.
These functions include a simple exponential, a power-law, and a first-order Bernstein polynomial.
Since the dimuon distribution in the signal region is statistically limited, simpler functions are preferred to avoid over-fitting the background.
For our comparison, we take the first-order Bernstein polynomial as the empirical model used by CMS.

We perform the same experiments conducted on the dijet dataset, as detailed in Sec.~\ref{sec:dijet}.
Starting from the original dimuon spectrum, we generate pseudodata by injecting a Gaussian signal centered at $m_{\mu\mu}=500$ GeV ($s_1$), with a width of 20 GeV ($2s_2$) and a signal strength of $s_0=350$.
To model the background, we blind the signal region by masking the $m_{\gamma\gamma}$ bins between 450 and 550 GeV in the pseudodata and perform the fits.

\begin{figure}[!t]
     \centering
     \begin{subfigure}[b]{0.495\textwidth}
         \centering
         \includegraphics[width=\textwidth]{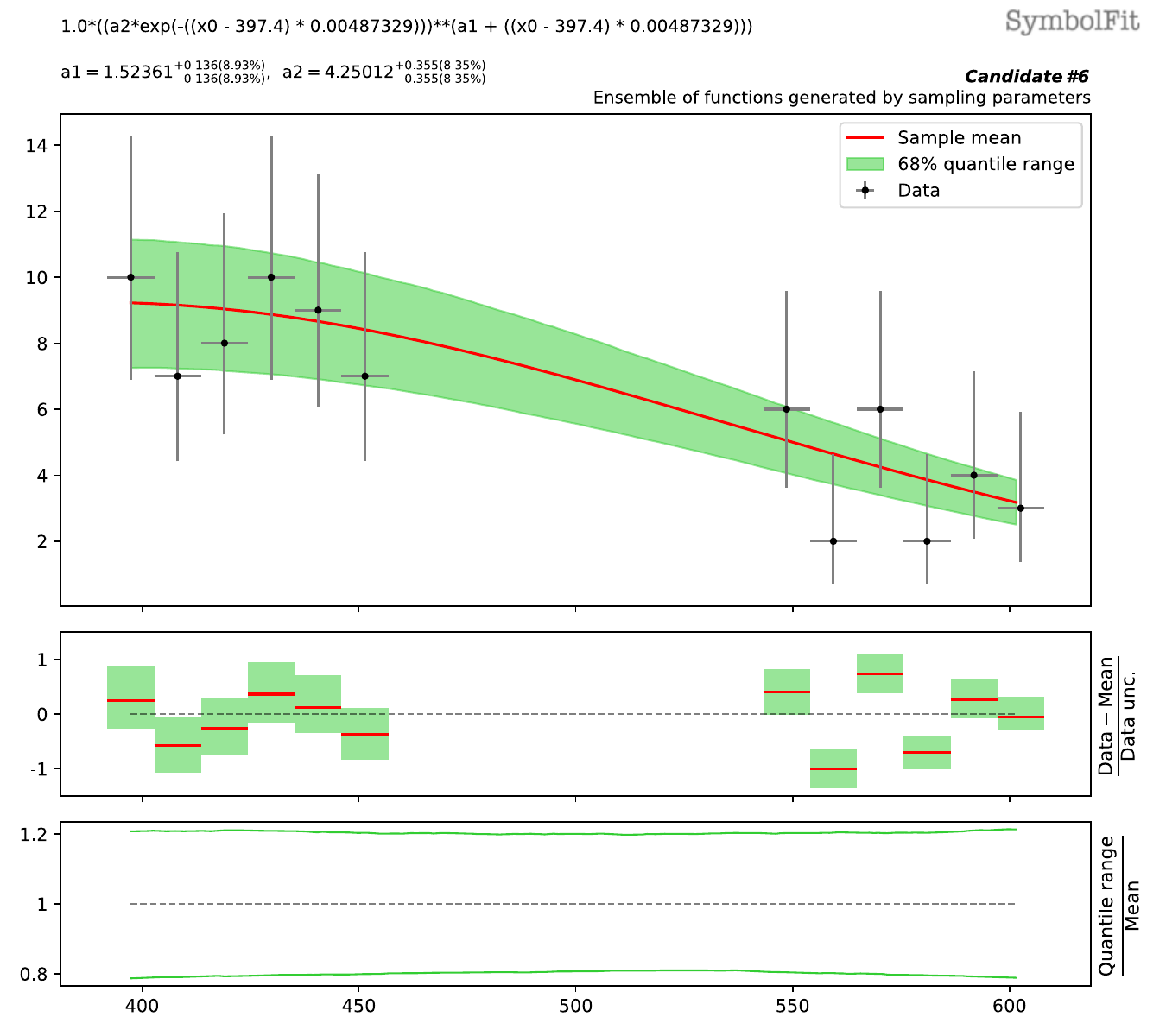}
         \caption{SR model 1.}
     \end{subfigure}
     \begin{subfigure}[b]{0.495\textwidth}
         \centering
         \includegraphics[width=\textwidth]{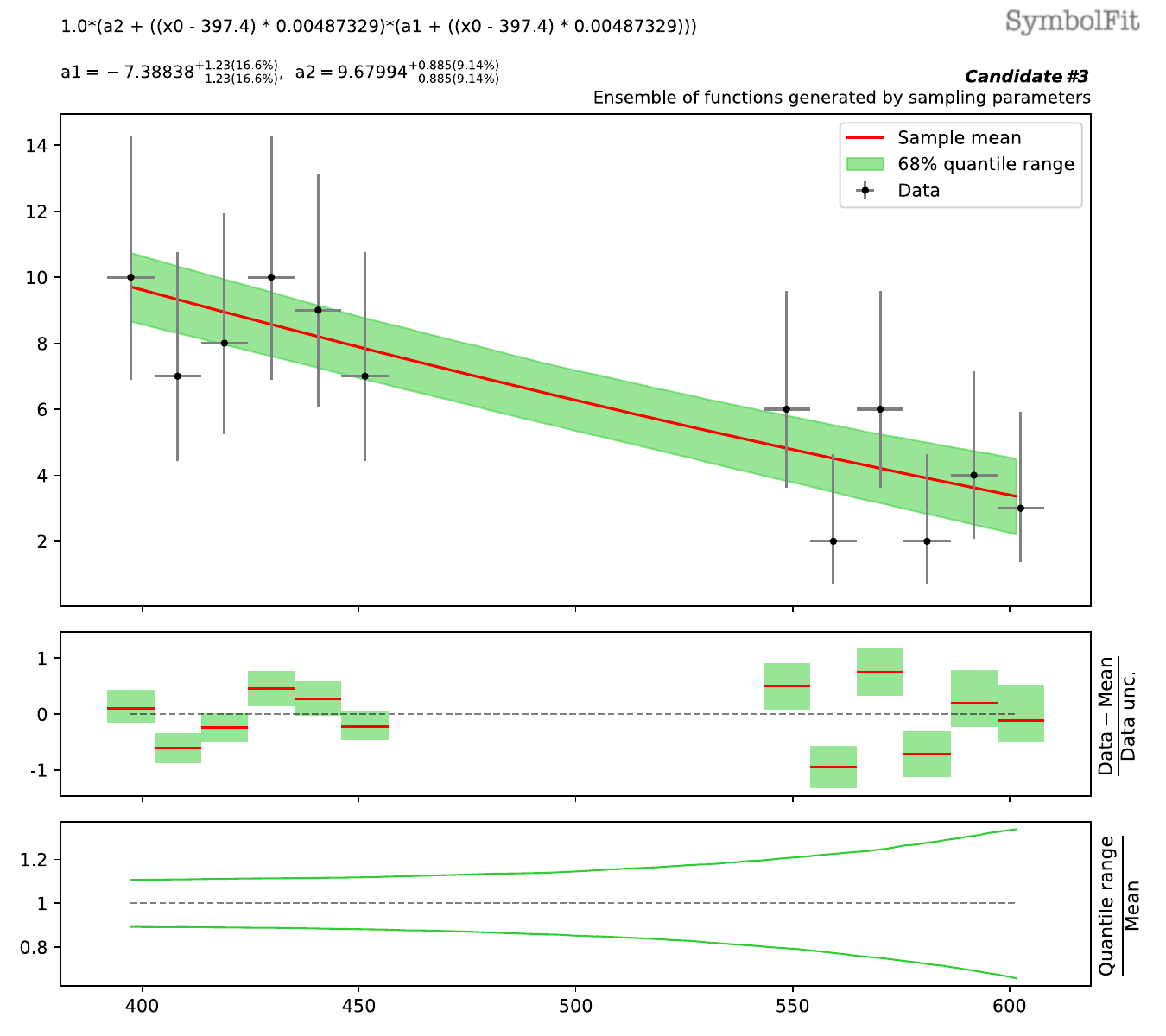}
         \caption{SR model 2.}
     \end{subfigure}\\\vspace{0.5cm}
     \begin{subfigure}[b]{0.495\textwidth}
         \centering
         \includegraphics[width=\textwidth]{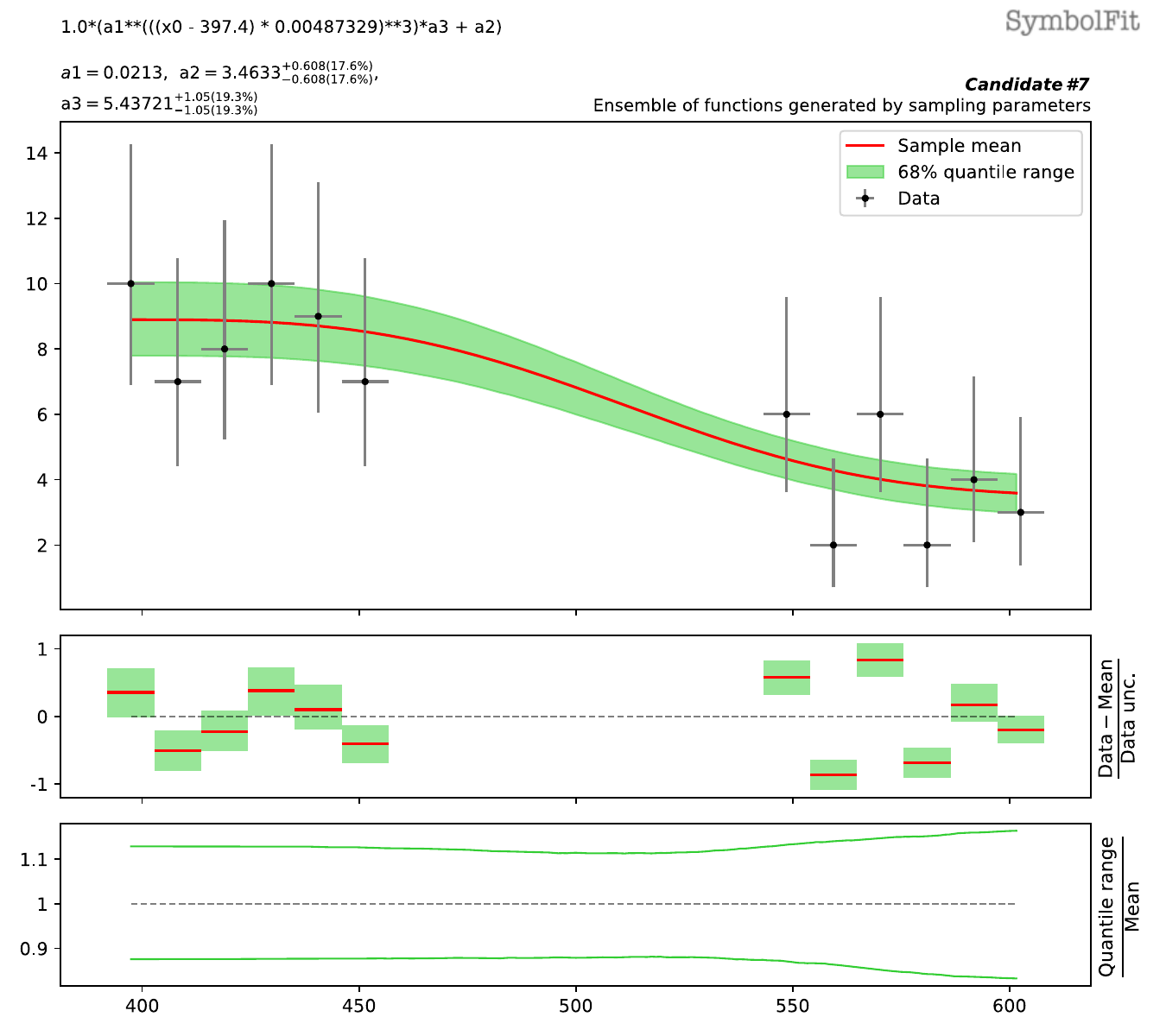}
         \caption{SR model 3.}
     \end{subfigure}\vspace{0.5cm}
     
    \caption{The three SR models fitted to the pseudodata of the dimuon spectrum with the signal region blinded (see Tab.~\ref{tab:dimuon_candidates}).
        To visualize the total uncertainty coverage of each candidate function, the green band in each subfigure represents the 68\% quantile range of functions obtained by sampling parameters, taking into account the best-fit values and the covariance matrix within a multidimensional normal distribution.
        The red line denotes the mean of the function ensemble.
        At the top of each subfigure, the candidate function and the fitted parameters are shown.
        The middle panel shows the weighted residual error: $\frac{\text{Data}-\text{Mean}}{\text{Data unc.}}$.
        The bottom panel shows the ratio of the 68\% quantile range to the mean.}
    \label{fig:dimuon_sampling}
\end{figure}

\begin{figure}[!t]
     \centering
     \begin{subfigure}[b]{1\textwidth}
         \centering
         \includegraphics[width=\textwidth]{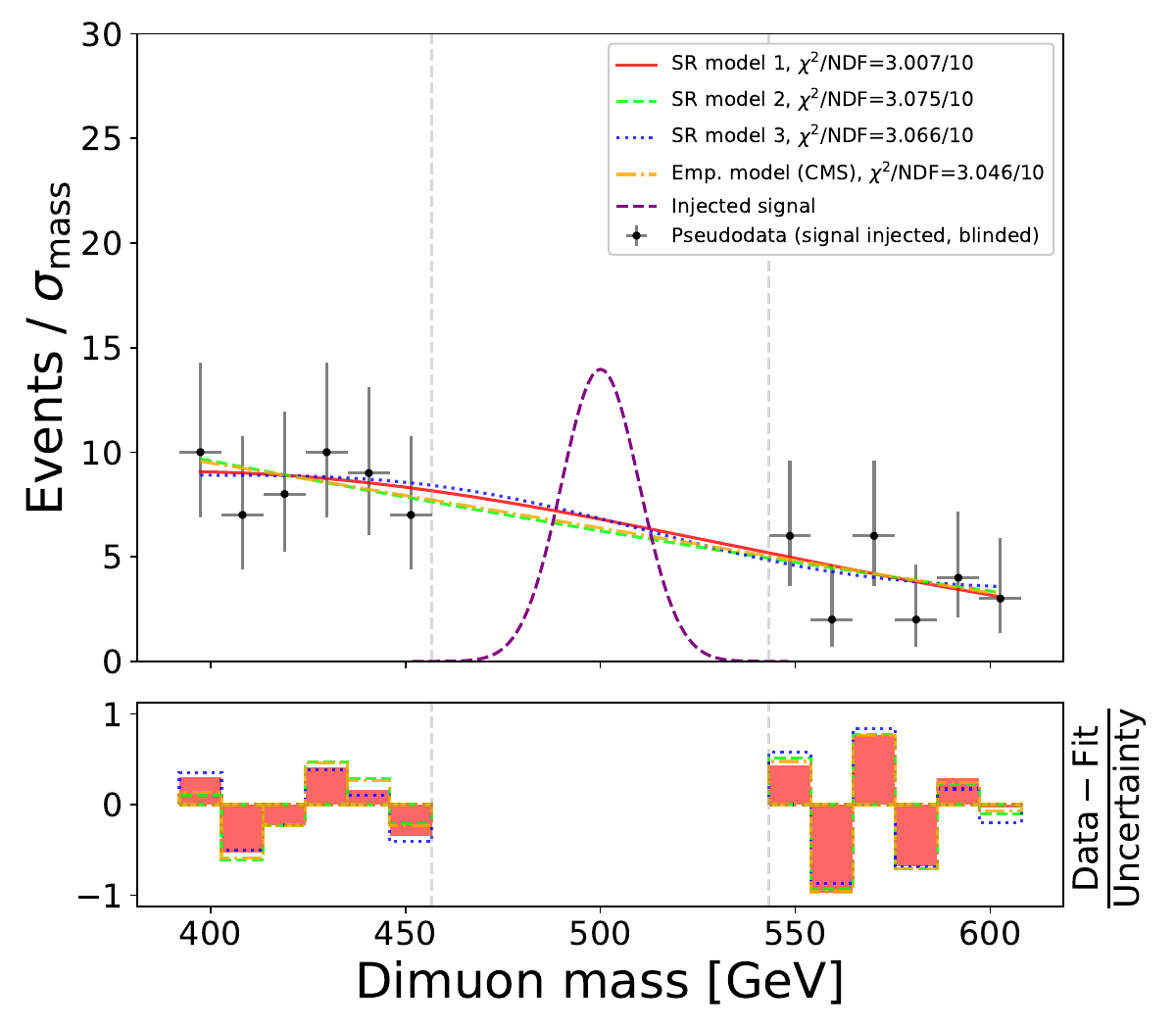}
     \end{subfigure}
     
    \caption{Pseudodata of the dimuon spectrum with the injected signal shown in the blinded signal region. The three SR models (see Tab.~\ref{tab:dimuon_candidates}) are compared against the empirical model used by CMS. The lower panel shows the residual error per bin, measured in units of the data uncertainty.}
    \label{fig:dimuon_blinded}
\end{figure}

Three $\tt{SymbolFit}$ runs using different random seeds are carried out, applying the same $\tt{PySR}$ configuration as used for the dijet dataset (see List.~\ref{config-lhc}), except that the maximum complexity is set at 20 instead of 80, since the diphoton distribution shape is less complex.
Tab.~\ref{tab:dimuon_candidates} lists the three SR models, each obtained from a fit initialized with a different random seed.
The $\chi^2/\text{NDF}$ scores improve significantly after the ROF step compared to the original functions returned by $\tt{PySR}$.
The three background models fit the blinded pseudodata well, as shown in Fig.~\ref{fig:dimuon_sampling} for the total uncertainty coverage and Fig.~\ref{fig:dimuon_blinded} for a comparison with the empirical model used by CMS.

\begin{table*}[!t]
\caption{The candidate functions are obtained from three fits using different random seeds, fitted to the pseudodata of the dimuon spectrum with the (injected) signal region blinded.
The fits were performed on a scaled dataset (to enhance fit stability and prevent numerical overflow), and the functions can be transformed back to describe the original spectrum using the transformation: $x\rightarrow 0.00487(x - 397.4)$.
These functions are plotted and compared with the blinded pseudodata in Fig.~\ref{fig:dimuon_blinded}.
Numerical values are rounded to three significant figures for display purposes.}
\label{tab:dimuon_candidates}
\centering
\resizebox{\textwidth}{!}{
\begin{tabular}{c|l|c|c|c|c}\hline
    & \textbf{Candidate function} & \textbf{\# param.} & \textbf{$\chi^2/\text{NDF}$} & \textbf{$\chi^2/\text{NDF}$} & \textbf{p-value} \\ 

    & (after ROF) & & (before ROF) & (after ROF) & (after ROF) \\ \hline
    &  & & & & \\
    SR model 1 & $(4.25\exp(-x))^{1.52 + x}$ & 2 & 3.469 / 10 = 0.3469 & 3.007 / 10 = 0.3007 & 0.9813 \\
    & $ $ & & & & \\ \hline
    &  & & & & \\
    SR model 2 & $9.68 + x(-7.39 + x)$ & 2 & 3.484 / 10 = 0.3484 & 3.075 / 10 = 0.3075 & 0.9796 \\
    & $ $ & & & & \\ \hline
    &  & & & & \\
    SR model 3 & $0.0213^{x^3}5.44 + 3.46$ & 2 & 3.456 / 10 = 0.3456 & 3.066 / 10 = 0.3066 & 0.9798 \\
    & $ $ & & & & \\ \hline
    
    \end{tabular}%
    }
\end{table*}

\begin{figure}[!t]
     \centering
     \begin{subfigure}[b]{0.495\textwidth}
         \centering
         \includegraphics[width=\textwidth]{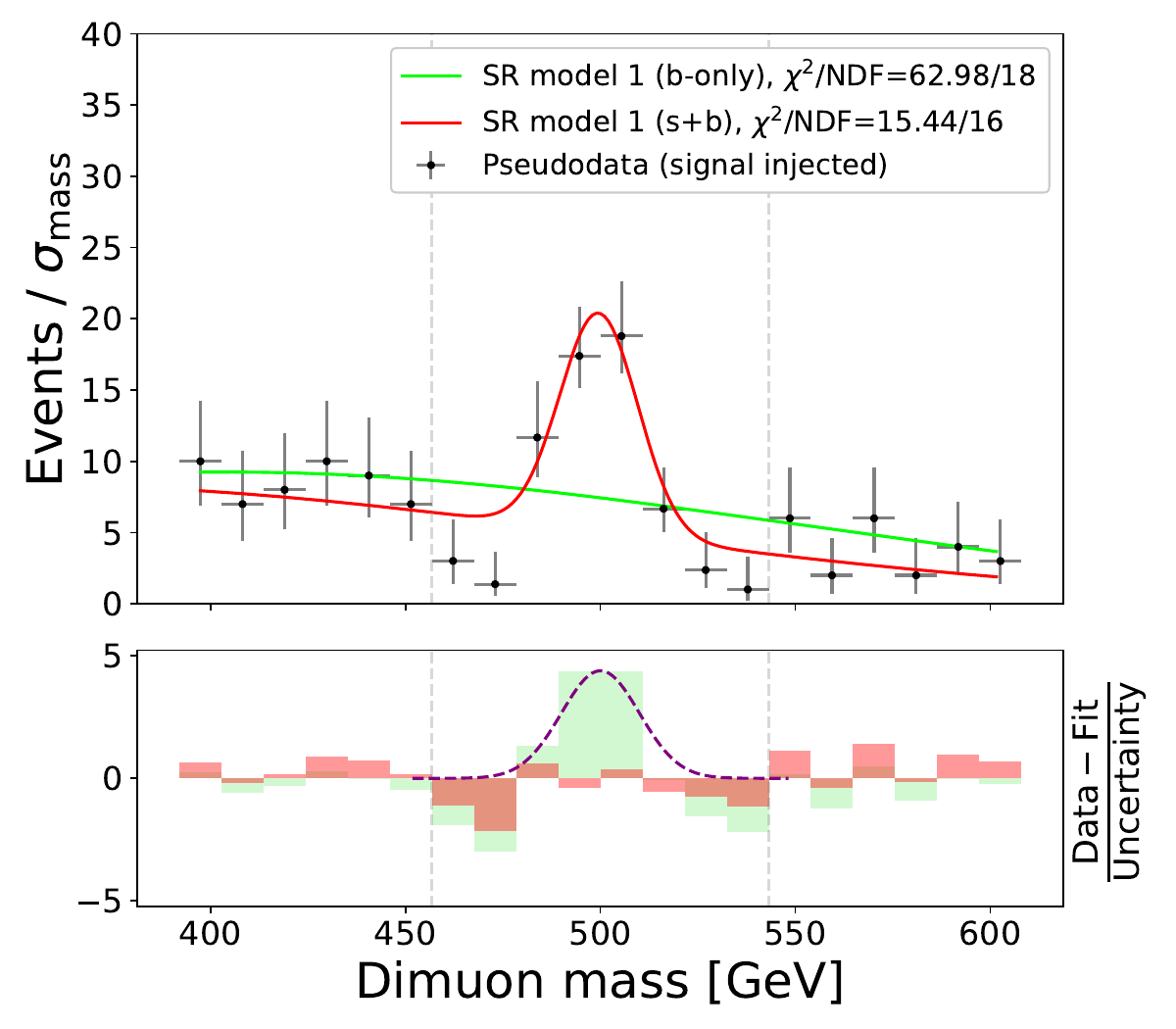}
         \caption{SR model 1.}
     \end{subfigure}
     \begin{subfigure}[b]{0.495\textwidth}
         \centering
         \includegraphics[width=\textwidth]{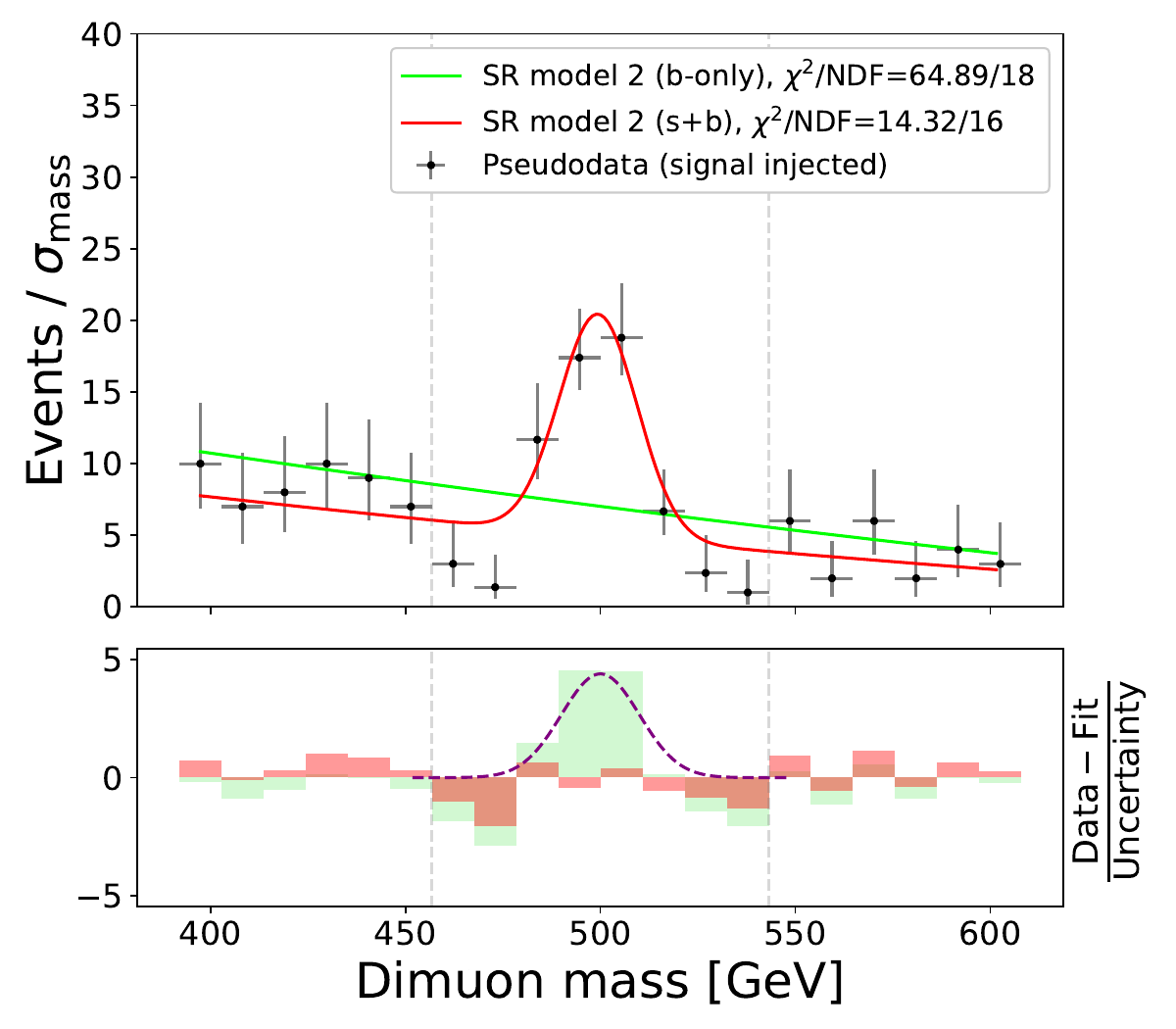}
         \caption{SR model 2.}
     \end{subfigure}\\\vspace{0.5cm}
     \begin{subfigure}[b]{0.495\textwidth}
         \centering
         \includegraphics[width=\textwidth]{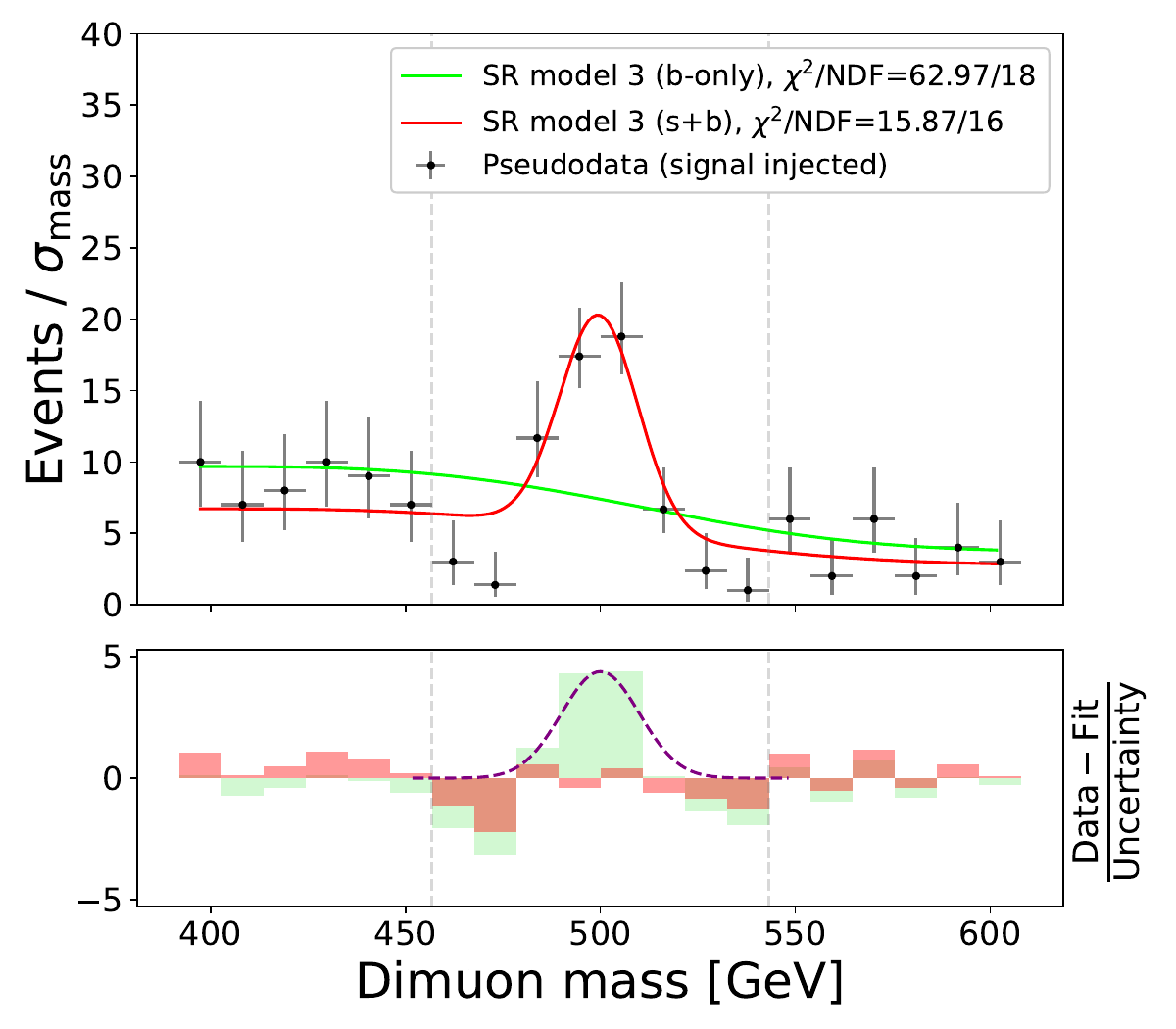}
         \caption{SR model 3.}
     \end{subfigure}
     \begin{subfigure}[b]{0.495\textwidth}
         \centering
         \includegraphics[width=\textwidth]{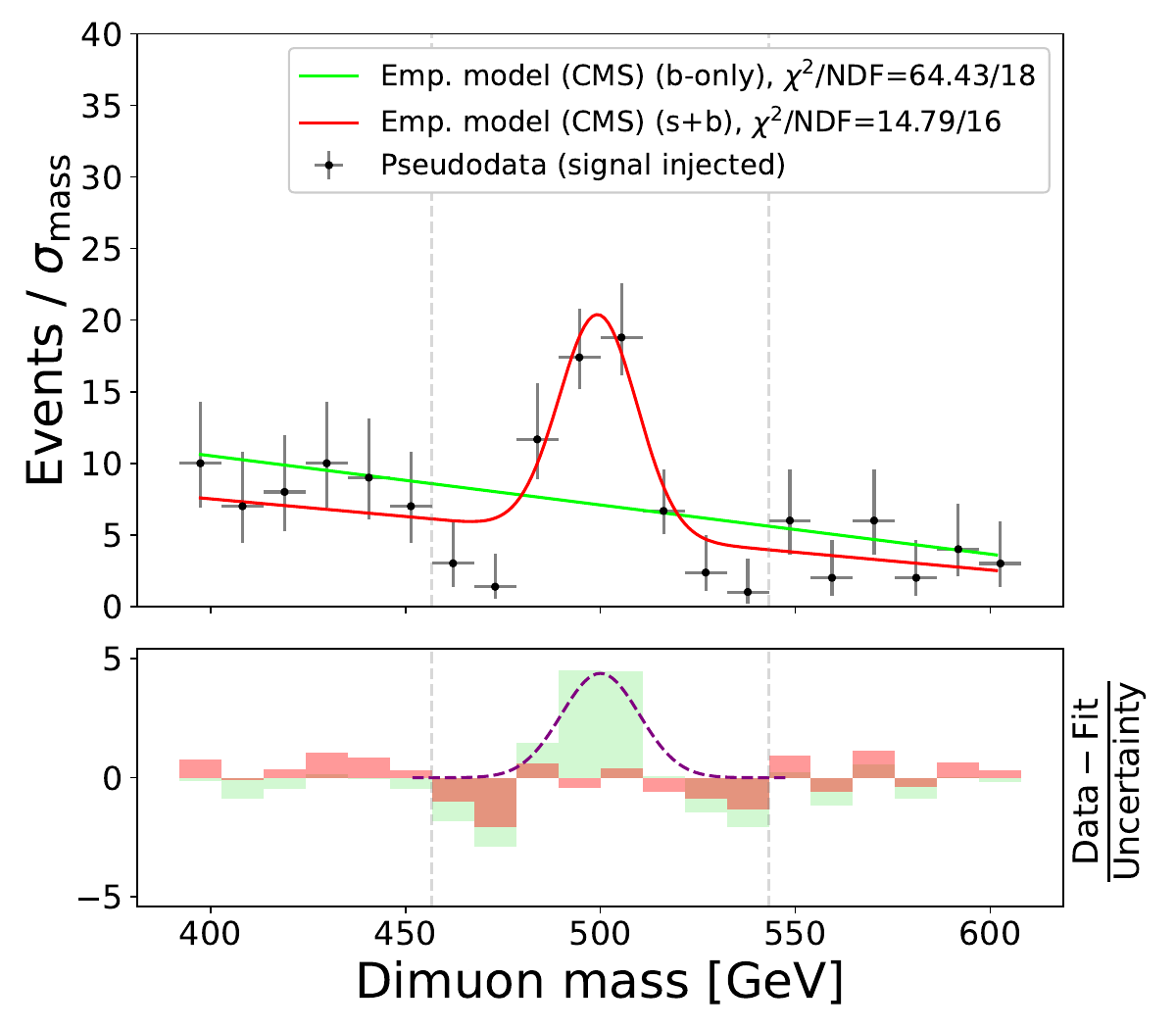}
         \caption{Empirical model (CMS).}
     \end{subfigure}\vspace{0.5cm}
     
    \caption{Comparison of the b-only fits and the s+b fits to the unblinded pseudodata of the dimuon spectrum. The lower panel shows the residual error per bin, measured in units of the data uncertainty. The shape of the injected signal is also shown.}
    \label{fig:dimuon_unblinded}
\end{figure}

Next, we unblind the pseudodata and perform b-only fits and s+b fits on the full pseudodata spectrum.
These results are shown in Fig.\ref{fig:dimuon_unblinded}.
In all three SR models, as well as the CMS empirical model, the excess of events over the background around the injected signal location observed in the b-only fits is reduced in the s+b fits, demonstrating that the models are sensitive to the injected signal.
Tab.~\ref{tab:dimuon_chi2} lists the $\chi^2/\text{NDF}$ scores for each model, showing the fit performance in response to the presence of the injected signal.

\begin{figure}[!t]
     \centering
     \begin{subfigure}[b]{0.6\textwidth}
         \centering
         \includegraphics[width=\textwidth]{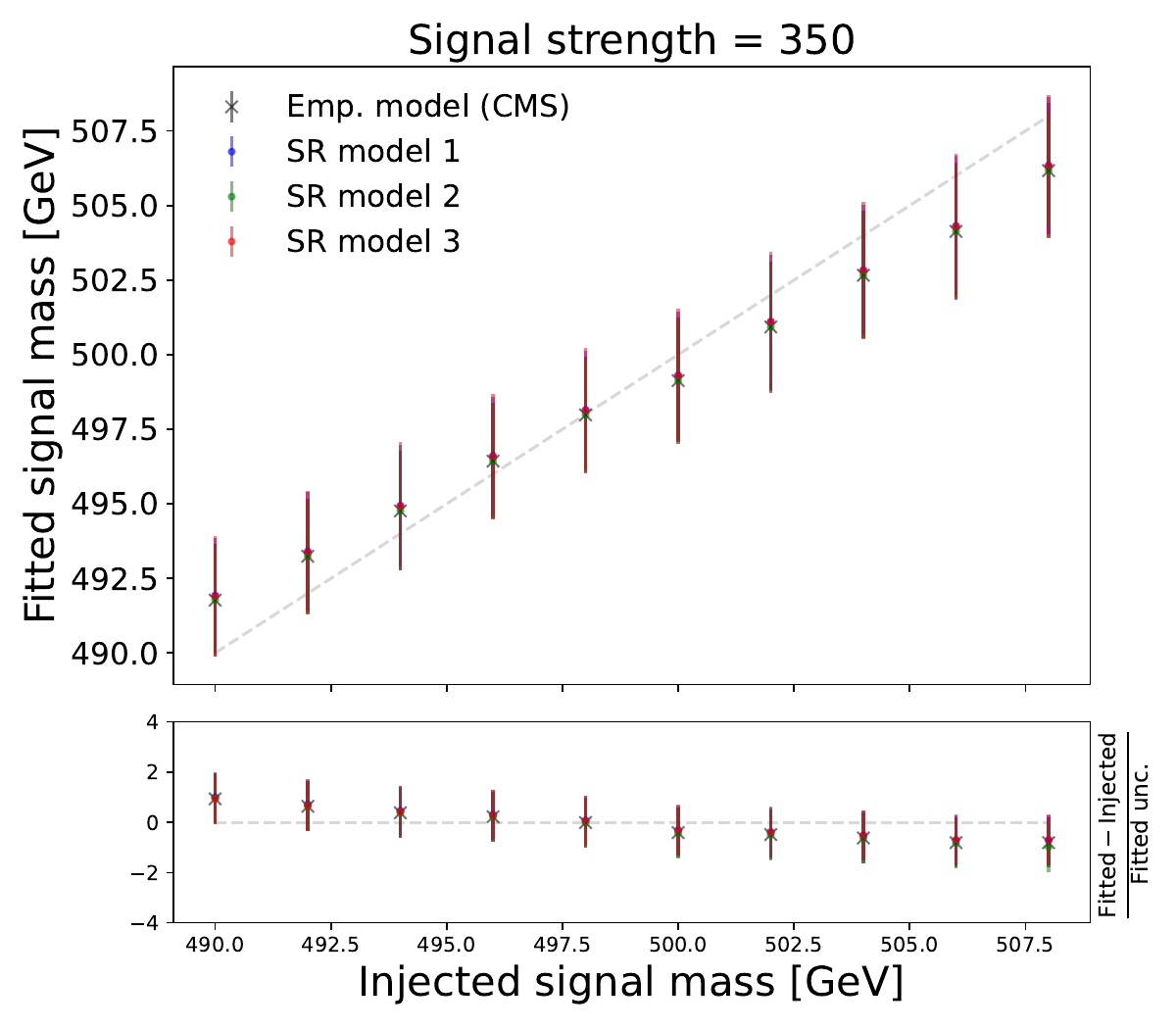}
         \caption{Fitted vs. injected signal mass at a specified signal strength.}
     \end{subfigure}\vspace{0.5cm}
     \begin{subfigure}[b]{0.6\textwidth}
         \centering
         \includegraphics[width=\textwidth]{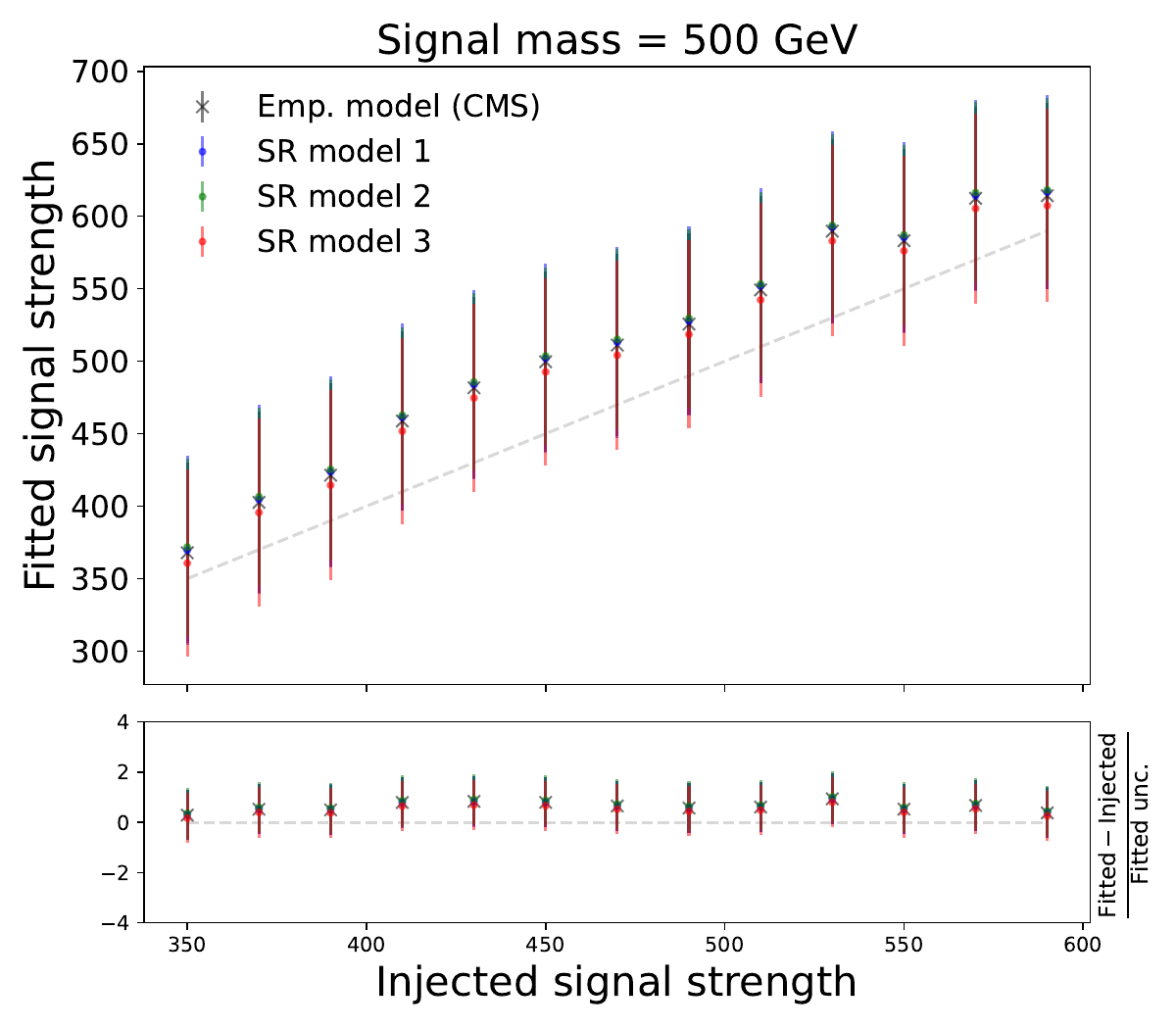}
         \caption{Fitted vs. injected signal strength at a specified signal mass value.}
     \end{subfigure}
     
    \caption{Fitted values vs. the true values of parameters of the injected signal in the dimuon dataset.
    The bottom panels show the residual error in units of the fitted uncertainty.}
    \label{fig:dimuon_scan}
\end{figure}

To assess whether the SR models can accurately extract the injected signals, we generate multiple sets of pseudodata by injecting Gaussian signals with different mean values ranging from 490 to 510 GeV and varying signal strength between 350 and 600.
We then perform the s+b fits to extract the corresponding signal parameters.
Fig.~\ref{fig:dimuon_scan} shows the extracted signal parameters plotted against their injected values.
All three SR models are capable of extracting the correct signal parameter values within reasonable uncertainties and are comparable to the empirical model used by CMS.

\begin{table*}[!t]
\caption{Comparison of the $\chi^2/\text{NDF}$ scores from three types of fits to the dimuon dataset: the b-only fits to the blinded pseudodata, b-only fits to the unblinded pseudodata, and s+b fits to the unblinded pseudodata. The background models used for the fits are listed in Tab.~\ref{tab:dimuon_candidates}, and the fits are shown in Fig.~\ref{fig:dimuon_blinded} (blinded) and Fig.~\ref{fig:dimuon_unblinded} (unblinded).}
\label{tab:dimuon_chi2}
\centering
\resizebox{\textwidth}{!}{
\begin{tabular}{l|c|c|c}\hline
    & \textbf{$\chi^2/\text{NDF}$ (b-only, blinded)} & \textbf{$\chi^2/\text{NDF}$ (b-only, unblinded)} & \textbf{$\chi^2/\text{NDF}$ (s+b, unblinded)} \\ \hline
    
    SR model 1 & 3.007 / 10 = 0.3007 & 62.98 / 18 = 3.499 & 15.44 / 16 = 0.965 \\
    & & & \\ \hline

    SR model 2 & 3.075 / 10 = 0.3075 & 64.89 / 18 = 3.605 & 14.32 / 16 = 0.895 \\
    & & & \\ \hline

    SR model 3 & 3.066 / 10 = 0.3066 & 62.97 / 18 = 3.498 & 15.87 / 16 = 0.9919 \\
    & & & \\ \hline

    Emp. model (CMS) & 3.046 / 10 = 0.3046 & 64.43 / 18 = 3.579 & 14.79 / 16 = 0.9244 \\
    & & & \\ \hline
    
    \end{tabular}%
    }
\end{table*}

\clearpage

\end{document}